\documentclass[a4paper,11pt]{article}
\usepackage[left=2.2cm,right=2.2cm,top=2cm,bottom=2.5cm]{geometry}
\usepackage{amssymb,amsmath,bm}
\usepackage{caption}
\usepackage{color}
\usepackage{slashed}
\usepackage{graphics}
\usepackage{graphicx}
\usepackage{cite}
\usepackage{xspace}
\usepackage{multirow}
\usepackage[utf8]{inputenc}
\usepackage{tabularx}
\usepackage[caption=false]{subfig}
\usepackage{hyperref}
\usepackage{url}
\usepackage{dsfont}
\usepackage{float} % pour le placement des images
\usepackage{cancel}
\usepackage[normalem]{ulem}
\usepackage{enumitem}

\newcommand{\refapp}[1]{Appendix~\ref{app:#1}}
\newcommand{\reffig}[1]{Fig.~\ref{fig:#1}}
\newcommand{\refeq}[1]{Eq.~(\ref{eq:#1})}

\newcommand{\refsec}[1]{Section~\ref{sec:#1}}
\newcommand{\refsubsec}[1]{Subsection~\ref{sec:#1}}
\newcommand{\reftab}[1]{Table~\ref{tab:#1}}
\newcommand{\GeV}{\,\text{GeV}}

\newcommand{\lamkin}{\lambda_{\text{kin}}}
\newcommand{\ccal}{{\cal C}^{\rm NP}}

\usepackage[dvipsnames]{xcolor}

\newcommand{\ab}[1]{{\color{Red}#1}}

\newcounter{TODO}

\begin{document}

\begin{flushright}

SI-HEP-2025-13\\
P3H-25-039 
\end{flushright}

%	\preprint{xx}

\vspace*{-2mm}
\begin{center}
	\fontsize{15}{20}\selectfont
	\bf
	\boldmath
 Impact on $L$-observables of a new combined analysis of $B_{d,s}\to K^{(*)}$ form factors
\end{center}

\vspace{2mm}

\begin{center}
	{Aritra Biswas$^{\,a}$, Nico Gubernari$^{\,b}$, Joaquim Matias$^{\,c}$ and Gilberto Tetlalmatzi-Xolocotzi$^{\,d}$}\\[5mm]
	{\it\small
		{$^{\, a,\, d}$} 
		Theoretische Physik 1, Center for Particle Physics Siegen (CPPS), Universit\"at Siegen, Walter-Flex-Str. 3, 57068 Siegen, Germany
		\\[2mm]
		{$^{\, b}$} 
		DAMTP, University of Cambridge, Wilberforce Road,\\
		Cambridge, CB3 0WA, United Kingdom
		\\[2mm]
		{$^{\, c}$} 
		Universitat Aut\`onoma de Barcelona, 08193 Bellaterra, Barcelona,\\
		Institut de F\'{i}sica d'Altes Energies (IFAE), The Barcelona Institute of Science and Technology, Campus UAB, 08193 Bellaterra (Barcelona)
		\\[2mm]
		E-mail: {$^{\,a}$\textnormal{\texttt{aritra1.biswas@gmail.com}}, 
        $^{\,b}$\textnormal{\texttt{nicogubernari@gmail.com}},\\
        $^{\,c}$\textnormal{\texttt{matias@ifae.es}}, 
        $^{\,d}$\textnormal{\texttt{gtx@physik.uni-siegen.de}}}
	}
\end{center}

\vspace{1mm}	
\begin{abstract}
\bigskip

\noindent
We explore the impact of a combined analysis of $B_{d,s}\to K^{(*)}$ form factors
on a set of $L$-observables. The $L$-observables are constructed from ratios of branching fractions in $B_{s}\to VV,PP,PV$ versus $B_d \to VV,PP,PV$ decays with $P=K^0,\bar{K}^0$ and $V=K^{*0},\bar{K}^{*0}$,  thereby partially reducing their hadronic uncertainties.
We show the change of the Standard Model predictions of the $L$-observables under different determinations of the ratio of the relevant form factors (with correlations) including lattice QCD data and a novel light-cone sum rule analysis. 
In addition, we provide precise results for all $B_{d,s} \to K^{(*)}$ form factors in machine-readable files.
We find that the inclusion of our up-to-date results, as well as the use or omission of lattice QCD data for the form factors, has a significant impact on the $L$-observables.  
We also discuss how the New Physics interpretation is affected by the updated form factors and present revised predictions for the mechanism identified in our analysis of $B \to VP$ decays, now employing more suitable new experimental observables defined in this paper.

\end{abstract}
%\maketitle

\section{Introduction}
\label{sec:intro}

$B$-Flavour anomalies in semileptonic decays have been observed systematically in angular observables of the 4-body decay distribution $B\to K^*(\to K\pi)\ell\ell$~\cite{LHCb:2020lmf,
CMS:2024atz,Belle:2016fev,
ATLAS:2018gqc,Descotes-Genon:2013vna,Capdevila:2017bsm, Descotes-Genon:2015uva}  and  in branching ratios of $b\to s \ell\ell$ governed decays~\cite{LHCb:2014cxe,LHCb:2016ykl,LHCb:2021zwz}. However, the signals of Lepton Flavour Universality Violation that were observed at LHCb in neutral decays, the so-called $R_K$ and $R_{K^*}$ observables~\cite{Hiller:2003js}, are, after the recent LHCb reanalysis of data~\cite{LHCb:2022vje,LHCb:2022qnv}, rather consistent with the Standard Model (SM) prediction. Instead, in charged decays governed by the transition $b\to c \ell\nu$, in the observables called $R_{D,D^*}$ a deviation from the SM prediction is still observed at a level of 3$\sigma$~\cite{BaBar:2012obs,
Belle:2017ilt,LHCb:2023zxo,LHCb:2023uiv}. The existence of these tensions with the SM calls for a more general search to identify other anomalies in $b\to s$ and/or $b\to d$
transitions beyond semileptonic decays.

%A natural expectation is to find anomalies in other $b\to s$ and possible $b\to d$ transitions.

In a series of papers \cite{Alguero:2020xca,Biswas:2023pyw,Biswas:2024bhn}  a new category of observables was proposed  to analyse a set of non-leptonic anomalies related to the U-spin related modes of $B_{d,s}\to \bar{K}K$ and $B_{d,s}\to \bar{K}^*K^*$ decays and possible explanations of these anomalies were proposed in \cite{Li:2022mtc,Alguero:2020xca,
Lizana:2023kei,Zheng:2024zjv}. These observables were called $L$-observables. These are  optimized ratios  constructed under the guidance of U-spin symmetry to reduce part of the hadronic sensitivity, for instance, to infrared divergences under the assumption of universality and working in the framework of QCD factorisation (QCDF)~\cite{Beneke:2003zv,Beneke:2006hg}.   However, it was found already in Refs.~\cite{Alguero:2020xca} that the main source of uncertainty comes not from the infrared divergences but from the ratio of form factors that control the observable, especially due to the lack of knowledge of the correlations among form factors. Therefore, progress in the knowledge and computation of form factors and their correlation is of utmost importance  for substantially improving  the SM predictions for the $L$-observables. Other interesting approaches to these decays assuming SU(3) flavour symmetry or in combination with QCDF can be found in Refs.~\cite{Bhattacharya:2022akr, Fleischer:2007hj, Huber:2021cgk,
BurgosMarcos:2025xja,
Berthiaume:2023kmp,Grossman:2024amc,Amhis:2022hpm,Bhattacharya:2025wcq}.

In this paper, given the crucial role played by form factors,  we evaluate, for the first time, the $L$-observables using not only the state-of-the-art lattice QCD computation of the form factors but also the relevant light-cone sum rules (LCSRs) calculations and their correlations that were unavailable in the past. 
This is achieved by performing a novel LCSR analysis of the $B_{d,s}\to K^{(*)}$ form factors.
Our goal is to explore the sensitivity of the SM predictions of the $L$-observables as a function of the chosen form factor determination.

The paper is structured as follows.
In \refsec{FFs} we perform a new LCSR analysis of the $B_{d,s}\to K^{(*)}$ form factors.
We then combine these results with available lattice QCD computations to perform the first combined analysis of $B_{d,s} \to K^{(*)}$ form factors.
We present results for these form factors within two different frameworks: a) using LCSR computations alone and b) LCSR computation combined with all lattice data\footnote{{For completeness we will sketch some results using only lattice data albeit  we prefer to focus on the comparison between LCSR and combined LCSR+lattice for the reasons discussed in \refsubsec{subdef}.}}.
In \refsec{Lobs} we discuss the structure of the $L$-observables, explore their sensitivity to each of the frameworks mentioned above and discuss a set of new promising observables that can be accessible at LHCb, including their SM predictions in the different frameworks. In addition, we briefly explore in \refsec{np} the New Physics (NP) sensitivity of all these observables as a function of the framework used, and identify an interesting mechanism that can provide a rather unique NP signal. 
We present the conclusions that follow from our analyses in \refsec{conl}.

%\bigskip

%\bigskip
%with emphasis in the precision that can be achieved for the extraction of the Wilson coefficients.

 %together with the experimental ones. To illustrate the method we generate two different toy data in presence of NP and compare with the predictions.

%Also we will present different geometrical bounds and internal consistency tests that will be applied first to the SM predictions to show their precision and second to 
% the data to test the robustness and internal consistency of the analysis done by LHCb. %And second, 
%we will present the results of our own theoretical global fit and compare with the data-driven analysis done by LHCb.
%Finally, we make emphasis on all the reasons that can break the main relation and what do we learnt from it.

\section{Form Factor calculations and combined analysis}\label{sec:FFs}

The $B_{d,s}\to K^{(*)}$ form factors play a pivotal role in $B_{d,s}(p)\to K^{(*)}(k) X(q)$ transitions.
Here, $X$ may be either a hadron, as in the processes considered in this paper, or a lepton-antilepton pair, as in semileptonic processes.
We define the $B_d\to K^{(*)}$ form factors following conventions of Ref.~\cite{Gubernari:2018wyi}:
\begin{align}
    \langle\bar  K(k) |  \bar{s}\, \gamma^\mu b | \bar{B}(p) \rangle &=
        \left[ (p + k)^\mu - \frac{m_{B_d}^2 - m_K^2}{q^2} q^\mu \right] f_+^{B_d\to K}
        + \frac{m_{B_d}^2 - m_K^2}{q^2} q^\mu f_0^{B_d\to K}, \\
    \langle\bar  K(k) | \bar{s}\, \sigma^{\mu \alpha}q_\alpha  b | \bar{B}(p) \rangle &=
        \frac{i f_T^{B_d\to K}}{m_{B_d} + m_K} \left[ q^2 (p + k)^\mu - (m_{B_d}^2 - m_K^2) q^\mu \right].
    \\
    \langle\bar  K^*(k, \eta) |  \bar{s}\, \gamma^\mu b |\bar{B}(p) \rangle &=
        \epsilon^{\mu\nu\rho\sigma} \eta_\nu^* p_\rho k_\sigma \frac{2 V^{B_d\to K^*}}{m_{B_d} + m_{K^*}}, \\
    \label{eq:th:param:A}
    \langle\bar  K^*(k, \eta) | \bar{s}\, \gamma^\mu \gamma_5 b |\bar{B}(p) \rangle &=
        i \eta_\nu^* \bigg[ g^{\mu\nu} (m_{B_d} + m_{K^*}) A_1^{B_d\to K^*}
        - (p + k)^\mu q^\nu  \frac{A_2^{B_d\to K^*}}{m_{B_d} + m_{K^*}} \nonumber \\
    & 
    \hspace*{-2.5cm}
    - 2 m_{K^*} \frac{q^\mu q^\nu}{q^2} 
    \left(
        \frac{m_{B_d} + m_{K^*}}{2 \, m_{K^*}} \, A_1^{B_d\to K^*} - \frac{m_{B_d} - m_{K^*}}{2 \, m_{K^*}} \, A_2^{B_d\to K^*} - A_0^{B_d\to K^*}
    \right) \bigg],\\
    \langle\bar  K^*(k, \eta) | \bar{s}\, \sigma^{\mu \alpha}q_\alpha  b | \bar{B}(p) \rangle &=
        \epsilon^{\mu\nu\rho\sigma} \eta_\nu^* p_\rho k_\sigma \, 2 T_1^{B_d\to K^*}, \\
    \langle\bar  K^*(k, \eta) | \bar{s}\, \sigma^{\mu \alpha}q_\alpha \gamma_5 |\bar{B}(p) \rangle &=
        i \eta_\nu^* \bigg[ \Big( g^{\mu\nu} (m_{B_d}^2 - m_{K^*}^2) - (p + k)^\mu q^\nu \Big) T_2^{B_d\to K^*} \nonumber \\
        & 
        + q^\nu \left( q^\mu - \frac{q^2}{m_{B_d}^2 - m_{K^*}^2} (p + k)^\mu \right) T_3^{B_d\to K^*} \bigg],
\end{align}
where $\eta$ is the polarisation four-vector of the $K^*$.
The definitions of the $B_s\to K^{(*)}$ form factors can be obtained with obvious replacements.
For brevity, we suppress the $q^2$ dependence of the form factors, showing it only when needed.
We also introduce the helicity form factors
\begin{align}
    A_{12}^{B_d\to K^*} &= \frac{(m_{B_d} + m_{K^*})^2 (m_{B_d}^2 - m_{K^*}^2 - q^2) \, A_1^{B_d\to K^*} - \lamkin \, A_2^{B_d\to K^*}}{16 \, m_{B_d} m_{K^*}^2 (m_{B_d} + m_{K^*})}\,, \\
    \label{eq:th:T23}
    T_{23}^{B_d\to K^*} &= \frac{(m_{B_d}^2 - m_{K^*}^2) (m_{B_d}^2 + 3 m_{K^*}^2 - q^2) \, T_2^{B_d\to K^*} -  \lamkin \, T_3^{B_d\to K^*}}{8 \, m_{B_d} m_{K^*}^2 (m_{B_d} - m_{K^*})}\,,
\end{align}
where $ \lamkin \equiv (m_{B_d}^2 - m_{K^{(*)}}^2 -q^2)^2-4q^2m_{K^{(*)}}^2$ is the Källén function.

Although the calculation of the $B_{d,s}\to \bar{K}K$ and $B_{d,s}\to \bar{K}^*K^*$ branching ratios within QCDF requires only the knowledge of $f_0^{B_{d,s}\to K}$ and $A_0^{B_{d,s}\to K^*}$ at $q^2\sim 0$, we nonetheless predict the complete set of $B_{d,s}\to K^{(*)}$ form factors.
This is important for two reasons.  
First, as mentioned above, these form factors play a crucial role in various decay processes and can therefore be used in other phenomenological studies.  
Second, when multiple calculations exist for the same transition, it is essential to ensure consistency across all form factors within the given process.  
To illustrate this, consider a hypothetical scenario where two collaborations, A and B, independently compute the complete set of $B_d\to K^*$ form factors, and we aim to combine their results using a specific parametrization.  
If, for instance, the calculation of $A_1^{B_d\to K^*}$ from collaboration A is inconsistent with that from collaboration B, one would need to rescale the correlation matrix of the parameters describing the form factors by a factor of $\sqrt{\chi^2/\text{d.o.f.}}$ following the PDG recommendation~\cite{ParticleDataGroup:2024cfk}.
As a consequence, this would also directly affect the predictions for $A_0^{B_d\to K^*}$.

In the rest of this section, we first conduct a new LCSR analysis of the $B_{d,s} \to K^{(*)}$ form factors and then summarize the available lattice QCD calculations for these form factors.
Next, we describe the parametrization used to combine these calculations and present our final form factor results.

\subsection{Light-cone sum rule calculations}
\label{sec:LCSR}

Light-cone sum rules (LCSRs) are a method for calculating form factors based on a light-cone operator product expansion (OPE) and semi-global quark-hadron duality~\cite{Colangelo:2000dp,Gubernari:2020zil}.  
The OPE leads to a series in which each term consists of a perturbatively calculable Wilson coefficient and a corresponding operator, with higher-order terms being power-suppressed.  
A key challenge in the OPE for LCSRs arises from the fact that the matrix elements of these operators --- namely, the light-cone distribution amplitudes (DAs) --- are often not well known.  
Although future lattice QCD calculations and experimental measurements may improve our understanding of the DAs~\cite{Beneke:2018wjp,Desiderio:2020oej}, the current theoretical precision remains limited, especially for $B$-meson DAs.
Furthermore, semi-global quark-hadron duality introduces a systematic uncertainty that can be estimated but is difficult to reduce~\cite{Colangelo:2000dp,Gubernari:2020zil}.

In this paper, we use the LCSRs with $B$-meson DAs~\cite{Khodjamirian:2005ea,Khodjamirian:2006st}.  
One of the main advantages of these LCSRs is that the structure of the analytical formulas depends only on the spin-parity of the interpolating current and the final-state meson.  
As a result, we can directly use the analytical expressions derived in Ref.~\cite{Gubernari:2018wyi} for the $B\to$~pseudoscalar and $B\to$~vector form factors. 
To apply these expressions to the $B_s \to K^{(*)}$ form factors, we need to modify the input parameters, such as the masses and decay constants, and perform a new Bayesian analysis to fix the LCSR threshold $s_0$ (see below). 
All the inputs used in this calculation are listed in \reftab{inputs}. 
Below, we highlight the inputs that are particularly important for the LCSR evaluation.

For the parameter $\lambda_{B_d}$ we use the value calculated in Ref.~\cite{Braun:2003wx}\footnote{Notice that this is not the same value we used in Ref.\cite{Biswas:2023pyw} and the prediction of some observables may have changed from our previous work when using this new value of the $\lambda_{B_d}$ parameter. See discussion in Subsection~\ref{sec:lambdas}. }: 
\begin{align} \label{lambdaBdeq}
    \lambda_{B_d}
    =
    0.460\pm 0.110 \GeV
    \,.
\end{align}
We obtain the value of $\lambda_{B_s}$ using the ratio
\begin{align}
    \frac{\lambda_{B_s}}{\lambda_{B_d}} = 1.19 \pm 0.14
    \,,
\end{align}
%
%%%% table
%
\begin{table}[!t]
	\begin{center}
%\tabcolsep=1.26cm\begin{tabular}{|c|c|c|}
%\hline\multicolumn{3}{|c|}{$B_{d,s}$ Distribution Amplitudes (at $\mu=1$ GeV)~\cite{Khodjamirian:2020hob,Braun:2003wx} \ng{can be removed, already in text}} \\ \hline
%$\lambda_{B_d} $ [GeV]&$\lambda_{B_s}/\lambda_{B_d}$& $\sigma_B$\\
%\hline
%$0.460\pm0.110$&$1.19\pm0.14$&$1.4\pm0.4$\\ \hline
%\end{tabular}

%\vskip 1pt

\tabcolsep=0.712cm\begin{tabular}{|c|c|c|c|}
\hline\multicolumn{4}{|c|}{$K^*$ Distribution Amplitudes (at $\mu=2$ GeV)~\cite{Ball:2007rt} }  \\ \hline
$\alpha_1^{K^*}$&
$\alpha_{1,\perp}^{K^*}$&
$\alpha_2^{K^*}$&
$\alpha_{2,\perp}^{K^*}$\\
\hline
$0.02\pm0.02$&$0.03\pm0.03$&$0.08\pm0.06$&$0.08\pm0.06$\\ \hline
\end{tabular}

\vskip 1pt

\tabcolsep=2.512cm\begin{tabular}{|c|c|}
\hline\multicolumn{2}{|c|}{$K$ Distribution Amplitudes (at $\mu=2$ GeV)~\cite{RQCD:2019osh}}  \\ \hline
$\alpha_1^{K}$&
%$\alpha_{1,\perp}^{K^*}$&
$\alpha_2^{K}$\\
%$\alpha_{2,\perp}^{K^*}$\\
\hline
$0.0525^{+31}_{-33}$&$0.106^{+15}_{-16}$\\ \hline
\end{tabular}

\vskip 1pt

\tabcolsep=2.1cm\begin{tabular}{|c|c|}
\hline\multicolumn{2}{|c|}{Decay Constants for $B$ mesons~\cite{FLAG:2024oxs}}\\ \hline  
$f_{B_d}$&$f_{B_s}/f_{B_d}$\\
\hline
$0.190\pm0.0013$&$1.209\pm0.005$\\ \hline
\end{tabular}

\vskip 1pt

\tabcolsep=1.0cm\begin{tabular}{|c|c|c|}
\hline\multicolumn{3}{|c|}{Decay Constants for Kaons (at $\mu=2$ GeV)~\cite{Workman:2022ynf,Allton:2008pn,Straub:2015ica}}  \\ \hline
$f_{K}$&$f_{K^*}$&$f^\perp_{K^*}/f_{K^*}$\\
\hline
$0.1557\pm 0.0003$&$0.204\pm0.007$&$0.712\pm0.012$\\ \hline
\end{tabular}
\vskip 1pt

\tabcolsep=2.162cm\begin{tabular}{|c|c|}
\hline\multicolumn{2}{|c|}
{B-meson lifetimes (ps)}  \\ \hline
%$A_0^{B_s}(q^2=0)$&$A_0^{B_d}(q^2=0)$ &
$\tau_{B_d}$ & $\tau_{B_s}$\\
\hline
%$0.314 \pm 0.048$ &$ 0.356 \pm 0.046$ & 
$1.519\pm0.004$ & $1.515\pm0.004$ \\ \hline
\end{tabular}

%\vskip 1pt

%\tabcolsep=2.162cm\begin{tabular}{|c|c|}
%\hline\multicolumn{2}{|c|}{$B_{d}\to K$~\cite{Parrott:2022rgu} and $B_s\to K$~\cite{Khodjamirian:2017fxg} form factors \ng{can be removed}}  \\ \hline
%$f_0^{B_s}(q^2=0)$&$f_0^{B_d}(q^2=0)$ \\
%\hline
%$0.336 \pm 0.023$ &$ 0.332 \pm 0.012$ \\ \hline
%\end{tabular}

\vskip 1pt

%\tabcolsep=0.33cm\begin{tabular}{|c|c|c|c|}
%\hline\multicolumn{4}{|c|}{\ab{$B_{d,s}\to K^{(*)}$ new form factors from Nico}} \\ \hline
%\ab{$A_0^{B_s}(q^2=0)$}&\ab{$A_0^{B_d}(q^2=0)$}&\ab{$f_0^{B_s}(q^2=0)$}&\ab{$f_0^{B_d}(q^2=0)$}\\
%\hline
%\ab{$0.3578 \pm 0.0203$} &\ab{$ 0.3498 \pm 0.0278$} & \ab{$0.2742\pm0.0170$} & \ab{$0.3333\pm0.0086$} \\ \hline
%\end{tabular}

\vskip 1pt

\tabcolsep=0.515cm\begin{tabular}{|c|c|c|c|c|}
\hline
\multicolumn{4}{|c|}{Wolfenstein parameters~\cite{Charles:2004jd} }  \\ \hline
$A$ & $\lambda$ & $\bar\rho$ &
$\bar\eta$\\ \hline
$0.8235^{+0.0056}_{-0.0145}$ &$	0.22484^{+0.00025}_{-0.00006}$&$0.1569^{+0.0102}_{-0.0061}$&$0.3499^{+0.0079}_{-0.0065}$\\
\hline
\end{tabular}

\vskip 1pt

\tabcolsep=1.436cm\begin{tabular}{|c|c|c|}
	\hline
	\multicolumn{3}{|c|}{QCD scale and quark masses [GeV] }\\
		\hline 
 $\bar{m}_b(\bar{m}_b)$  & $m_b/m_c$ &$\Lambda_{{\rm QCD}}$
      \\ \hline
      $4.2$  & $4.577\pm0.008$  &$0.225$
      \\ \hline
\end{tabular}
\vskip 1pt
\tabcolsep=1.193cm\begin{tabular}{|c|c|c|c|}
	\hline
	\multicolumn{4}{|c|}{Hadron masses [GeV] }\\
		\hline 
 $m_{B_d} $& $m_{B_s} $& $m_{K^*} $ & $m_K$ 
      \\ \hline
      $5.280$ & $5.367$&$0.892$& 0.496
      \\ \hline
\end{tabular}
\vskip 1pt

\tabcolsep=0.50cm\begin{tabular}{|c|c|c|c|c|c|}
\hline
	\multicolumn{6}{|c|}{SM Wilson Coefficients (at $\mu=4.2$ GeV) }\\
		\hline 
${\cal C}_1$ &  ${\cal C}_2$ & ${\cal C}_3$ & ${\cal C}_4$ &  ${\cal C}_5$ & ${\cal C}_6$
      \\ \hline
 1.082 & -0.191 & 0.013 & -0.036 & 0.009 &  -0.042
      \\ \hline
 ${\cal C}_{7}/\alpha_{em}$ & ${\cal C}_{8}/\alpha_{em}$ & ${\cal C}_{9}/\alpha_{em}$ &  ${\cal C}_{10}/\alpha_{em}$ & ${\cal C}^{\rm eff}_{7\gamma}$ &  ${\cal C}^{\rm eff}_{8g}$
      \\ \hline
 -0.011 & 0.058 & -1.254 & 0.223 & -0.318 & -0.151 
      \\ \hline
		\end{tabular}
		\caption{Summary of input parameters used in this paper. }
		\label{tab:inputs}
	\end{center}
\end{table}
%%%%%
%
calculated in Ref.~\cite{Khodjamirian:2020hob}.
For the parameters $\lambda_{B_d,E}^2$ and $\lambda_{B_d,H}^2$ we use the calculation of Ref.~\cite{Nishikawa:2011qk}:
\begin{align}
    \lambda_{B_d,E}^2 & = 0.03 \pm 0.02 \,\GeV^2 \,, \\ 
    \lambda_{B_d,H}^2 & = 0.06 \pm 0.03 \,\GeV^2 \,.
\end{align}
We do not use the calculation from Ref.~\cite{Rahimi:2020zzo}, as the OPE exhibits signs of very slow convergence, in contrast to the findings of Ref.~\cite{Nishikawa:2011qk}.
This discrepancy is presumably due to the higher dimensionality of the correlator used to compute the OPE in Ref.~\cite{Rahimi:2020zzo} compared to Ref.~\cite{Nishikawa:2011qk}.
Furthermore, Ref.~\cite{Nishikawa:2011qk} includes additional $\alpha_s$ corrections compared to Ref.~\cite{Rahimi:2020zzo}, which have been shown to be essential for the stability of the QCD sum rule.
As in Ref.~\cite{Bordone:2019guc}, we take $\lambda_{B_d,E}^2 = \lambda_{B_s,E}^2$ and $\lambda_{B_d,H}^2 = \lambda_{B_s,H}^2$ as these parameters only contribute at subleading power and hence these $SU(3)_F$ symmetry-breaking effects have a negligible impact on the LCSRs predictions.

In addition, one has to determine the LCSR threshold $s_0$ for each form factor considered.
Following Ref.~\cite{Gubernari:2018wyi}, for all the $B_d\to K$ form factors we use $s_0= 1.05 \pm 0.10$, taken from Ref.~\cite{Khodjamirian:2003xk}.
For the $B_s \to K$ form factors, we adopt the same values of $s_0$ and add a $20\%$ uncorrelated systematic uncertainty to account for $SU(3)_F$ symmetry-breaking effects.
For each $B_{d,s} \to K^*$ form factor, we determine $s_0$ using the procedure outlined in Ref.~\cite{SentitemsuImsong:2014plu} and applied in Ref.~\cite{Gubernari:2018wyi} to the $B \to \text{vector}$ form factors.
This involves deriving a daughter LCSR that provides a constraint, allowing for the determination of $s_0$.
The union of the various $68\%$ confidence intervals for the individual $s_0$ values obtained in these analyses is given by $[1.48, 1.64]$ for the $B_d \to K^*$ form factors and $[1.39, 1.61]$ for the $B_s \to K^*$ form factors.
This Bayesian analysis has been performed using the EOS software~\cite{EOSAuthors:2021xpv} in version 1.0.13~\cite{EOS:v1.0.13}.
In this analysis, we obtain a notably high value for the logarithm of the Bayesian evidence, $\log Z = 72.0$, which indicates decisive support for the model under consideration, given the LCSR formulas and the specified prior distributions.

\begin{table}[!t]
\begin{center}
    \renewcommand{\arraystretch}{1.4}
    \tabcolsep=1.26cm\begin{tabular}{|c|c|c|}
    \hline
    Form factor                   &
    Central val. $\pm$ unc.       &
    Correlation
    \\ \hline
    $f_0^{B_d \to K}(q^2 = 0)$    &
    $0.278 \pm 0.079$             &
    \multirow{2}{*}{$0.86$}
    \\ 
    $f_0^{B_s \to K}(q^2 = 0)$    &
    $0.276 \pm 0.095$             &
    \\ \hline
    $A_0^{B_d \to K^*}(q^2 = 0)$  &
    $0.361 \pm 0.064$             &
    \multirow{2}{*}{$0.61$}
    \\
    $A_0^{B_s \to K^*}(q^2 = 0)$  &
    $0.309 \pm 0.066$             &
    \\ \hline
    \end{tabular}
\end{center} 
\caption{Summary of  form factor predictions (and correlation) relevant to the $L$-observables obtained using LCSRs.}\label{tab:ff}\end{table}
%\vspace*{-0.4cm}
To evaluate the LCSRs, we must choose an appropriate window for the Borel parameter $M^2$ so that higher-order corrections in the OPE, as well as contributions from the continuum and excited states, are sufficiently suppressed.
For the processes considered in this work, we determine that the suitable window is $M^2 \in [0.5,1.5]$, which is consistent with Refs.~\cite{Gubernari:2018wyi,Khodjamirian:2006st}.
Within this window, the LCSRs exhibit relative stability.
To account for the spurious $M^2$ dependence, we include the following systematic uncertainties in our form factor LCSR results:
\begin{align}
    \label{eq:M2dep}
    &
    B_{d,s} \to K : 5\% \,,
    &&
    B_{d,s} \to K^* : 8\% \,,
    &
\end{align}
which are obtained by varying $M^2$ in the chosen window while keeping the other parameters fixed. 
We also account for the finite-width effects of the $K^*$, following Ref.~\cite{Descotes-Genon:2019bud}.  
As a result, we scale the central values of the $B_{d,s} \to K^*$ form factors by a factor of $1.09$.

Finally, using the inputs listed in this section and in \reftab{inputs}, we obtain predictions for the $B_{d,s} \to K^{(*)}$ form factors.
We propagate the uncertainties to the final result by drawing samples from the posterior distribution, thereby accounting for all the parametric uncertainties.
The model uncertainties associated with the $M^2$ dependence are summarized in \refeq{M2dep}. Regarding semi-global quark-hadron duality, we verify that the contribution from the tail of the OPE (i.e., the region above $s_0$) remains small compared to the total OPE integrated over the full range for each form factor. 
Specifically, since none of the OPE tails exceed 25\% of the integrated OPE, we conclude that the associated uncertainty from semi-global quark-hadron duality is negligible compared to the parametric uncertainties~\cite{Straub:2015ica}.%see bottom of page 11

We provide results at five different $q^2$ points: $q^2 =\{-15,-10,-5,0,+5\}\GeV^2$.
Since these results will be used to predict the $L$-observables, we provide the correlations between all $B_d \to K^{(*)}$ and $B_s \to K^{(*)}$ form factors, respectively.
These correlations, along with the central values for the $q^2$ points considered, are provided in machine-readable format as ancillary files attached to the arXiv version of this paper. 
These files are named 
\begin{align*}
     \texttt{LCSR\_Bq\_to\_K.yaml} 
 \quad \texttt{and}     \quad \texttt{LCSR\_Bq\_to\_Kstar.yaml}
\end{align*}
%\begin{align*}
%    & \texttt{LCSR\_Bq\_to\_K.yaml} %\\
%    & \texttt{LCSR\_Bq\_to\_Kstar.yaml}
%\end{align*}
Our results for the $B_d \to K^{(*)}$ form factors agree with those of Ref.~\cite{Gubernari:2018wyi}, as the analytical expressions are identical. Minor differences arise from the updated input parameters used in our calculation.
To the best of our knowledge, the results for the $B_s \to K^{(*)}$ form factors presented here are the first obtained using $B$-meson DA LCSRs.
For the predictions of the $L$-observables, only $f_0^{B_d \to K}(q^2 = 0)$, $f_0^{B_s \to K}(q^2 = 0)$, $A_0^{B_d \to K^*}(q^2 = 0)$, and $A_0^{B_s \to K^*}(q^2 = 0)$ are required.
The LCSR results for these form factors are presented in \reftab{ff}.

We conclude this subsection by noting that the form factors considered here can also be calculated using LCSRs with light-meson DAs, as in, e.g., Refs.~\cite{Khodjamirian:2017fxg,Straub:2015ica}.
These LCSRs typically yield smaller uncertainties compared to those based on $B$-meson DAs, owing to the reduced parametric uncertainties of the light-meson DAs.
However, the correlations of all the $B_d \to K^{(*)}$ and $B_s \to K^{(*)}$ form factors have not been published, nor have the complete analytical expressions required to derive them. 
Consequently, it is not possible to reconstruct these correlations without repeating the entire calculation from scratch. 
We therefore encourage the authors to make their formulas publicly available in future works.
Moreover, it remains unclear how to go beyond the narrow-width approximation for the $K^*$ within the framework of LCSRs with light-meson DAs.

\subsection{Lattice QCD datasets}
\label{sec:LQCD}

Besides LCSR calculations, several lattice QCD computations exist for the $B_{d,s} \to K^{(*)}$ form factors.  
We include all these computations in our analysis and provide a list below for the reader's convenience, along with details on their implementation.  
\begin{description}
    \item[{\boldmath$B_d\to K$}]
    There are three lattice QCD calculations for these form factors:  
    one by HPQCD in 2013~\cite{Bouchard:2013eph},  
    one by the Fermilab Lattice and MILC collaborations (FNAL/MILC) in 2015~\cite{Bailey:2015dka},  
    and another by HPQCD in 2022~\cite{Parrott:2022rgu}.
    The results of these calculations are given as multivariate Gaussian distributions for the parameters of the BCL expansion~\cite{Bourrely:2008za}.  
    To facilitate their use and fitting to a different parametrization, these results must be rephrased in terms of $q^2$ data points, ensuring that the number of degrees of freedom remains unchanged.  
    For the first two calculations, we use $q^2 = \{17, 20, 23\} \, \GeV^2$, while for the last, we use $q^2 = \{0, 12, 22.9\} \, \GeV^2$ for each form factor, namely $f_+^{B_d \to K}$, $f_0^{B_d \to K}$, and $f_T^{B_d \to K}$~\footnote{
        For this last calculation we exclude the point $f_0^{B_d \to K}$ at $q^2=0$, due to the end-point relation $f_+^{B_d \to K}(q^2=0) = f_0^{B_d \to K}(q^2=0)$.
    }.
    These choices are based on the range of validity of the respective calculations.  
    The corresponding constraints are implemented in the EOS software~\cite{EOSAuthors:2021xpv}.

    \item[{\boldmath$B_s\to K$}]
    There are three lattice QCD calculations for these form factors:
    one by HPQCD in 2014~\cite{Bouchard:2014ypa},
    one by FNAL/MILC in 2019~\cite{FermilabLattice:2019ikx},
    and another by RBC/UKQCD in 2023~\cite{Flynn:2023nhi} (this calculation supersedes the one of Ref.~\cite{Flynn:2015mha}).
    As discussed in the FLAG Review 2024~\cite{FLAG:2024oxs}, these calculations are not mutually compatible.
    Therefore, the correlation matrix of the BCL coefficients obtained from the combined fit is rescaled by a factor of $\sqrt{\chi^2/\text{d.o.f.}}$.
    Starting from this correlation matrix and the central values provided in Table 40 of Ref.~\cite{FLAG:2024oxs} we obtain values at $q^2 =\{17.5,19.5,21.5,23.5\}\,\GeV^2$ for $f_+^{B_s \to K}$ and $f_0^{B_s \to K}$, as there is no lattice QCD calculation for $f_T^{B_s \to K}$.
    We drop the data point $f_0^{B_s \to K}$ at $q^2 = 21.5 \,\GeV^2$ to account for the end-point relation $f_+^{B_s \to K}(q^2=0) = f_0^{B_s \to K}(q^2=0)$.

    \item[{\boldmath$B_{d,s}\to K^*$}]
    There is only one lattice QCD calculation for these form factors.
    This is given by Ref.~\cite{Horgan:2015vla}, which is an updated version of Ref.~\cite{Horgan:2013hoa}.
    For the $B_d \to K^*$ form factors, we use the twelve $q^2$ data points provided by the authors for each of the seven form factors, as implemented in the EOS software.  
    For the $B_s \to K^*$ form factors, we obtain values at $q^2 = \{16.0, 20.0\} \, \GeV^2$ for each of the seven form factors from the coefficients given in Ref.~\cite{Horgan:2015vla}.  
    Note that the correlations between the $B_d \to K^*$ and $B_s \to K^*$ form factors are not provided.  
    We encourage lattice QCD collaborations to include these correlations in future studies.  
\end{description}

\subsection{Combined analysis}
\label{sec:comb}

\begin{table}[t!]
\addtolength{\arraycolsep}{3pt}
\renewcommand{\arraystretch}{1.4}
$$
\begin{array}{|l|c|c|c|}
    \hline
    F_i & J^P &  m_{R,i}^{b \to d}\, [\GeV] & m_{R,i}^{b \to s}\, [\GeV]  \\ \hline
    A_0 & 0^-  & 5.279 & 5.366 \\
    f_0 & 0^+  & 5.540 & 5.630 \\
    f_+, f_T, V, T_1 & 1^- & 5.325 & 5.415 \\
    A_{1} , A_{12}, T_{2}, T_{23} & 1^+ & 5.726 & 5.829\\
    \hline
\end{array}
$$ 
%\vspace*{-20pt}
\caption{
    The masses of the resonances, $m_{R,i}$, appearing in \refeq{BSZ} are organized according to their spin and parity $J^P$, as well as the quark transitions $b \to \{d, s\}$. %
    All values have been taken from Ref.~\cite{ParticleDataGroup:2024cfk}, except for the $0^+$ resonance masses, which have been chosen consistently with the EOS implementation.  
}
\label{tab:res}
\end{table}

We combine the LCSR results obtained in \refsubsec{LCSR} with the LQCD results listed in \refsubsec{LQCD} using a simplified series expansion.
We use the expansion in the exact same form as proposed in Ref.~\cite{Straub:2015ica}:
\begin{equation}
    F_i(q^2) = \frac{1}{1-q^2/m_{R,i}^2} \sum_{n=0}^N \alpha_n^i \left[z(q^2)-z(0)\right]^n\,,
    \label{eq:BSZ}
\end{equation}
where
\begin{equation}
    z(q^2) = \frac{\sqrt{t_+ - q^2}-\sqrt{t_+ - t_0^{\phantom{1}}}}{\sqrt{t_+ -  q^2}+\sqrt{t_+ - t_0^{\phantom{1}}}} \,,
    \qquad
    \text{and}
    \qquad
    t_0 = t_+\left(1-\sqrt{1-\frac{t_-}{t_+}}\right)
    \,.
\end{equation}
For $B_d \to K$ processes, for instance, we define $t_\pm \equiv (m_{B_d}\pm m_K)^2$.
The definition of $t_\pm$ for the other processes can be obtained with obvious replacements.
The values of the masses of the resonances $m_{R,i}$ are given in \reftab{res}.

We truncate the expansion in \refeq{BSZ} after three terms, i.e., for $N = 2$.
As demonstrated in Refs.~\cite{Straub:2015ica, Bharucha:2010im, Gubernari:2023puw}, this truncation order provides a sufficiently precise description of the form factors.  
We perform a Bayesian analysis with flat priors for all fitted coefficients.  
Two separate analyses are carried out: one for the $B_{d,s} \to K$ form factors and another for the $B_{d,s} \to K^*$ form factors.  
This allows us to account for the correlations between the $K$ and $K^*$ final states.  
The likelihoods used in our analyses are detailed in \refsubsec{LCSR} and \refsubsec{LQCD}.  
To generate all posterior samples we use the nested sampling algorithm~\cite{Higson:2018}, as implemented in the \texttt{dynesty} package~\cite{Speagle:2020,dynesty:v2.0.3}.
The outcomes of our fits are summarized in \reftab{FFfits}. 
It is evident from  \reftab{FFfits} that both fits exhibit excellent quality, indicating that \refeq{BSZ} with $N=2$ provides an accurate description of the data.

\begin{table}[t!]
    \newcommand{\pp}{\phantom{+}}
    \centering
    \renewcommand{\arraystretch}{1.30}
    \begin{tabular}{|c|c|c|c|c|}
        \hline
            Processes              &
            $\chi^2$               &
            d.o.f.                 &
            $p$-value [\%]         &
            $\log Z$                \\
            \hline
            $B_{d,s} \to K$        &
            17.7                   &
            45                     &
            99.9                   &
            147.1                  \\

            $B_{d,s} \to K^*$      &
            64.3                   &
            126                    &
            99.9                   &
            321.6                  \\
        \hline
    \end{tabular}
    \caption{%
        \label{tab:FFfits}
        Summary of fit quality metrics for the processes considered.
        Here $\log Z$  refers to the logarithm of the Bayesian evidence.
    }
\end{table}

\begin{figure}[t!]
    \centering
    \subfloat%[First Subfigure]
    {
        \includegraphics[width=0.45\textwidth]{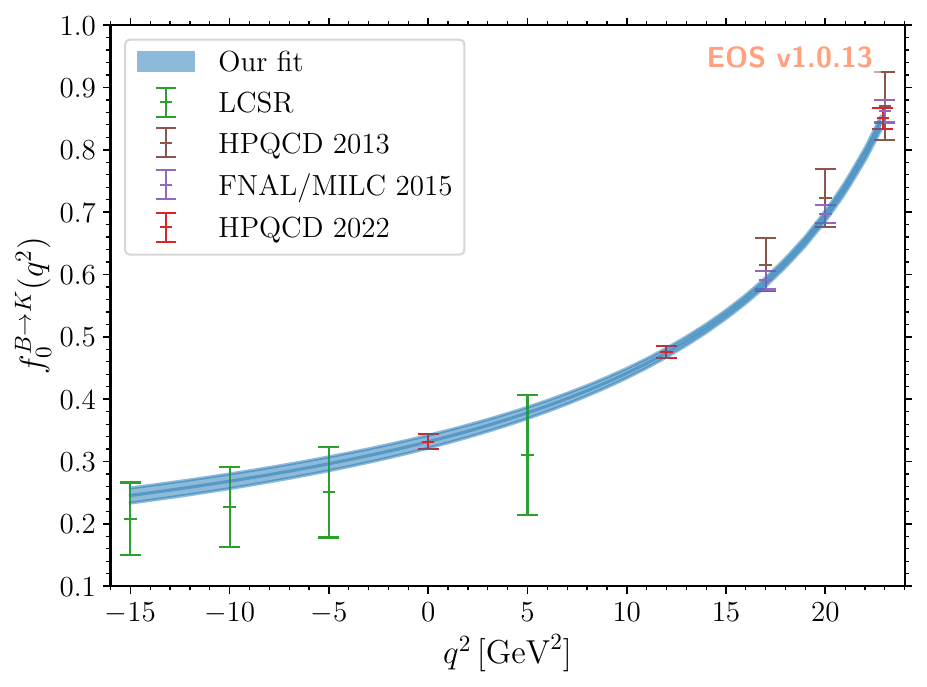}
    }
    \hfill
    \subfloat%[Second Subfigure]
    {
        \includegraphics[width=0.45\textwidth]{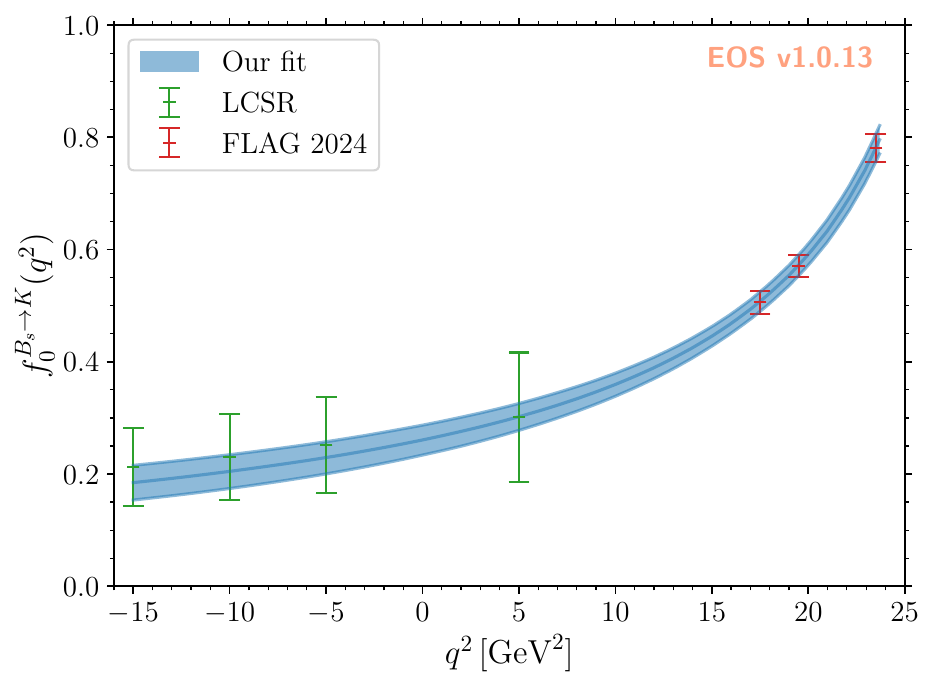}
    }
    
    %\vskip 1em % Adds vertical space between rows
    
    \subfloat
    %[Third Subfigure]
    {
        \includegraphics[width=0.45\textwidth]{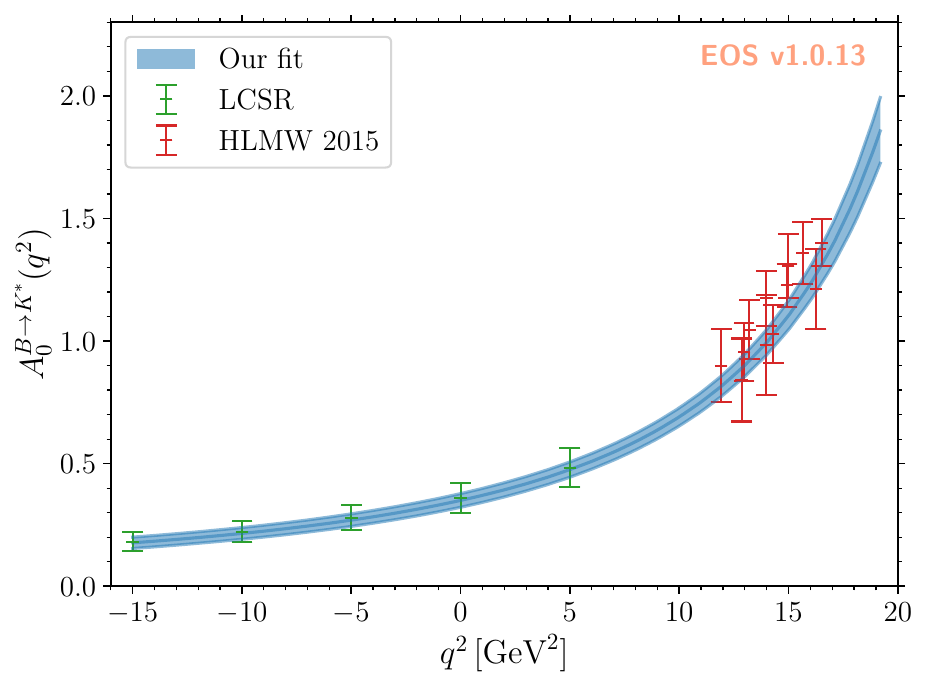}
    }
    \hfill
    \subfloat%[Second Subfigure]
    {
        \includegraphics[width=0.45\textwidth]{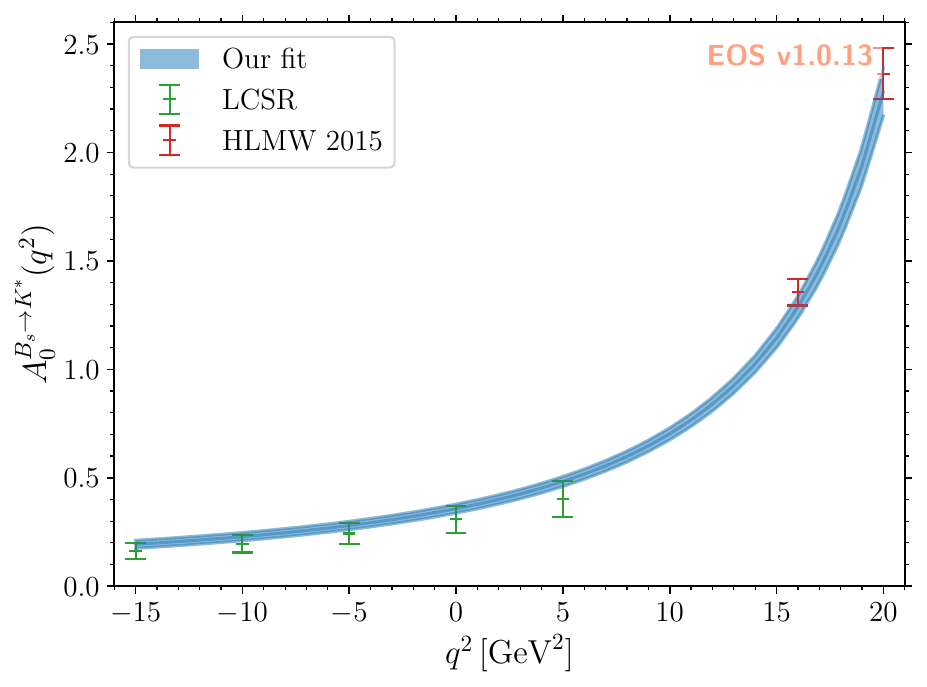}
    }
    
    \caption{Plot of the $f_0^{B_d\to K}$, $f_0^{B_s\to K}$, $A_0^{B_d\to K^*}$, $A_0^{B_s\to K^*}$ form factors as a function of $q^2$.
    The data points show the central values and associated uncertainties of our LCSR (see \refsubsec{LCSR}), HPQCD\,2013~\cite{Bouchard:2013eph}, FNAL/MILC\,2015~\cite{Bailey:2015dka}, HPQCD\,2022~\cite{Parrott:2022rgu}, HLMW2015~\cite{Horgan:2015vla}, and FLAG2024~\cite{FLAG:2024oxs}.
    }
    \label{fig:FFplots}
\end{figure}

We show the plots of the $f_0^{B_d \to K}$, $f_0^{B_s \to K}$, $A_0^{B_d \to K^*}$, and $A_0^{B_s \to K^*}$ form factors as a function of $q^2$ in \reffig{FFplots}.
The values of these form factors at $q^2 = 0$ are presented in \reftab{combined}.

\begin{table}[t]
\begin{center}
    \renewcommand{\arraystretch}{1.4}
    \tabcolsep=1.26cm\begin{tabular}{|c|c|c|}
    \hline
    Form factor                   &
    Central val. $\pm$ unc.       &
    Correlation
    \\ \hline
    $f_0^{B_d \to K}(q^2 = 0)$    &
    $0.3333 \pm 0.0086$           &
    \multirow{2}{*}{$0.056$}
    \\ 
    $f_0^{B_s \to K}(q^2 = 0)$    &
    $0.2742 \pm 0.0170$           &
    \\ \hline
    $A_0^{B_d \to K^*}(q^2 = 0)$  &
    $0.350 \pm 0.028$             &
    \multirow{2}{*}{$0.031$}
    \\
    $A_0^{B_s \to K^*}(q^2 = 0)$  &
    $0.356 \pm 0.020$             &
    \\ \hline
    \end{tabular}
\end{center}
\caption{Summary of  form factor predictions (and correlation) relevant to the $L$-observables obtained using a combined analysis of LCSR and LQCD results.}\label{tab:combined}
\end{table}
%\vspace*{-0.4cm}

%\noindent
The plots of the remaining form factors are included as ancillary files attached to the arXiv version of this paper. 
We also provide the central values and covariance matrices of the posterior distributions for the $\alpha_n^i$ parameters of the $z$-expansion in the ancillary files
\begin{align*}
    & \texttt{BSZ\_Bq\_to\_K.yaml} \quad \texttt{and} \quad
     \texttt{BSZ\_Bq\_to\_Kstar.yaml}
\end{align*}

To assess the impact of the HPQCD 2022 results~\cite{Parrott:2022rgu} on our analysis, we repeat the fit excluding them. 
For example, we find that the central value of $f_0^{B_d \to K}(q^2 = 0)$ remains nearly unchanged, while the associated uncertainties increase by approximately 50\% compared to those reported above.
We conclude that the HPQCD 2022 inputs play a significant role in constraining the form factors and enhancing the overall precision of the fit. 
Nevertheless, they are largely consistent with the other inputs and have only a minimal impact on the central values.

We also perform a fit using only lattice QCD inputs, excluding the LCSR results. 
As expected, we find that LCSRs play a crucial role in constraining the $B_{d,s} \to K^*$ form factors, while their impact on the $B_{d,s} \to K$ form factors is marginal but not negligible.
For instance we find $f_0^{B_d \to K}(q^2 = 0) = 0.3318 \pm 0.0101$ and $f_0^{B_s \to K}(q^2 = 0) = 0.2605 \pm 0.0265$.
Since the fits excluding the HPQCD 2022 inputs and LCSRs were performed solely to test the stability of our results, we do not present their outcomes. 
We conclude this section by noting that the uncertainties of the present analysis could be further reduced by including unitarity constraints~\cite{Boyd:1997kz,Gubernari:2023puw}.
Such an analysis should take into account the subthreshold branch cuts appearing in these decays, and hence use the procedure outlined in Ref.~\cite{Gopal:2024mgb}.

\section{Updated predictions for the $L$-observables}
\label{sec:Lobs}

In this Section, we present updated SM predictions for a set of $L$-observables, based on the form factor results discussed in \refsec{FFs}. These predictions incorporate not only improved form factor calculations but also, for the first time, their correlations—previously unavailable in our earlier works~\cite{Alguero:2020xca,Biswas:2023pyw}. We place particular emphasis on how the predictions vary with the different form factor determinations discussed in this paper.

%In this section, we also analyze the internal structure of the $L$-observables to show how U-spin breaking enters and how the different determinations of the size of this breaking impact the predictions of the $L$-observables.
%the importance of a precise determination of form factors.
%
%Finally, we perform a comparison between the different predictions of the $L$-observables from the perspective of U-spin breaking.

%This section is organized as follows. In subsection~\ref{sec:subdef} we collect the main results concerning the updated predictions for the $L$-observables. In  subsection~\ref{sub:uspin} we  
%analyze the internal structure of the $L$-observables to show how U-spin breaking enters these observables.  And, finally, in subsection~\ref{sub:results} we discuss how the different determinations of form factors that imply different size of U-spin breaking impact the predictions of the $L$-observables found in subsection~\ref{sec:subdef}. 

This Section is structured as follows. In \refsubsec{subdef}, we present the main results for the updated predictions of the $L$-observables. \refsubsec{uspin} explores the internal structure of these observables, highlighting the role of U-spin breaking. In \refsubsec{results}, we examine how different form factor determinations, each implying a different size of U-spin breaking, affect the predictions introduced in \refsubsec{subdef}. We conclude this Section with a discussion on how different determinations of the inverse moments of the distribution amplitudes impact our predictions.

%Finally, in subsection~\ref{sub:results} we  discuss the predictions presented in subsection~\ref{sec:subdef} from this perspective.

%interpret the observables as U-spin tests 
%focusing on the different sources of U-spin breaking. 

%and in subsection\ref{sub:results} we will discuss the  presented predictions from this perspective.

\bigskip

\subsection{Definitions of the $L$-observables: present and updated SM predictions}\label{sec:subdef}

In this subsection, we recall the definitions of the $L_{{K}^* \bar{K}^*}$ and $L_{K\bar{K}}$ observables and 
introduce two new observables, which are  modified versions of those in Ref.~\cite{Biswas:2023pyw},
to make them more experimentally accessible~\cite{privatedavide}.
For %completeness and 
comparison, we provide  their SM predictions using updated form factors under two different determinations. We also include the earlier predictions from Ref.~\cite{Biswas:2023pyw}, which did not account for form factor correlations, along with available experimental measurements.

We label the updated predictions for the $L$-observables according to the form factor determinations used: ``I" refers to predictions based solely on LCSR, while ``II" corresponds to a combined analysis incorporating both LCSR and lattice QCD data. 
%For the specific case of the $B_{d,s} \to K^0\bar{K}^0$ decays, 
Note that the results of determination ``III'', that rely on the $B_{d,s} \to K$ form factors, are closely aligned with those of ``II'', since, as discussed in \refsec{FFs}, these form factor predictions are dominated by lattice QCD data.
In contrast, the results of determination ``III'' that rely on the $B_{d,s} \to K^*$ form factors cannot be fully trusted, as lattice QCD calculations in these cases suffer from large uncertainties and are only available at high $q^2$. Consequently, one must rely on a rather uncertain extrapolation.
For these reasons, our analysis primarily focuses on determinations ``I" and ``II".

%\begin{itemize}
%\item I: We use only Light-cone sum rule calculations to compute the form factors. 

%\item II: We use a combined analysis of light-cone sum rules together with lattice datasets.
% \end{itemize}

The set of $L$-observables relevant for our discussion are:
\begin{itemize}
\item $L_{K^*\bar{K}^*}$  defined by \cite{Alguero:2020xca,Biswas:2023pyw}
\begin{eqnarray}\label{eq:LKstarKstar}
L_{K^*\bar{K}^*}&=&\rho(m_{K^{*0}},m_{K^{*0}})\frac{{\cal B}({\bar{B}_s \to K^{*0} {\bar K^{*0}}})}{{\cal B}({\bar{B}_d \to K^{*0} {\bar K^{*0}})}}\frac{ f_L^{B_s}}{ f_L^{B_d}}  \cr
&=&\frac{|A_0^s|^2+ |\bar A_0^s|^2}{|A_0^d|^2+ |\bar A_0^d|^2}\,.
\end{eqnarray}
See Ref.~\cite{Alguero:2020xca} for definitions of the phase space factor $\rho$, the longitudinal polarization fractions $f_{L}^{B_{s,d}}$ and the corresponding longitudinal amplitudes $A_0^{s,d}$.
The SM prediction for this observable within the QCD Factorization (QCDF) framework,
using the input
parameters listed in 
 \reftab{inputs} 
  and the previous form factor set (without correlations) from \reftab{table6},
is given by \cite{Biswas:2023pyw}\footnote{Notice that, as mentioned before, there are tiny differences with the SM predictions in Ref.\cite{Biswas:2023pyw} using the previous form factors due to the use of the updated value for $\lambda_{B_d}$. }
 \begin{equation} 
L_{K^*\bar{K}^*}^{\rm SM}=
19.57^{+9.42}_{-6.70}\,,\label{eq:SMKstarKstarprev}\end{equation}
  while the experimental measurement~\cite{BaBar:2007wwj,LHCb:2019bnl} is given by
 \begin{equation} \label{eq:expdataLKstKst}
L_{K^*\bar{K}^*}^{\rm exp}=
4.43\pm 0.92 \,,
 \end{equation}
where a 7\% relative uncertainty has been added to the experimental number corresponding to the correction due to the $B_s$-mixing effect (see 
Ref.~\cite{Descotes-Genon:2011rgs}). This effect is negligible for $B_d$ but can be significant for $B_s$. It comes from the $B_s$ mesons originating from a  $b\bar{b}$ pair incoherent production at LHCb that decay into CP-eigenstates.

 Now with the new inputs and correlations the updated SM prediction  becomes for each form factor determination:
 \begin{equation} \label{eq:SMKstarKstarnew}
  L_{K^*\bar{K}^*}^{I, \, \rm SM}=
18.34^{+7.47}_{-5.83} \,, \quad \quad  \quad L_{K^*\bar{K}^*}^{II, \, \rm SM}=
26.08^{+5.70}_{-4.72} \,.
\end{equation}
{For completeness, we also provide the SM prediction for $L_{K^*\bar{K}^*}$  when only the lattice inputs for the form factors from ref.~\cite{Horgan:2015vla} are considered (albeit without the complete information on the corresponding correlations)\footnote{{Notice the increase of uncertainties in this latter case driven by the significant increase of the uncertainty of the lattice form factors.}
}:}
 \begin{equation} \label{eq:SMKK}
  {L_{K^*\bar{K}^*}^{III, \, \rm SM}=22.88^{+15.35}_{-8.46}}.
\end{equation}
Eq.~\ref{eq:SMKstarKstarnew}  implies that the previously observed tension of 2.6$\sigma$~\footnote{We should mention that following Refs~\cite{Biswas:2023pyw,Biswas:2024bhn}  this 2.6$\sigma$ tension was obtained using $1/L_{K^*\bar{K}^*}$  whose p.d.f. for the difference between theory and experiment is more symmetric. If $L_{K^*\bar{K}^*}$ is used the deviation comes down to 2.3$\sigma$. All other observables, included $L_{K^*\bar{K}^*}$ with new form factors, are not so asymmetric and can  be calculated as usual. } between the SM prediction and experimental data, based on our earlier form factor set, slightly decreases to 2.3$\sigma$ under determination ``I", due to a lower central value. In contrast the tension significantly increases to 4.4$\sigma$ under determination ``II". Using the Wilson coefficients from the effective Hamiltonian describing these transitions (see \refapp{WET}), one can express the explicit dependence of $L_{{K}^*\bar{K}^*}$ on the NP Wilson coefficients in the three determinations\footnote{Eq.~(\ref{eq:L_NP_cen}) and all subsequent equations showing the dependence of the various observables discussed in this paper on the NP Wilson Coefficients $C_{id,s}^{NP}, i=4,6,8g$ have been obtained by substituting the central values of the various relevant inputs provided in Table~\ref{tab:inputs} in the corresponding expressions for the observable(s) under consideration. Furthermore, notice that the small discrepancy between the SM prediction in \refeq{SMKstarKstarnew} compared to \refeq{L_NP_cen} and \refeq{L_NP_cen2} arises because the former is derived from the median of the distribution, while the latter uses central values of the input parameters.} 
\begin{align}\label{eq:L_NP_cen}
\!\!L_{K^* \bar{K}^*}^{I} = 
&\; (70.71 - 3401.42\; {\cal C}_{4s}^{\rm NP} + 51471.40 \;({\cal C}_{4s}^{\rm NP})^2 + 196.92\; {\cal C}_{6s}^{\rm NP} + 261.12\; ({\cal C}_{6s}^{\rm NP})^2 
\nonumber \\
& + 176.60\; {\cal C}_{8gs}^{\rm NP}
  + 130.50\; ({\cal C}_{8gs}^{\rm NP})^2 - 2446.62\; {\cal C}_{4s}^{\rm NP}\; {\cal C}_{6s}^{\rm NP}
  - 5172.27\; {\cal C}_{4s}^{\rm NP}\; {\cal C}_{8gs}^{\rm NP} \nonumber \\ & + 
 145.70\; {\cal C}_{6s}^{\rm NP}\; {\cal C}_{8gs}^{\rm NP})\;  /\;  (3.87 - 184.99\; {\cal C}_{4d}^{\rm NP} + 2802.64 \;({\cal C}_{4d}^{\rm NP})^2 + 12.09\; {\cal C}_{6d}^{\rm NP} 
\nonumber \\
& + 15.68\; ({\cal C}_{6d}^{\rm NP})^2  + 10.11\; {\cal C}_{8gd}^{\rm NP}  + 7.87\; ({\cal C}_{8gd}^{\rm NP})^2 - 169.76\; {\cal C}_{4d}^{\rm NP}\; {\cal C}_{6d}^{\rm NP}
  - 296.34\; {\cal C}_{4d}^{\rm NP}\; {\cal C}_{8gd}^{\rm NP} \nonumber \\ & + 
 10.32\; {\cal C}_{6d}^{\rm NP}\; {\cal C}_{8gd}^{\rm NP})\,,
 \end{align}
 \begin{align}\label{eq:L_NP_cen2}\!\!L_{K^* \bar{K}^*}^{II} = 
&\; (94.83 - 4535.03\; {\cal C}_{4s}^{\rm NP} + 68302.20 \;({\cal C}_{4s}^{\rm NP})^2 + 278.14\; {\cal C}_{6s}^{\rm NP} + 362.95\; ({\cal C}_{6s}^{\rm NP})^2 
\nonumber \\
& + 237.13\; {\cal C}_{8gs}^{\rm NP}
  + 175.54\; ({\cal C}_{8gs}^{\rm NP})^2 - 3657.50\; {\cal C}_{4s}^{\rm NP}\; {\cal C}_{6s}^{\rm NP}
  - 6910.15\; {\cal C}_{4s}^{\rm NP}\; {\cal C}_{8gs}^{\rm NP} \nonumber \\ & + 
 215.95\; {\cal C}_{6s}^{\rm NP}\; {\cal C}_{8gs}^{\rm NP})\;  /\;  (3.63 - 173.88\; {\cal C}_{4d}^{\rm NP} + 2637.57 \;({\cal C}_{4d}^{\rm NP})^2 + 11.26\; {\cal C}_{6d}^{\rm NP} 
\nonumber \\
& + 14.63\; ({\cal C}_{6d}^{\rm NP})^2  + 9.49\; {\cal C}_{8gd}^{\rm NP}  + 7.39\; ({\cal C}_{8gd}^{\rm NP})^2 - 157.02\; {\cal C}_{4d}^{\rm NP}\; {\cal C}_{6d}^{\rm NP}
  - 278.56\; {\cal C}_{4d}^{\rm NP}\; {\cal C}_{8gd}^{\rm NP} \nonumber \\ & + 
 9.55\; {\cal C}_{6d}^{\rm NP}\; {\cal C}_{8gd}^{\rm NP})\,  \,, \end{align}  and
 \begin{align}\label{eq:L_NP_cen3}\!\!{L_{K^* \bar{K}^*}^{III} =} 
&\; {(83.24 - 3990.54\; {\cal C}_{4s}^{\rm NP} + 60222.90 \;({\cal C}_{4s}^{\rm NP})^2 + 238.88\; {\cal C}_{6s}^{\rm NP} + 313.68\; ({\cal C}_{6s}^{\rm NP})^2} 
\nonumber \\
& {+ 208.03\; {\cal C}_{8gs}^{\rm NP}
  + 153.88\; ({\cal C}_{8gs}^{\rm NP})^2 - 3069.99\; {\cal C}_{4s}^{\rm NP}\; {\cal C}_{6s}^{\rm NP}
  - 6075.21\; {\cal C}_{4s}^{\rm NP}\; {\cal C}_{8gs}^{\rm NP}} \nonumber \\ & {+ 
 181.86\; {\cal C}_{6s}^{\rm NP}\; {\cal C}_{8gs}^{\rm NP})\;  /\;  (3.66 - 175.06\; {\cal C}_{4d}^{\rm NP} + 2655.02 \;({\cal C}_{4d}^{\rm NP})^2 + 11.35\; {\cal C}_{6d}^{\rm NP}} 
\nonumber \\
& {+ 14.74\; ({\cal C}_{6d}^{\rm NP})^2  + 9.56\; {\cal C}_{8gd}^{\rm NP}  + 7.44\; ({\cal C}_{8gd}^{\rm NP})^2 - 158.36\; {\cal C}_{4d}^{\rm NP}\; {\cal C}_{6d}^{\rm NP}
  - 280.44\; {\cal C}_{4d}^{\rm NP}\; {\cal C}_{8gd}^{\rm NP}} \nonumber \\ & {+ 
 9.63\; {\cal C}_{6d}^{\rm NP}\; {\cal C}_{8gd}^{\rm NP})\,.} 
 \end{align}

In the latter expression, the sensitivity to the NP Wilson coefficients, reflected in the prefactors multiplying each coefficient, is enhanced in determination ``II" compared to our previous result in Ref.~\cite{Biswas:2023pyw}. This enhancement is consistent with expectations and is approximately proportional to the ratio of the SM central values between the new and the old determination.

\item $L_{K\bar{K}}$  defined by \cite{Biswas:2023pyw}
\begin{eqnarray}\label{eq:LKtKt}
L_{K\bar{K}}&=&\rho(m_{K^0},m_{K^0})\frac{{\cal B}({\bar{B}_s \to K^{0} {\bar K^{0}}})}{{\cal B}({\bar{B}_d \to K^{0} {\bar K^{0}}})} \nonumber \cr &=&\frac{|A^s|^2+ |\bar A^s|^2}{|A^d|^2+ |\bar A^d|^2}\,.
\end{eqnarray}

The SM prediction for this observable  obtained within QCDF \cite{Beneke:2003zv}  %and taking as inputs of App.~\ref{app:QCDF} 
using the previous set of form factors (with no correlations) in Table.~\ref{tab:table6} is
 \begin{equation} 
L_{K\bar{K}}^{\rm SM}=25.90^{+3.90}_{-3.60}\,.\end{equation}

On the experimental side, 
combining the data of Refs.~\cite{Belle:2012dmz,BaBar:2006enb,LHCb:2020wrt, Belle:2015gho}
following PDG~\cite{Workman:2022ynf} and adding a 7\% uncertainty due to $B_s$-mixing effect one finds
for the experimental value of this observable
\begin{equation}
L_{K\bar{K}}^{\rm exp}=14.58\pm 3.37 \,.\end{equation}

Using the newly updated and correlated form factor inputs, the SM predictions for the two determinations are
 \begin{equation} \label{eq:SMKstarKstar}
  L_{K\bar{K}}^{I, \, \rm SM}=
25.21^{+8.64}_{-7.67} \,, \quad \quad  \quad L_{K\bar{K}}^{II, \, \rm SM}=
17.88^{+2.55}_{-2.42} \,.\end{equation}
The previous tension of 2.4$\sigma$ (2.3$\sigma$ with the new $\lambda_{B_d}$) observed in our earlier determination is now reduced. With the updated inputs, both determinations ``I" and ``II" yield predictions that are consistent with the corresponding experimental measurement, within 1$\sigma$ for determination ``II", and slightly above 1$\sigma$ for determination ``I". If only lattice inputs are considered, the SM prediction becomes
 \begin{equation} \label{eq:SMKK}
  L_{K\bar{K}}^{III, \, \rm SM}=
16.51^{+3.50}_{-3.27} \,,\end{equation}
which is even closer to the experimental measurement. The explicit dependence on the NP Wilson coefficients of determination ``I" is in this case
\begin{align}
\!\!L_{K \bar{K}}^{I} = &\; (33.00 - 479.04\; {\cal C}_{4s}^{\rm NP} + 2032.66\; ({\cal C}_{4s}^{\rm NP})^2 - 809.44\; {\cal C}_{6s}^{\rm NP} + 5528.73\; ({\cal C}_{6s}^{\rm NP})^2
\nonumber \\
&+ 39.68\; {\cal C}_{8gs}^{\rm NP} 
  + 12.56\; ({\cal C}_{8gs}^{\rm NP})^2 + 
 6690.55\; {\cal C}_{4s}^{\rm NP}\; {\cal C}_{6s}^{\rm NP} - 315.29\; {\cal C}_{4s}^{\rm NP}\; {\cal C}_{8gs}^{\rm NP} \nonumber \\
& - 524.47\; {\cal C}_{6s}^{\rm NP}\; {\cal C}_{8gs}^{\rm NP})\;  /\;  (1.31 - 19.24\; {\cal C}_{4d}^{\rm NP} + 82.71 \;({\cal C}_{4d}^{\rm NP})^2 - 32.02\; {\cal C}_{6d}^{\rm NP} 
\nonumber \\
& + 218.33\; ({\cal C}_{6d}^{\rm NP})^2  + 1.72\; {\cal C}_{8gd}^{\rm NP}  + 0.60\; ({\cal C}_{8gd}^{\rm NP})^2 + 268.21\; {\cal C}_{4d}^{\rm NP}\; {\cal C}_{6d}^{\rm NP}
  - 13.86\; {\cal C}_{4d}^{\rm NP}\; {\cal C}_{8gd}^{\rm NP} \nonumber \\ & - 
 22.71\; {\cal C}_{6d}^{\rm NP}\; {\cal C}_{8gd}^{\rm NP}) \,,\end{align}
and for determination ``II"
\begin{align} 
\!\!L_{K \bar{K}}^{II} = &\; (32.70 - 474.64\; {\cal C}_{4s}^{\rm NP} + 2013.23\; ({\cal C}_{4s}^{\rm NP})^2 - 802.41\; {\cal C}_{6s}^{\rm NP} + 5481.57\; ({\cal C}_{6s}^{\rm NP})^2
\nonumber \\
&+ 39.30\; {\cal C}_{8gs}^{\rm NP} 
  + 12.43\; ({\cal C}_{8gs}^{\rm NP})^2 + 
 6630.02\; {\cal C}_{4s}^{\rm NP}\; {\cal C}_{6s}^{\rm NP} - 312.19\; {\cal C}_{4s}^{\rm NP}\; {\cal C}_{8gs}^{\rm NP} \nonumber \\
& - 519.58\; {\cal C}_{6s}^{\rm NP}\; {\cal C}_{8gs}^{\rm NP})\;  /\;  (1.82 - 26.96\; {\cal C}_{4d}^{\rm NP} + 117.05 \;({\cal C}_{4d}^{\rm NP})^2 - 44.19\; {\cal C}_{6d}^{\rm NP} 
\nonumber \\
& + 299.83\; ({\cal C}_{6d}^{\rm NP})^2  + 2.44\; {\cal C}_{8gd}^{\rm NP}  + 0.85\; ({\cal C}_{8gd}^{\rm NP})^2 + 373.93\; {\cal C}_{4d}^{\rm NP}\; {\cal C}_{6d}^{\rm NP}
  - 19.77\; {\cal C}_{4d}^{\rm NP}\; {\cal C}_{8gd}^{\rm NP} \nonumber \\ & - 
 31.91\; {\cal C}_{6d}^{\rm NP}\; {\cal C}_{8gd}^{\rm NP}) \,,\end{align}
where we observe a decrease in the absolute value of all prefactors compared to our previous expression in Ref.~\cite{Biswas:2023pyw}  close to the value obtained by the ratio of SM central value predictions. {For completeness we also provide the NP WC dependence for determination ``III":}
\begin{align} 
\!\!{L_{K \bar{K}}^{III} =} &\; {(29.87 - 432.63\; {\cal C}_{4s}^{\rm NP} + 1828.04\; ({\cal C}_{4s}^{\rm NP})^2 - 735.29\; {\cal C}_{6s}^{\rm NP} + 5030.74\; ({\cal C}_{6s}^{\rm NP})^2}
\nonumber \\
&{+ 35.70\; {\cal C}_{8gs}^{\rm NP} 
  + 11.22\; ({\cal C}_{8gs}^{\rm NP})^2 + 
 6052.22\; {\cal C}_{4s}^{\rm NP}\; {\cal C}_{6s}^{\rm NP} - 282.64\; {\cal C}_{4s}^{\rm NP}\; {\cal C}_{8gs}^{\rm NP} }\nonumber \\
& {- 472.92\; {\cal C}_{6s}^{\rm NP}\; {\cal C}_{8gs}^{\rm NP})\;  /\;  (1.81 - 26.73\; {\cal C}_{4d}^{\rm NP} + 116.054 \;({\cal C}_{4d}^{\rm NP})^2 - 43.84\; {\cal C}_{6d}^{\rm NP} }
\nonumber \\
& {+ 297.45\; ({\cal C}_{6d}^{\rm NP})^2  + 2.41\; {\cal C}_{8gd}^{\rm NP}  + 0.85\; ({\cal C}_{8gd}^{\rm NP})^2 + 370.84\; {\cal C}_{4d}^{\rm NP}\; {\cal C}_{6d}^{\rm NP}
  - 19.59\; {\cal C}_{4d}^{\rm NP}\; {\cal C}_{8gd}^{\rm NP}} \nonumber \\ & {- 
 31.64\; {\cal C}_{6d}^{\rm NP}\; {\cal C}_{8gd}^{\rm NP}). } 
\end{align}

%Comment on results: and reduction of theoretical uncertainties by $35\% $....

\end{itemize}

\noindent
In our previous work, Ref.~\cite{Biswas:2023pyw}, we introduced two observables, $\hat{L}_{K^*}$ and $\hat{L}_K$ defined according to the meson that carries the spectator quark.
\begin{itemize}
\item ${\hat L}_{{K}^{*}}$, where the vector 
$K^{*0}$ 
      collects the spectator quark, and it is defined by:
    \begin{eqnarray}\label{eq:LKst-def}
{\hat L}_{{K}^{*}}&=&\rho(m_{K^0},m_{K^{*0}})\frac{{\cal B}({{\bar B}_s\to {{ K^{*0}}\bar K}^{0})}
}{{\cal B}({{\bar B}_d \to {\bar K}^{*0} { K^{0}})}} \nonumber \cr &=&\frac{|A^s|^2+ |\bar A^s|^2}{|A^d|^2+ |\bar A^d|^2}\,.
\end{eqnarray} 
 The SM prediction for this observable, based on our previous form factor set, was reported in Ref.~\cite{Biswas:2023pyw}
 %is obtained from \refeq{:Lgeneraldiscussion} simply changing $P_{d,s}$ and $\Delta_q$ by the corresponding ones of these modes:
     \begin{equation}
   {\hat L}^{\rm SM}_{{K}^{*}} =21.55^{+7.16}_{-6.23} \,.    \end{equation} 

\item ${\hat L}_{K}$, where the pseudoscalar 
$K^0$  collects the spectator quark and it is defined by:
    \begin{eqnarray}\label{eq:LK-def}
{\hat L}_{K}  &=&\rho(m_{K^0},m_{K^{*0}})\frac{{\cal B}({{\bar B}_s \to 
{ K^{0}}{\bar K}^{*0} )}}{{\cal B}({{\bar B}_d \to {\bar K}^{0} { K^{*0}})}} \nonumber \cr
&=&\frac{|A^s|^2+ |\bar A^s|^2}{|A^d|^2+ |\bar A^d|^2}\,,
\end{eqnarray}   
and the corresponding  SM prediction using the same form factors  is \cite{Biswas:2023pyw}:
   \begin{equation}
    {\hat L}^{\rm SM}_{K}= 25.64^{+4.71}_{-4.84} \,.
    \end{equation}
\end{itemize}
When considered together, the observables $\hat{L}_{K^*}$ and $\hat{L}_K$ exhibit a distinctive behaviour in the presence of the NP required to explain the anomalies in $L_{K^*\bar{K}^*}$ and $L_{K\bar{K}}$. Specifically, $\hat{L}_{K^*}$ becomes significantly enhanced, while $\hat{L}_K$ is suppressed. However, due to experimental challenges in disentangling the decay modes appearing in the numerators and denominators of $\hat{L}_K$ and $\hat{L}_{K^*}$, in Ref.~\cite{Biswas:2023pyw} alternative observables called $L_K$ and $L_{K^*}$ were proposed where the $B_d$ mode is not flavour-tagged.
In this work, we do not update those observables. Instead, following a similar strategy, we define two new observables. The motivation for these new definitions is twofold~\cite{privatedavide}: (i) the $B_d$ meson has a much lower oscillation frequency compared to the $B_s$, making it more likely to decay before oscillating when produced and tagged; and (ii) a time-integrated analysis is feasible for $B_d$ due to its slow oscillation, whereas the rapid oscillation of $B_s$ complicates such an analysis. Consequently, flavour tagging is more effective for $B_d$ than for $B_s$.
Therefore, the new observables we define and predict in the following are:

%These two observables, when considered together, exhibit a unique mechanism, namely  in presence of the NP required to explain $L_{K^*\bar{K}^*}$ and $L_{KK}$ they behave oppositely, i.e., one gets largely enhanced ($\hat{L}_{K^*}$) while the other one gets suppressed ($\hat{L}_K$). However, due to the experimental difficulty of disentangling the two modes  in the numerators (denominators), respectively, of $\hat{L}_K$ and in $\hat{L}_{K^*}$ it was proposed in \cite{Biswas:2023pyw} to introduce  a set of other observables ${L}_K$ and ${L}_{K^*}$
%where the $B_d$ mode is not tagged. We will, however, not update these observables but following 
% a similar strategy we define two new observables. 
% The rationale behind the definition of these new observables is the following~\cite{privatedavide}: i) $B_d$ has a much lower oscillation frequency with respect to $B_s$ and, therefore, after it is produced and tagged, the probability of it decays before oscillation is high, ii) a time-integrated analysis is viable for $B_d$ precisely because the oscillation frequency is low, while $B_s$ has a high oscillation frequency that makes it difficult to perform the time integral. Consequently it is better to flavour tag on the $B_d$ than on the $B_s$ due to the different oscillation frequency.
% Therefore the new observables that we will define and update in the following, are:
%also update the SM prediction of these observables but in order to facilitate the contact with experiment we will also introduce a modified version of these observables:

\begin{itemize}

\item 
$\tilde{L}_{K^*}$ defined by 
\begin{eqnarray} \label{eq:newLtildeKstar}
\tilde L_{K^*}= \frac{1}{\rho(m_{K^0},m_{K^*})} \frac{{\cal B}(\bar{B_d}\to\bar{K}^{*0} K^0)}{{\cal B}(\bar{B_s}\to K^{*0} \bar{K}^0)+{\cal B}(\bar{B_s}\to K^0\bar{K}^{*0})}=\frac{1}{L_{\rm total}}\frac{1}{(1+1/R_d)} \,,\end{eqnarray}
where the $\bar{K}^{*0}$ in the decay of the  numerator collects the spectator quark. We also provide in \refeq{newLtildeKstar}
the
relation of the newly defined  observable in terms of the observables $L_{\rm total}$   and $R_d$ introduced  in Ref.~\cite{Biswas:2023pyw}. Both of these observables are defined below and their updated predictions are presented accordingly. The observable $R_d$ is defined by
\begin{equation}
    R_d=\frac{{\cal B}({{\bar B}_d \to {\bar K}^{*0} {{ K}^{0}})}}{{\cal B}({{\bar B}_d \to {\bar K}^{0} { K^{*0}})}},\,
 \end{equation} 
 whose value using the previous form factor set is given by\footnote{
 The observable 
$R_d$
  is found to be particularly sensitive to the value of 
$\lambda_{B_d}$
  employed in the analysis. In addition to this sensitivity, we identified an issue in our previous computation of $R_d$
  presented in Ref.~\cite{Biswas:2023pyw}, where an incorrect treatment of the renormalization scale uncertainty and the annihilation contributions led to an inaccurate prediction for 
$R_d$ that we corrected here.
 }

\begin{equation}
R^{\rm SM}_d=0.87^{+0.33}_{-0.23} \,,\end{equation}

with updated values
\begin{equation}
\quad\quad  R_{d}^{I,\;\rm SM}=1.25^{+1.51}_{-0.56}\,,\quad \quad \quad\quad R_{d}^{II,\;\rm SM}=0.84^{+0.26}_{-0.19}\,,\quad \quad \quad\quad {R_{d}^{III,\;\rm SM}=0.84^{+0.42}_{-0.28}} \,.\end{equation}
The observable $R_{d}$ exhibits a significant sensitivity to the size of the uncertainties in the form factors that grows as the central value rises. %In  particular $R_{d}^{I,\;\rm SM}$ has a larger uncertainty in comparison with $R_{d}^{II,\;\rm SM}$.
Indeed, if we artificially reduce the uncertainties for the purely LCSR determination of the  form factors by a factor of $75\%$\footnote{Notice that the uncertainty of the $B_d \to K$ form factor using LCSR or combined differ by one order of magnitude.} we get a reduction of 60\% on the larger error of $R_{d}^{I,\;\rm SM}$.
%$R_{d, new}^{I,\;\rm SM}=1.26^{+0.61}_{-0.35}$.  
%
The explicit dependence of $R_d$ on the NP Wilson coefficients, for {all the three} form factor determinations, is given by 
\begin{align} \label{eq:RdI}
\quad\quad  \!\!R_d^I = &\; (3.95 + 206.06\; {\cal C}_{4d}^{\rm NP} + 2692.17\; ({\cal C}_{4d}^{\rm NP})^2 - 293.76\; {\cal C}_{6d}^{\rm NP} + 5491.88\; ({\cal C}_{6d}^{\rm NP})^2
\nonumber \\
&- 6.33\; {\cal C}_{8gd}^{\rm NP} 
  + 2.55\; ({\cal C}_{8gd}^{\rm NP})^2 - 
 7684.07\; {\cal C}_{4d}^{\rm NP}\; {\cal C}_{6d}^{\rm NP} - 165.59\; {\cal C}_{4d}^{\rm NP}\; {\cal C}_{8gd}^{\rm NP} \nonumber \\
& + 236.55\; {\cal C}_{6d}^{\rm NP}\; {\cal C}_{8gd}^{\rm NP})\;  /\;  (3.35 - 165.28\; {\cal C}_{4d}^{\rm NP} + 2288.16 \;({\cal C}_{4d}^{\rm NP})^2 + 10.54\; {\cal C}_{6d}^{\rm NP} 
\nonumber \\
& + 13.62\; ({\cal C}_{6d}^{\rm NP})^2  + 9.08\; {\cal C}_{8gd}^{\rm NP}  + 6.65\; ({\cal C}_{8gd}^{\rm NP})^2 - 191.88\; {\cal C}_{4d}^{\rm NP}\; {\cal C}_{6d}^{\rm NP}
  - 246.23\; {\cal C}_{4d}^{\rm NP}\; {\cal C}_{8gd}^{\rm NP} \nonumber \\ &  
 + 11.31\; {\cal C}_{6d}^{\rm NP}\; {\cal C}_{8gd}^{\rm NP}) \,,\end{align}
\begin{align} \label{eq:RdII}
\quad\quad  \!\!R_d^{II} = &\; (3.76 + 195.15\; {\cal C}_{4d}^{\rm NP} + 2533.37\; ({\cal C}_{4d}^{\rm NP})^2 - 278.65\; {\cal C}_{6d}^{\rm NP} + 5184.39\; ({\cal C}_{6d}^{\rm NP})^2
\nonumber \\
&- 5.99\; {\cal C}_{8gd}^{\rm NP} 
  + 2.39\; ({\cal C}_{8gd}^{\rm NP})^2 - 
 7242.35\; {\cal C}_{4d}^{\rm NP}\; {\cal C}_{6d}^{\rm NP} - 155.65\; {\cal C}_{4d}^{\rm NP}\; {\cal C}_{8gd}^{\rm NP} \nonumber \\
& + 222.70\; {\cal C}_{6d}^{\rm NP}\; {\cal C}_{8gd}^{\rm NP})\;  /\;  (4.85 - 237.02\; {\cal C}_{4d}^{\rm NP} + 3250.14 \;({\cal C}_{4d}^{\rm NP})^2 + 16.14\; {\cal C}_{6d}^{\rm NP} 
\nonumber \\
& + 20.79\; ({\cal C}_{6d}^{\rm NP})^2  + 13.10\; {\cal C}_{8gd}^{\rm NP}  + 9.56\; ({\cal C}_{8gd}^{\rm NP})^2 - 299.13\; {\cal C}_{4d}^{\rm NP}\; {\cal C}_{6d}^{\rm NP}
  - 351.95\; {\cal C}_{4d}^{\rm NP}\; {\cal C}_{8gd}^{\rm NP} \nonumber \\ &  
 + 17.64\; {\cal C}_{6d}^{\rm NP}\; {\cal C}_{8gd}^{\rm NP}) \,,\end{align}
{and}
\begin{align} \label{eq:RdIII}
\quad\quad  \!\!{R_d^{III} =} &\; {(3.78 + 196.31\; {\cal C}_{4d}^{\rm NP} + 2550.15\; ({\cal C}_{4d}^{\rm NP})^2 - 280.25\; {\cal C}_{6d}^{\rm NP} + 5216.91\; ({\cal C}_{6d}^{\rm NP})^2}
\nonumber \\
&{- 6.03\; {\cal C}_{8gd}^{\rm NP} 
  + 2.41\; ({\cal C}_{8gd}^{\rm NP})^2 - 
 7289.05\; {\cal C}_{4d}^{\rm NP}\; {\cal C}_{6d}^{\rm NP} - 156.705\; {\cal C}_{4d}^{\rm NP}\; {\cal C}_{8gd}^{\rm NP}} \nonumber \\
& {+ 224.16\; {\cal C}_{6d}^{\rm NP}\; {\cal C}_{8gd}^{\rm NP})\;  /\;  (4.81 - 234.90\; {\cal C}_{4d}^{\rm NP} + 3221.87 \;({\cal C}_{4d}^{\rm NP})^2 + 15.97\; {\cal C}_{6d}^{\rm NP}} 
\nonumber \\
& {+ 20.58\; ({\cal C}_{6d}^{\rm NP})^2  + 12.98\; {\cal C}_{8gd}^{\rm NP}  + 9.48\; ({\cal C}_{8gd}^{\rm NP})^2 - 295.93\; {\cal C}_{4d}^{\rm NP}\; {\cal C}_{6d}^{\rm NP}
  - 344.84\; {\cal C}_{4d}^{\rm NP}\; {\cal C}_{8gd}^{\rm NP}} \nonumber \\ &  
 {+ 17.45\; {\cal C}_{6d}^{\rm NP}\; {\cal C}_{8gd}^{\rm NP}).} 
\end{align}

 %
 % will help to understand the behaviour of $\tilde{L}_{K^*}$ in presence of New Physics that will be discussed in Sec.\ref{sec:np}.
  %
 The SM prediction of $\tilde{L}_{K^*}$ using the previous form factors is %(\qm{with central value corresponding, as in the previous cases, to the median of the distribution}) 
 \begin{equation}
{\tilde L}_{K^*}^{\rm SM}=0.019^{+0.007}_{-0.005} \,, \end{equation}
where the central value corresponds, as in the previous cases, to the median of the distribution. The corresponding SM prediction using 
the {three} form factor determinations are
\begin{equation}
\quad\quad  \tilde{L}_{K^*}^{I,\;\rm SM}=0.024^{+0.013}_{-0.008}\,,\quad \quad  \quad \tilde{L}_{K^*}^{II,\;\rm SM}=0.021^{+0.005}_{-0.004}\,,\quad \quad  \quad {\tilde{L}_{K^*}^{III,\;\rm SM}=0.023^{+0.011}_{-0.007}} \,.\end{equation}

\item 
$\tilde{L}_K$ defined by 
\begin{equation}
\tilde L_{K}=\frac{1}{\rho(m_{K^0},m_{K^*})} \frac{{\cal B}(\bar{B_d}\to\bar{K}^{0} K^{*0})}{{\cal B}(\bar{B_s}\to K^{*0} \bar{K}^0)+{\cal B}(\bar{B_s}\to K^0\bar{K}^{*0})}=\frac{1}{L_{\rm total}}\frac{1}{(1+R_d)} \,,\end{equation}
where in this case is the $\bar{K}^{0}$ in the decay of the  numerator the meson that collects the spectator quark.
The SM prediction of this observable using the previous form factors is
 \begin{equation}
{\tilde L}_{K}^{\rm SM}=0.022^{+0.005}_{-0.003}
\,,\end{equation}
and the corresponding SM prediction using 
the {three} form factor determinations are
\begin{equation}
\quad\quad  \tilde{L}_{K}^{I,\;\rm SM}=0.019^{+0.008}_{-0.007} \,,\quad \quad \quad\tilde{L}_{K}^{II,\;\rm SM}=0.025^{+0.005}_{-0.003}\,,\quad \quad \quad {\tilde{L}_{K}^{III,\;\rm SM}=0.027^{+0.008}_{-0.005}}.
\end{equation}

\end{itemize}
In these observables, we have inverted the ratio to place the suppressed mode, typically associated with higher statistical uncertainty, in the numerator. This choice improves the statistical behaviour of the observable, making it more Gaussian. Moreover, this new pair of observables exhibits a mechanism analogous to that of the $\hat{L}_{K^*}$–$\hat{L}_K$ pair, as will be discussed in Section~\ref{sec:np}.

Finally, to complete this Subsection, we update the prediction for $L_{\rm total}$, an observable that does not require tagging of either the $B_d$ or the $B_s$ meson. The definition is \cite{Biswas:2023pyw}
 \begin{eqnarray}
    {L}_{\rm total} &=& \rho(m_{K^0},m_{K^{*0}}) \left(\frac{{\cal B}({\bar{B}_s \to K^{*0} {\bar K^{0}})}+ {\cal B}({\bar{B}_s \to K^{0} {\bar K^{*0}}})}{{\cal B}({\bar{B}_d \to {\bar K}^{*0} { K^{0}})}+ {\cal B}({\bar{B}_d \to {\bar K}^{0} { K^{*0}})}}\right)\nonumber
    \\&=&
    \frac{1}{\tilde{L}_{K^*}+\tilde{L}_K} \,,    %=
   % \frac{\hat{L}_{K}+ \hat{L}_{K^*} R^d}{1+R^d}
\end{eqnarray}
whose SM prediction is:
    \begin{equation}
       {L}^{\rm SM}_{\rm total} = 23.74^{+4.27}_{-4.32} \,,    \end{equation}
and the updated value using the {three} form factor determinations is:
\begin{equation}
\quad\quad  {L}_{\rm total}^{I,\;\rm SM}=22.69^{+5.32}_{-5.32}  \,, \quad  \quad\quad {L}_{\rm total}^{II,\;\rm SM}=21.38^{+3.18}_{-3.06} \,, \quad  \quad\quad {{L}_{\rm total}^{III,\;\rm SM}=19.55^{+4.83}_{-4.53}}\,.\end{equation}
Finally, the explicit dependence of the observable on the NP Wilson coefficients is given by
\begin{align}
\quad\quad  \!\!L_{\rm total}^I = &\; (6.76 - 2.92\; {\cal C}_{4s}^{\rm NP} + 4408.03\; ({\cal C}_{4s}^{\rm NP})^2 - 230.67\; {\cal C}_{6s}^{\rm NP} + 4366.71\; ({\cal C}_{6s}^{\rm NP})^2
\nonumber \\
&+ 4.62\; {\cal C}_{8gs}^{\rm NP} 
  + 7.58\; ({\cal C}_{8gs}^{\rm NP})^2 - 
 6171.87 \; {\cal C}_{4s}^{\rm NP}\; {\cal C}_{6s}^{\rm NP} - 346.90\; {\cal C}_{4s}^{\rm NP}\; {\cal C}_{8gs}^{\rm NP} \nonumber \\
& + 164.00\; {\cal C}_{6s}^{\rm NP}\; {\cal C}_{8gs}^{\rm NP})\;  /\;  (0.30 + 1.71\; {\cal C}_{4d}^{\rm NP} + 208.33 \;({\cal C}_{4d}^{\rm NP})^2 - 11.85\; {\cal C}_{6d}^{\rm NP} 
\nonumber \\
& + 230.30\; ({\cal C}_{6d}^{\rm NP})^2  + 0.11\; {\cal C}_{8gd}^{\rm NP}  + 0.38\; ({\cal C}_{8gd}^{\rm NP})^2 - 329.45\; {\cal C}_{4d}^{\rm NP}\; {\cal C}_{6d}^{\rm NP}
  - 17.23\; {\cal C}_{4d}^{\rm NP}\; {\cal C}_{8gd}^{\rm NP} \nonumber \\ & + 
 10.37\; {\cal C}_{6d}^{\rm NP}\; {\cal C}_{8gd}^{\rm NP}) \,,\end{align}
\begin{align}
\quad\quad  \!\!L_{\rm total}^{II} = &\; (7.52 + 45.84\; {\cal C}_{4s}^{\rm NP} + 5063.91\; ({\cal C}_{4s}^{\rm NP})^2 - 296.15\; {\cal C}_{6s}^{\rm NP} + 5689.06\; ({\cal C}_{6s}^{\rm NP})^2
\nonumber \\
&+ 3.29\; {\cal C}_{8gs}^{\rm NP} 
  + 7.99\; ({\cal C}_{8gs}^{\rm NP})^2 - 
 8063.62 \; {\cal C}_{4s}^{\rm NP}\; {\cal C}_{6s}^{\rm NP} - 380.29\; {\cal C}_{4s}^{\rm NP}\; {\cal C}_{8gs}^{\rm NP} \nonumber \\
& + 213.88\; {\cal C}_{6s}^{\rm NP}\; {\cal C}_{8gs}^{\rm NP})\;  /\;  (0.36 - 1.75\; {\cal C}_{4d}^{\rm NP} + 241.92 \;({\cal C}_{4d}^{\rm NP})^2 - 10.98\; {\cal C}_{6d}^{\rm NP} 
\nonumber \\
& + 217.73\; ({\cal C}_{6d}^{\rm NP})^2  + 0.30\; {\cal C}_{8gd}^{\rm NP}  + 0.50\; ({\cal C}_{8gd}^{\rm NP})^2 - 315.46\; {\cal C}_{4d}^{\rm NP}\; {\cal C}_{6d}^{\rm NP}
  - 21.23\; {\cal C}_{4d}^{\rm NP}\; {\cal C}_{8gd}^{\rm NP} \nonumber \\ & + 
 10.05\; {\cal C}_{6d}^{\rm NP}\; {\cal C}_{8gd}^{\rm NP}) \,,\end{align}
and 
\begin{align}
\quad\quad  \!\!{L_{\rm total}^{III} =} &\; {(6.81 + 39.40\; {\cal C}_{4s}^{\rm NP} + 4522.42\; ({\cal C}_{4s}^{\rm NP})^2 - 266.14\; {\cal C}_{6s}^{\rm NP} + 5054.20\; ({\cal C}_{6s}^{\rm NP})^2}
\nonumber \\
&{+ 3.02\; {\cal C}_{8gs}^{\rm NP} 
  + 7.16\; ({\cal C}_{8gs}^{\rm NP})^2 - 
 7133.39 \; {\cal C}_{4s}^{\rm NP}\; {\cal C}_{6s}^{\rm NP} - 340.40\; {\cal C}_{4s}^{\rm NP}\; {\cal C}_{8gs}^{\rm NP}} \nonumber \\
& {+ 188.64\; {\cal C}_{6s}^{\rm NP}\; {\cal C}_{8gs}^{\rm NP})\;  /\;  (0.36 - 1.61\; {\cal C}_{4d}^{\rm NP} + 241.44 \;({\cal C}_{4d}^{\rm NP})^2 - 11.05\; {\cal C}_{6d}^{\rm NP}} 
\nonumber \\
& {+ 219.08\; ({\cal C}_{6d}^{\rm NP})^2  + 0.29\; {\cal C}_{8gd}^{\rm NP}  + 0.50\; ({\cal C}_{8gd}^{\rm NP})^2 - 317.28\; {\cal C}_{4d}^{\rm NP}\; {\cal C}_{6d}^{\rm NP}
  - 21.15\; {\cal C}_{4d}^{\rm NP}\; {\cal C}_{8gd}^{\rm NP}} \nonumber \\ & {+ 
 10.11\; {\cal C}_{6d}^{\rm NP}\; {\cal C}_{8gd}^{\rm NP})} \,.\end{align}

Notice that the sensitivity to
NP decreases with the requirement of flavour tagging. As a result, the highest sensitivity is achieved with the $\hat{L}$ observables, followed by a slight reduction in the
$\tilde{L}$ observables, and is lowest for $L_{\rm total}$.

%An approximate exercise shows that while in the SM the two observables differ approximately by $40\%$ in presence of NP the difference can be up to $900\%$, assuming $L_{total}$ to be approximately constant.

%\end{itemize}

\subsection{Understanding the $L_{K^*\bar{K}^*}$ and $L_{K\bar{K}}$ observables: U-spin breaking}\label{sec:uspin}

To understand the physical significance of the $L$-observables as probes of U-spin symmetry breaking, it is useful to recast the decay amplitudes in a more compact form, as outlined below. We begin with the case of $\bar{B}_{d,s} \to K^{*0} \bar{K}^{*0}$ decays. However, the reasoning is general and applies to any process whose amplitude can be expressed with the same structure.

%its form in terms of the $\lambda_u$ and $\lambda_t$ components of the amplitudes.

The amplitude for a $\bar B_{d,s}$ meson decaying  via a penguin-mediated process into  a $K^{*0} \bar{K}^{*0}$ state with a definite polarisation (longitudinal one in our case) through a $b\to q$ transition ($q=d,s$), can be decomposed in two pieces:
\begin{equation}
\bar{A}_f\equiv A(\bar{B}_{d,s}\to K^{*0} \bar{K}^{*0})
  =\lambda_u^{(q)} T_q + \lambda_c^{(q)} P_q=\lambda_u^{(q)} \Delta_q-\lambda_t^{(q)} P_q \,,
\label{eq:dec}
\end{equation}
with the CKM factors $\lambda_U^{(q)}=V_{Ub} V^*_{Uq}$ and where the structures $T_q$ and $P_q$ are matrix elements both associated to penguin topologies.  
%In order to understand better the structure of the $L$-observable  
We have introduced  an infrared finite quantity $\Delta_q=T_q-P_q$~\cite{
Descotes-Genon:2006spp,Descotes-Genon:2011rgs} on the RHS of \refeq{dec} using the unitarity relation $\lambda_u^{(q)}+\lambda_c^{(q)}+\lambda_t^{(q)}=0$.

If we define the quantity
\begin{equation}
\alpha^q=\lambda_u^{(q)}/(\lambda_c^{(q)}+\lambda_u^{(q)}) \,,\end{equation}
we can re-express the amplitude in a compact form by
\begin{equation} \label{eq:compact}
\bar{A}_f
  =(\lambda_u^{(q)} + \lambda_c^{(q)}) \left[ P_q +\alpha^q \Delta_q \right]\,,\end{equation}
and the corresponding CP conjugate amplitude is obtained by conjugating the weak phases in $\alpha^q$.
Finally, the structure of the observable $L_{K^* \bar{K}^*}$ becomes clear using \refeq{LKstarKstar} together with the compact expression of \refeq{compact} to find~\cite{Alguero:2020xca}
\begin{equation}\label{eq:LKstKstDeltaP}
\!\! L_{K^*\bar{K}^*}=  \kappa \left|\frac{P_s}{P_d}\right|^2 
 \left[\frac{1+\left|\alpha^s\right|^2\left|\frac{\Delta_s}{P_s}\right|^2
 + 2 {\rm Re} \left( \frac{ \Delta_s}{P_s}\right) {\rm Re}(\alpha^s) 
 }{1+\left|\alpha^d\right|^2\left|\frac{\Delta_d}{P_d}\right|^2
  + 2 {\rm Re} \left( \frac{ \Delta_d}{P_d}\right) {\rm Re}(\alpha^d)} \right]\,,
 \end{equation}
where
\begin{equation}
\kappa=\left|\frac{\lambda^s_u+\lambda^s_c}{\lambda^s_u+\lambda^s_c} \right|^2=22.92^{+0.52}_{-0.30}. \end{equation}
See Ref.~\cite{Alguero:2020xca} for the numerical values of the $\alpha$'s, $\Delta_q$ and $P_q$ in this expression.
The advantage of \refeq{LKstKstDeltaP} lies in its ability to decompose the observable into distinct structural components. This decomposition is crucial for gaining a clearer understanding of what the $L$-observables are actually probing. Specifically, we identify three main contributions: the parameter $\kappa$, the ratio $P_s/P_d$, and a more complex term enclosed in parentheses. The latter is numerically close to unity, with an uncertainty at the level of approximately 1\%, primarily due to the smallness of the $\alpha$ parameters and a lower sensitivity to power corrections in our modelling of the divergent power corrections in the term $\Delta_q=T_q- P_q$. This suggests that, to a good approximation, this term is largely insensitive to NP, provided there are no significant NP-induced weak phases, and thus plays a subleading role in $L_{{K}^*\bar{K}^*}$. Consequently, this observable effectively serves as a probe of $\kappa$, modulated by the degree of U-spin breaking between the penguin amplitudes $P_q$ in the $b \to s$ and $b \to d$ transitions.

\begin{table}[htbp]
\centering
\begin{tabular}%{|c|c|c|p{1cm}p{0.5cm}p{0.5cm}p{0.5cm}p{1cm}p{1cm}p{1cm}|}
{|c|c|c|c|c|p{1cm}p{0.5cm}p{0.5cm}p{0.5cm}p{1cm}p{1cm}p{1cm}|}
\hline
\ab{} & $B_s$ Form Factor  & $B_d$ Form Factor & $\rho_1^2$ & Observable 
 %\multicolumn{7}{|c|}{F}  
 \\[0.5mm] \hline
 & & &  &\\[1mm]\multirow{ 3}{*}{Previous  } & $B_s \to K^*$  & $B_d \to K^*$ 
        &  &  $L_{{K}^*\bar{K}^*}$ \\[2.5mm] & $0.314\pm 0.048$  & $0.356\pm 0.046$  & 0.83 & $19.57^{+9.42}_{-6.70}$ (2.6$\sigma$) \\[0.5mm]
& LCSR  & LCSR  & &\\[2.5mm]      \hline
 & & & &\\[0.5mm] \multirow{ 3}{*}{Previous  } & $B_s \to K$  & $B_d \to K$ & 
& $L_{K\bar{K}}$
\\[2.5mm] & $0.336\pm 0.023$ &   $0.332\pm 0.012$ & 1.09
& $25.90^{+3.90}_{-3.60}$ (2.3$\sigma$) \\[0.5mm]
& LCSR &  Lattice & 
&  \\[2.5mm] %& & & &\\[0.5mm] %\hline
\hline
 %& B & C & D & \multicolumn{7}{|c|}{F} 
 \\[0.5mm] \hline & & & &\\[0.5mm]\multirow{ 8}{*}{This work} & $B_s \to K^*$   & $B_d \to K^*$   &  & 
$L_{K^*\bar{K}^*}$ 
\\[2.5mm] & $0.309\pm 0.066$  & $0.361\pm 0.064$& 0.78 & 
$18.34^{+7.47}_{-5.83}$ ($2.3\sigma$) \\[0.5mm]
& LCSR &  LCSR & & \\[0.5mm] 
\multirow{ 18}{*}{This work } &   &   &  &  
\\[2.5mm] & $0.356\pm 0.020$  & $0.350\pm 0.028$  & 1.10 &  
$26.08^{+5.70}_{-4.72}$ ($4.4\sigma$) \\[0.5mm]
& LCSR+f.w.e.+Lat.  & LCSR+f.w.e.+Lat.  & & \\[2.5mm] 
 &   &   &  &  \\
 & { $0.335 \pm 0.039$ }  & {$0.351 \pm 0.074$ }  & {0.97} &  
{$22.88^{+15.35}_{-8.46}$ ($2.2\sigma$)} \\[0.5mm]
& {Lattice }  & {Lattice }  & & \\[2.5mm] \hline
& & & & \\[0.5mm]
& $B_s \to K$   & $B_d \to K$  &  & $L_{K\bar{K}}$
\\[2.5mm]   & $0.276\pm 0.095$  & $0.278\pm 0.079$  & 1.05 &
$25.21^{+8.64}_{-7.67}$ ($1.2\sigma$) \\[0.5mm]
& LCSR  & LCSR  & & \\[0.5mm] 
 & & &  &
\\[0.5mm] & $0.2605\pm 0.0265$  & $0.3318\pm 0.0101$ & 0.66 & $16.51^{+3.50}_{-3.27}$ ($0.4\sigma$)\\[0.5mm]
& Lattice  & Lattice & & \\[0.5mm] 
 & & & & \\[0.5mm]
 & $0.2742\pm 0.0170$  & $0.3333\pm 0.0086$ & 0.72 & $17.88^{+2.55}_{-2.42}$ ($0.8\sigma$)\\[0.5mm]
& LCSR+Lat.  & LCSR+Lat. & & \\[0.5mm] 
 & &  & &
\\ 
%& $B_s \to K$  & $B_d \to K$  & 0.70 & $L_{KK}=\ab{17.80^{+2.73}_{-2.61}}$ (\ab{$0.73\sigma$})
%\\[0.5mm] 
%& \ab{$0.3340\pm 0.0135$}  & \ab{$0.2738\pm 0.0172$} & &\\[0.5mm]
%& \ab{LCSR+no-HPQCD}  & \ab{LCSR+no-HPQCD} & & \\[0.5mm] 
% & &  & &
\hline
\end{tabular}
\caption{Summary of all relevant form factors considered in this work. This includes our previous form factor set used in Ref.~\cite{Biswas:2023pyw} as well as those  obtained from 
new LCSR determinations, lattice QCD, and combined LCSR+lattice QCD approaches.
 For the vector 
  case, finite width effects (f.w.e.) are also incorporated. Additionally, SM predictions for the  $L_{K^*\bar{K}^*}$ and $L_{K\bar{K}}$ observables, as well as the U-spin breaking parameter 
$\rho_1^2$
  (central value only) defined in the text, are presented for each scenario. 
  }
%\caption{Summary of all relevant form factors in our previous work Ref.~\cite{Biswas:2023pyw} and in this work using LCSR, only lattice, combined LCSR+lattice and finite width effects (f.w.e.) in the vector $K^*$ case, together with the SM predictions for $L_{\bar{K}^*K^*}$ and $L_{KK}$ and the U-spin breaking parameter $\rho_1^2$  (only central value is given) for each case. \qm{1 or 2 digits for significance??}}
\label{tab:table6}
\end{table}

Moreover, we can disentangle the dominant sources of U-spin breaking in the SM in this observable using the explicit form of the longitudinal $P_q$ amplitudes:
\begin{eqnarray}\label{main}
&&\!\!\! P_d\equiv P(\bar{B}_d\to \bar{K}^{*0}K^{*0})=A_{\bar{K}^*K^*}^d
   [\alpha_4^c-\frac{1}{2}\alpha_{4,EW}^c+\beta_3^c+\beta_4^c-\frac{1}{2}\beta_{3,EW}^c-\frac{1}{2}\beta_{4,EW}^c] \nonumber \\
   && \qquad \qquad \qquad \qquad \, \, +A_{K^*\bar{K}^*}^d[\beta_4^c-\frac{1}{2}\beta_{4,EW}^c]\,,\nonumber \\
&&\!\!\! P_s \equiv P(\bar{B}_s\to \bar{K}^{*0}K^{*0})=A_{\bar{K}^*K^*}^s[\beta_4^c-\frac{1}{2}\beta_{4,EW}^c]\nonumber\\
&& \qquad \qquad \qquad \qquad \, \,  +A_{K^*\bar{K}^*}^s[\alpha_4^c-\frac{1}{2}\alpha_{4,EW}^c+\beta_3^c+\beta_4^c-\frac{1}{2}\beta_{3,EW}^c-\frac{1}{2}\beta_{4,EW}^c] \,, 
%&&T(\bar{B}_d\to \bar{K}^{*0}\phi)=A_{\bar{K}^*\phi}
%   [\alpha_3^u+\alpha_4^u-\frac{1}{2}\alpha_{3,EW}^u-\frac{1}{2}\alpha_{4,EW}^u+%\beta_3^u-\frac{1}{2}\beta_{3,EW}^u]\\
%&&P(\bar{B}_d\to \bar{K}^{*0}\phi)=A_{\bar{K}^*\phi}
 %  [\alpha_3^c+\alpha_4^c-\frac{1}{2}\alpha_{3,EW}^c-\frac{1}{2}\alpha_{4,EW}^c+%\beta_3^c-\frac{1}{2}\beta_{3,EW}^c]
 \end{eqnarray}
where the explicit definition of the parameters $\alpha_i^{c}$ and $\beta_i^c$ can be found in Ref.~\cite{Beneke:2003zv}.

\noindent
For the purpose of the discussion, we recall the definition of the normalization factor associated with the longitudinal amplitude
\begin{equation}
    A_{V_1 V_2}^q={ i}\frac{G_F}{\sqrt{2}}m^2_{B_q}f_{V_2}A_0^{B_q\to V_1}(0)\,,
\end{equation}
and the relation between  the weak annihilation parameters  $\beta_i^p$ and $b_i^p$
\begin{equation}
\beta_i^p=\frac{B(V_1 V_2)}{A(V_1 V_2)} b_i^p
\,,
\end{equation}
with %$B(V_1 V_2)= i G_F f_{B_q} f_{V_1} f_{V_2} $
\begin{equation} 
\label{eq:annihilation}
B(V_1 V_2)= i G_F f_{B_q} f_{V_1} f_{V_2} \,,\end{equation}
which differs between a $B_d$ and a $B_s$ decay. 
By combining all previous expressions, one can identify three distinct sources of U-spin breaking in the SM prediction of the observable $L_{K^*\bar{K}^*}$, which, in order of decreasing importance, are:
\begin{itemize}
\item[i)] The dominant source of U-spin breaking that we will denote here by $\rho_1$, arises primarily from the ratio of form factors between $B_s$ and $B_d$ decays; with a subleading contribution coming from $B$-meson mass differences. Based on the previous discussion, the observables $L_{K^*\bar{K}^*}$ can be approximated as
\begin{equation}
L_{K^*\bar{K}^*}\simeq \kappa \rho_1^2 \,,\end{equation}
where
\begin{equation}
\rho_1=\frac{m_{B_s}^2 A_0^{B_s \to K^*}(0)}{m_{B_d}^2 A_0^{B_d \to K^*}(0)} \,.\end{equation}
The limit $\rho_1=1$
 corresponds to exact U-spin symmetry, i.e, absence of U-spin breaking that would point to a value for $L_{K^*\bar{K}^*}\simeq \kappa$ in this limit. For convenience, we define the degree of U-spin breaking through the quantity 
$\Delta U=1-\rho_1^2$, where 
a small (large) and positive  value of $\Delta U$ indicates a small (large) deviation from exact U-spin symmetry. If instead $\Delta U$ becomes negative, its absolute value reflects the magnitude of U-spin breaking, with larger absolute values implying a stronger breaking.
A comparison between the SM prediction for $L_{{K}^*\bar{K}^*}$ 
%(and similarly for $L_{K\bar{K}}$)
and the corresponding experimental measurement reveals a preference of data for a positive and sizeable U-spin breaking, i.e., a large  positive $\Delta U$, which corresponds to a suppressed value of the 
$\rho_1^2$ factor.

%the sign of $\Delta U$
%indicates whether the breaking is positive or negative. A comparison between the SM predictions for $L_{{K}^*\bar{K}^*}$ (same for $L_{K\bar{K}}$)
%with the corresponding experimental measurements reveals a preference for a negative and sizable (in absolute value) U-spin breaking, which corresponds to a suppressed value of the 
%$\rho_1$ factor.

%The main source of U-spin breaking denoted by $\rho_1$ comes from the form factor ratio  between $B_s$ and $B_d$ and mildly from mass differences. Following the previous discussion we can simplify the observable to the following form
%\begin{equation}
%L_{K^*\bar{K}^*}\simeq \kappa \rho_1^2
%\end{equation}
%where
%\begin{equation}
%\rho_1=\frac{m_{B_s}^2 A_0^{B_s \to K^*}(0)}{m_{B_d}^2 A_0^{B_d \to K^*}(0)}
%\end{equation}
%with $\rho_1=1$ implying exact U-spin.  Conventionally, we will define a positive or negative amount of U-spin breaking given by the sign of the quantity: $\Delta {\rm U}=1-\rho_1^2$. If one compares the SM predictions for $L_{\bar{K}^*K^*}$ and $L_{KK}$ with the experimental measurements one finds that data prefers a negative and large (in absolute value) U-spin breaking corresponding to a small $\rho$ factor.

%{\bf The relevant message here is that the smaller the U-spin breaking the larger the tension with data. Data prefers a negative and large (in absolute value) U-spin breaking, induced by the $\rho$ factor, the secondary source discussed below or by New Physics as we will discuss in the next section. }

\item[ii)] 
A secondary source of U-spin breaking denoted by $\rho_2$ arises from annihilation contributions. The normalization  of the annihilation amplitude (see 
\refeq{annihilation})
depends on the decay constants, which differ between $B_s$ and $B_d$ mesons. Therefore,
using the ratio 
$f_{B_s}/f_{B_d}=1.209 \pm 0.005$
and taking into account 
 that the relative weight of the annihilation and the  dominant penguin contributions inside $|P_s/P_d|$ differs between 
$B_s$ and $B_d$
decays, one obtains the relation:
\begin{equation} \label{eq:uspinbreaking2}
\left| \frac{P_s}{P_d} \right|^2\simeq \rho_1^2 \rho_2^2 \simeq \rho_1^2 \left|\frac{\alpha_4^c(B_s \to K^*)}{\alpha_4^c(B_d \to K^*)}\right|^2
\left|\frac{1+
\sqrt{2}
\frac{f_{Bs} f_{K^*}}{m_{Bs}^2 A_0^{B_s \to K^*}}
\frac{b_4^c (B_s \to K^*)}{\alpha_4^c (B_s \to K^*)}}{1+
\sqrt{2}
\frac{f_{Bd} f_{K^*}}{m_{Bd}^2 A_0^{B_d \to K^*}}
\frac{b_4^c (B_d \to K^*)}{\alpha_4^c (B_d \to K^*)}} \right|^2 \,.\end{equation}
This constitutes an additional source of U-spin breaking, typically of the order of a few percent, which adds to the leading contribution. Accordingly the observable can be approximated as
\begin{equation}
L_{K^*\bar{K}^*}\simeq \kappa \rho_1^2\rho_2^2
\,,
\end{equation}
and the total U-spin breaking is quantified by $\Delta {U}=1-\rho_1^2 \rho_2^2$. In Table~\ref{tab:table7rho2old} and Table~\ref{tab:table7rho2nou} the values of $\rho_2^2$ for the different form factor determinations are provided. 
By considering only these two sources and assuming Gaussian uncertainties in the estimation, we find that the central values reported in \reftab{table6} can be reproduced with an accuracy of approximately 3\%.

%and the total amount of U-spin breaking becomes $\Delta {\rm U}=1-\rho_1^2 \rho_2^2$.
%Taking into account only these two sources and assuming gaussianity for this estimate we can reproduce the central values in the Table~\ref{table2} within a $\simeq 3\% $ error except for the new Lattice and LCSR+Lattice for $B\to K$ where the error increases to $\simeq 6\%$.

%This implies a further U-spin breaking of order $5\%$ that adds on top of the leading terms. In summary one gets $L_{K^*K^*} \simeq 22.92 \times 0.80 \times 1.06 \simeq 19.43$ in very good agreement with $L_{K^*K^*}=19.49^{+9.41}_{-6.69}$. In the case of $KK$ one has
%\begin{equation}
%\left| \frac{P_s}{P_d} \right|^2\simeq %\left|\frac{1+
%\sqrt{2}
%\frac{f_{Bs} f_{K}}{m_{Bs}^2 f_0^{B_s \to K}}
%\frac{b_4^c (B_s \to K)}{\alpha_4^c (B_s \to K)}}{1+
%\sqrt{2}
%\frac{f_{Bd} f_{K}}{m_{Bd}^2 f_0^{B_d \to K}}
%\frac{b_4^c (B_d \to K)}{\alpha_4^c (B_d \to K)}} \right|^2 \simeq 1.01 
%\end{equation}
%that implies $L_{KK}\simeq 22.92 \times  1.06 \times 1.01 \simeq 24.5$ close to $L_{KK}=25.90$

\item[iii)] Finally, there are additional minor sources of U-spin breaking arising from small differences in subleading contributions, such as the subdominant $\alpha_i$ and in our modelling of the $\beta_i$ terms. These include form factor and inverse moments differences between $B_s$ and $B_d$ decays 
 (see \refsubsec{lambdas}), etc.

\end{itemize}

In the presence of NP additional sources of U-spin breaking can be introduced via distinct contributions to the Wilson coefficients governing the 
$b \to s$
and $b\to d$
 transitions. These differences propagate through the 
$\alpha_i$
  and 
$\beta_i$
  terms, potentially inducing significant U-spin breaking effects. This form of U-spin breaking, arising specifically from NP contributions, is precisely the scenario of interest in our analysis.

The same discussion, equations and results can be easily extended  to the pseudoscalar $\bar{B}_{d,s}\to K^0 \bar{K}^0$ decays which contribute to
the observable $L_{K\bar{K}}$.
The reason is that the structure of \refeq{LKstKstDeltaP} is very general
and applies also to the $L_{K\bar{K}}$ observable by substituting $P_q$, $\Delta_q$, form factors, and normalizations  by the corresponding ones in the pseudoscalar case. 
However, when employing the new lattice or lattice+LCSR determinations for $B\to K$ form factors, the approximate result deviates by approximately a 6\% when compared with the exact result in \reftab{table6}. This suggests that subleading sources of U-spin breaking play a more significant role in the pseudoscalar case.

%and indeed one can follow the same steps 
% for ${\bar B}_{d,s} \to K^0 \bar{K}^0$ that will lead to the same expression only substituting $P_q$, $\Delta_q$ and the $A^{B_q\to K^*}$ FFs  by the corresponding ones in the pseudoscalar case.

\subsection{Discussion of results}\label{sec:results}

The set of SM predictions for the $L_{{K}^*\bar{K}^*}$ and $L_{K\bar{K}}$ observables, based on both the new and previous form factor determinations, is summarized in 
\reftab{table6}. 
Building on the discussion of U-spin breaking sources in the previous section, we aim to guide the reader in understanding the rationale behind the significant spread observed in the SM predictions. In particular, the two dominant sources of U-spin breaking, as discussed in \refsubsec{uspin}, provide a framework to approximately reproduce the predicted values.

As experimental measurements become more precise, the degree of consistency between the two observables $L_{K^*\bar{K}^*}$ and $L_{K\bar{K}}$ may serve as a valuable diagnostic tool for assessing the reliability of the different form factor determinations. For instance, the impossibility to find a consistent explanation between these two rather close observables, once any alternative explanation has been explored, may help us identify 
 potential issues or inconsistencies in the various  form factor determinations.

%We will check if they provide a consistent picture for the two $L$-observables. 

%Understanding the sources and the size of U-spin breaking in the SM is key to guide the reader to understand the rationale behind the predictions of the $L$-observables under the different form factor determinations.

%Table.\ref{table2} summarizes the different predictions of FFs and their impact on the $L$-observables. We observe an important spread of results for the predictions of the $L$-observables depending on the FF determination used, implying different sizes of U-spin breaking.

The following remarks {concerning LCSR, LCSR+lattice and lattice (only for  $B_{d,s} \to K$) determinations } are in order:
\begin{itemize}
\item The previous LCSR determinations of the form factors $A_0^{B_q\to K^*}$ ($q=d,s$) and the new ones
based solely on LCSR are largely compatible,  
showing only a modest increase in the leading U-spin breaking (a decrease in $\rho_1^2$). The subleading U-spin breaking remains unchanged, resulting in a shift of the total U-spin breaking from 
$\Delta U=13\%$ to
$\Delta U=18\%$, and leading to a relatively small difference of approximately 6\% between the old and the new LCSR determination of $L_{K^*\bar{K}^*}$.
As discussed in the previous Subsection, the central value of  $L_{{K}^*\bar{K}^*}$ can be well approximated by considering only the leading and first subleading sources of U-spin breaking in the SM:
$
L_{{K}^*\bar{K}^*}^{\text{SM}} \simeq \kappa \rho_1^2 \rho_2^2,
$
yielding values of approximately 19.97 (previous LCSR) and 18.77 (new LCSR), respectively, and rather close to the exact results in \reftab{table6}.
It is worth noting that even if the central value of $L_{{K}^*\bar{K}^*} $ obtained using the new LCSR form factors for $A_0^{B_{d,s} \to K^*}$ is very close to the result reported in our earlier work as mentioned above, the current uncertainties are smaller, despite the fact that the uncertainties in the form factors themselves are larger. This improvement is due to the strong correlations among the new $ A_0^{B_{d,s} \to K^*} $ form factors, which were not incorporated in our previous analysis. In contrast, for the scalar form factors $ f_0^{B_q \to K} $, the new LCSR-only determinations are uniformly shifted downward for both $B_d$ and $B_s$ compared to the previous determinations, which combined LCSR (for  $B_s \to K$) and lattice (for  $B_d \to K$). Due to this common shift, the amount of U-spin breaking remains small and similar between the two determinations, changing from  $\Delta \text{U} = -11\% $ to $-8\%$. As a result, the central value of $L_{K\bar{K}}$ is only reduced by about 3\%. We also find that the central value of $L_{K\bar{K}}$ can be approximately reproduced using the simplified expression $L_{K\bar{K}} \simeq \kappa \rho_1^2 \rho_2^2$, yielding values of approximately 25.23 (previous form factors) and 24.79 (new LCSR) to be compared with the exact results in \reftab{table6}. However, the uncertainty in this observable increases significantly, by roughly a factor of two, due to the larger uncertainties in the new form factors (see discussion in \refsubsec{LCSR}): a factor of four for $f_0^{B_s \to K}$ and close to seven for  $f_0^{B_d \to K}$. This increase is only partially mitigated by the strong correlations among the form factors. Consequently, the statistical significance of the tension between the SM prediction and the experimental measurement of $L_{K\bar{K}}$ is reduced to slightly above $1\sigma$ when using the new LCSR form factor determination.

%However the amount of U-spin breaking is in this case rather small and almost  the same (from 6\% to 2\%) so the central value is only slightly shifted but the uncertainties due to the use of only LCSR are larger, implying a reduction in significance in the tension of the observable $L_{KK}$.

\item 
The situation changes significantly when using the new combined LCSR+lattice determination of the  $A_0^{B_q \to K^*}$ form factors. In this case, the amount of U-spin breaking shifts drastically from  $\Delta U = 18\%$ (using only the new LCSR determination) to $\Delta U = -12\%$ (using the new LCSR+lattice results). As a consequence, the tension between the SM prediction and the experimental data for  $L_{{K}^*\bar{K}^*}$ increases substantially, from 2.3$\sigma$ (new LCSR) to 4.4$\sigma$ (new LCSR+lattice). This increase in significance is driven not only by the change in the sign of $\Delta U$, but also by the reduction in uncertainties in the LCSR+lattice form factor determination by a factor of three compared to the LCSR-only case. Once again, we can approximately reproduce the value of $L_{{K}^*\bar{K}^*}$ using the two main sources of U-spin breaking and the new combined LCSR+lattice form factors, obtaining $\simeq 25.72$.
In contrast, for the $B_q \to K$ transitions, the situation is reversed. The U-spin breaking changes from $ \Delta U = -8\%$ (new LCSR) to $+27\% $ (new LCSR+lattice), primarily due to an upward shift in the $ f_0^{B_d \to K}$ form factor between the two new determinations (still compatible within 1$\sigma$).
Instead, we observe a significant downward shift in 
the $ f_0^{B_s \to K}$ form factor between the old and the new (LCSR+lattice) determination only marginally compatible at the level of more than 2$\sigma$ that explains the significant reduction of the tension in $L_{K\bar{K}}$.
Using the new LCSR+lattice computation, the approximation $L_{K\bar{K}} \simeq \kappa \rho_1^2 \rho_2^2$ becomes slightly less precise, yielding a value of $L_{K\bar{K}} \simeq 16.83$, which is about 6\% below the true value. Consequently, the previous tension of 2.3$\sigma$ in the observable $L_{K\bar{K}}$ is, as mentioned above, reduced to below the 1$\sigma$ level, accompanied by a significant reduction of the order of 31\%, in the central value of $L_{K\bar{K}}$.

\item The use of lattice QCD alone in the determination of the  $f_0^{B_q \to K}$ form factors leads to a  shift in the central value of the $f_0^{B_d \to K}$ form factor compared to the new determination based solely on LCSR, as discussed previously. However, due to the large uncertainty associated with the new LCSR determination, both results remain consistent within less than 1$\sigma$.
It is worth noting that the form factor determinations obtained using only lattice or the combined LCSR+lattice approach are quite similar, owing to the dominant contribution from the lattice input. Consequently, the corresponding predictions for $L_{K\bar{K}}$ in both cases differ by only $\sim 8\%$.

%or the combination of LCSR+lattice as expected mainly affects the uncertainty, with a reduction in the second case but rather close central values.

\end{itemize}

In summary, the new form factor determinations based solely on LCSR yield a picture that is broadly consistent with our previous results for the $L$-observables, albeit with larger uncertainties for $L_{K\bar{K}}$, where the uncertainty roughly doubles. As in our earlier analysis, a large U-spin breaking is observed in $B \to K^*$ and a smaller one in $B \to K$, leading to a similar level of tension with the data in $L_{K^*\bar{K}^*}$ observable as before, but a reduced tension in  $L_{K\bar{K}}$ due to the increased uncertainty.
In contrast, the inclusion of lattice data in combination with LCSR significantly changes the overall picture. In this case, a small U-spin breaking is found in $B \to K^*$ and a large one in $B \to K$, effectively reversing the pattern observed in the LCSR-only scenario. This results in an almost complete interchange of the SM predictions for $L_{{K}^*\bar{K}^*}$ and  $L_{K\bar{K}}$, with a substantial increase in the tension for $L_{{K}^*\bar{K}^*}$ and a marked reduction for $L_{K\bar{K}}$ compared to our previous findings.

%However, the tension for $L_{KK}$ is practically the same using LCSR or LCSR+lattice due to a significant reduction of the uncertainties (more than half) in the latter case.

%\begin{figure}[t]\centering
%\subfloat[]{\label{fig:c4_KstK}\includegraphics[width=0.5\textwidth,height=0.45\textwidth]{figures/pltc4ds_Lat_LCSR.png}}~~~
%\subfloat[]{\label{fig:c8_KstK}\includegraphics[width=0.5\textwidth,height=0.45\textwidth]{figures/pltc8gds_Lat_LCSR.png}}
%\caption{LCSR+Lattice. (left) $C_{4d}-C_{4s}$ and (right) $C_{8gd}-C_{8gs}$}
%\label{fig:BsKstK_effect}
%\end{figure}

%\begin{figure}[h]%\centering
%\subfloat[]{\label{fig:c4_KstK}\includegraphics[width=0.5\textwidth,height=0.45\textwidth]{figures/pltc4ds_LCSR.png}}~~~
%\subfloat[]{\label{fig:c8_KstK}\includegraphics[width=0.5\textwidth,height=0.45\textwidth]{figures/pltc8gds_LCSR.png}}
%
%\subfloat[]
%{\label{fig:c4_KstK}\includegraphics[width=0.5\textwidth,height=0.45\textwidth]{figures/pltc4ds_Lat_LCSR.png}}~~~
%\subfloat[]
%{\label{fig:c8_KstK}\includegraphics[width=0.5\textwidth,height=0.45\textwidth]{figures/pltc8gds_Lat_LCSR.png}}
%\caption{LCSR+Lattice. (left) $C_{4d}-C_{4s}$ and (right) $C_{8gd}-C_{8gs}$}
%\caption{Only LCSR.}
%\label{fig:BsKstK_effect}
%\end{figure}

%\subsection{New SM predictions for the $L$-observables under two/three scenarios}

%%%%%%%%%%%%%%%%%%%%%%%%%%%%%

%
\begin{table}[t!]
%\caption {\bf{Previous work}}
\begin{center}
\begin{tabular}%{|c|c|c|p{1cm}p{0.5cm}p{0.5cm}p{0.5cm}p{1cm}p{1cm}p{1cm}|}
{|c|c|c|c|c|}
\hline
Form Factors &~$\alpha_4^c$-values ~&~$b_4^c$-values ~&~$\rho_2^2$  \\[0.5mm] \hline
                      &                &            &           \\
 $B_s \to K^*$        &                &            &           \\[1mm] 
 $0.314\pm 0.048$&    $-0.033^{+0.001}_{-0.001}-0.014^{+0.001}_{-0.001}I$            &   $-0.52^{+0.33}_{-0.34}+0.00^{+0.33}_{-0.33}I$         &     %$-0.003^{+0.002}_{-0.002}+0.000^{+0.002}_{-0.002}I$  
 \\[0.5mm]
 LCSR 
 &                &            &     1.05      \\[0.5mm] 
        %              &                &            &           \\[1mm]
 $B_d \to K^*$        &             &          &   %$-0.002^{+0.001}_{-0.001}+0.000^{+0.001}_{-0.001}I$   
 \\[1mm] 
 $0.356\pm 0.046$ &$-0.033^{+0.001}_{-0.001}-0.014^{+0.001}_{-0.001}I$   &$-0.52^{+0.33}_{-0.34}+0.00^{+0.33}_{-0.34}I$  &           \\[0.5mm]
 LCSR 
 &                &            &           \\[0.5mm]
 \hline\hline
                      &                &            &           \\[0.5mm] 
 $B_s \to K$          &               &   &    %$-0.002^{+0.002}_{-0.002}+0.000^{+0.002}_{-0.002}I$        
 \\[0.5mm] 
$0.336\pm 0.023$ &$-0.099^{+0.003}_{-0.003}-0.026^{+0.001}_{-0.001}I$ &$-0.66^{+0.43}_{-0.45}+0.00^{+0.44}_{-0.44}I$ &           \\[0.5mm]
LCSR    
% &                &            &           \\[0.5mm]
                      &                &            &    1.01       \\[1mm]
 $B_d \to K$          &                &           &    %$-0.002^{+0.001}_{-0.001}+0.000^{+0.001}_{-0.001}$      
 \\[0.5mm]
 $0.332\pm 0.012$&$-0.099^{+0.003}_{-0.003}-0.026^{+0.001}_{-0.001}I$ &$-0.66^{+0.44}_{-0.43}+0.00^{+0.44}_{-0.43}I$ &           \\[0.5mm]
 Lattice   
&                &            &           \\[0.5mm]   
\hline
 %& B & C & D & \multicolumn{7}{|c|}{F} 
\end{tabular}
\caption{Numerical values in the SM of the 
coefficients $\alpha_i^c$ and $\beta_i^c$ (see \cite{Beneke:2003zv} for definitions) that involve form factors and convolutions of light-cone distribution amplitudes with perturbative kernels together with the Wilson coefficients of the effective Hamiltonian. These values are determined using the form factors we used in Ref.~\cite{Biswas:2023pyw}. $\rho_2^2$ stands for the subleading U-spin breaking induced by these coefficients as shown in \refeq{uspinbreaking2}.}
\label{tab:table7rho2old}\end{center}
\end{table}

\begin{table}[!ht]
%\caption {\bf{This work}}
\small
\begin{center}
\begin{tabular}%{|c|c|c|p{1cm}p{0.5cm}p{0.5cm}p{0.5cm}p{1cm}p{1cm}p{1cm}|}
{|c|c|c|c|c|}
\hline
Form Factors &~$\alpha_4^c$ values~&~$b_4^c$ values~&~$\rho_2^2$  \\[0.5mm] \hline
                       &                &            &           \\[1mm]
 $B_s \to K^*$         & &  &           \\[1mm] 
 $0.309\pm 0.066$&   $-0.033^{+0.001}_{-0.001}-0.014^{+0.001}_{-0.001}I$             &  $-0.52^{+0.33}_{-0.34}+0.00^{+0.33}_{-0.33}I$ & %$-0.003^{+0.002}_{-0.002}+0.000^{+0.002}_{-0.002}I$          
 \\[0.5mm]
LCSR  
 &                &            &    1.05       \\[1mm] 
  %                     &                &   %         &    1.05      \\[0.5mm]
 $B_d \to K^*$         &                &            &           \\[1mm] 
 $0.361\pm 0.064$&    $-0.033^{+0.001}_{-0.001}-0.014^{+0.001}_{-0.001}I$            & $-0.52^{+0.33}_{-0.34}-0.00^{+0.33}_{-0.33}I$            &   %$-0.002^{+0.001}_{-0.001}+0.000^{+0.001}_{-0.001}I$      
 \\[0.5mm]
 LCSR 
 &                &            &         \\[0.5mm]
       %                 &                &            &           \\[0.5mm] 
 $B_s \to K^*$          &                &            &           \\[0.5mm] 
$0.356\pm 0.020$ &    $-0.033^{+0.001}_{-0.001}-0.014^{+0.001}_{-0.001}I$            & $-0.52^{+0.33}_{-0.34}-0.00^{+0.33}_{-0.33}I$            &   %$-0.002^{+0.001}_{-0.002}+0.000^{+0.001}_{-0.001}I$      
\\[0.5mm]
 LCSR+f.w.+latt 
% &                &            &           \\[0.5mm]
                        &                &            &      1.02     \\[1mm]
 $B_d \to K^*$          &               &          &   %$-0.002^{+0.001}_{-0.001}+0.000^{+0.001}_{-0.001}I$     
 \\[0.5mm]
 $0.350\pm 0.028$&$-0.033^{+0.001}_{-0.001}-0.014^{+0.001}_{-0.001}I$ &$-0.52^{+0.33}_{-0.34}-0.00^{+0.33}_{-0.33}I$  &           \\[0.5mm]
 LCSR+f.w.+latt 
 &                &            &         \\[0.5mm]\hline
                       &                &            &           \\[1mm]
 $B_s \to K$           &                &            &           \\[1mm] 
 $0.276\pm 0.095$&    $-0.099^{+0.003}_{-0.003}-0.026^{+0.001}_{-0.001}I$            &  $-0.66^{+0.43}_{-0.46}+0.00^{+0.44}_{-0.44}I$          &    %$-0.002^{+0.002}_{-0.002}+0.000^{+0.002}_{-0.002}I$  
 \\[0.5mm]
 LCSR 
% &                &            &           \\[0.5mm] 
                       &                &            &     1.03     \\[1mm]
 $B_d \to K$           &              &          &    %$-0.002^{+0.001}_{-0.002}+0.000^{+0.001}_{-0.001}I$     
 \\[1mm] 
 $0.278\pm 0.079$
 &$-0.098^{+0.003}_{-0.003}-0.026^{+0.002}_{-0.002}I$  &$-0.66^{+0.43}_{-0.45}+0.00^{+0.43}_{-0.44}I$  &           \\[0.5mm]
 LCSR    
 %&                &            &         \\[0.5mm]
                        &                &            &           \\[0.5mm] 
 $B_s \to K$            &                &            &           \\[0.5mm] 
$0.2605\pm 0.0265$ &    $-0.099^{+0.003}_{-0.003}-0.026^{+0.001}_{-0.001}I$             & $-0.66^{+0.44}_{-0.45}+0.00^{+0.43}_{-0.44}I$           &  %$-0.002^{+0.002}_{-0.002}+0.000^{+0.002}_{-0.002}I$     
\\[0.5mm]
Lattice
% &                &            &           \\[0.5mm]
                        &                &            &    1.03        \\[1mm]
 $B_d \to K$            &                &            &           \\[0.5mm]
 $0.3318\pm 0.0101$&    $-0.099^{+0.003}_{-0.003}-0.026^{+0.002}_{-0.002}I$            &   $-0.66^{+0.43}_{-0.45}+0.00^{+0.44}_{-0.43}I$         &   %$-0.002^{+0.001}_{-0.001}+0.000^{+0.001}_{-0.001}I$        
 \\[0.5mm]
 Lattice 
 &                &            &       \\[0.5mm]\hline
                        &                &            &           \\[1mm]
 $B_s \to K$            &                &            &           \\[1mm] 
 $0.2742\pm 0.0170$&   $-0.099^{+0.003}_{-0.003}-0.026^{+0.001}_{-0.001}I$             &  $-0.66^{+0.43}_{-0.45}+0.00^{+0.44}_{-0.44}I$          &   %$-0.002^{+0.002}_{-0.002}+0.000^{+0.002}_{-0.002}I$      
 \\[0.5mm]
 LCSR+lat 
 %&                &            &           \\[0.5mm] 
                        &                &            &    1.02       \\[1mm]
 $B_d \to K$            &                &            &           \\[1mm] 
 $0.3333\pm 0.0086$&  $-0.099^{+0.003}_{-0.003}-0.026^{+0.002}_{-0.002}I$              &   $-0.66^{+0.44}_{-0.45}+0.00^{+0.44}_{-0.44}I$          &   %$-0.002^{+0.001}_{-0.001}+0.000^{+0.001}_{-0.001}I$       
 \\[0.5mm]
 LCSR+lat  
% &                &            &         \\[0.5mm]
                        &                &            &           \\[0.5mm] 
\hline
 %& B & C & D & \multicolumn{7}{|c|}{F} 
\end{tabular}
\end{center}
\caption{Corresponding SM values of $\alpha_4^c $ and $\beta_4^c$ using the new determinations of form factors. }
\label{tab:table7rho2nou}
\end{table}
%
%These $\tilde{L}$ observables can be related to the previous ones by the following relation
%\begin{eqnarray}
%\tilde L_{K^*}= \frac{1}{\rho(m_K^0,m_{K^*})} \frac{{\cal B}(\bar{B_d}\to\bar{K}^{*0} K^0)}{{\cal B}(\bar{B_s}\to K^{*0} \bar{K}^0)+{\cal B}(\bar{B_s}\to K^0\bar{K}^{*0})}=\frac{1}{L_{total}}\frac{1}{(1+1/R_d)}
%
%\tilde L_K= 2 rho [B(bar{Bs}->K*0 \bar{K}0)+B(\bar{Bs}->K0\bar{K}*0)]/B(\bar{Bd}->\bar{K}0 K*0)=Ltotal*(1+Rd)
%\end{eqnarray}
%
\begin{figure}[!h]%\centering
\includegraphics[width=0.5\textwidth,height=0.44\textwidth]{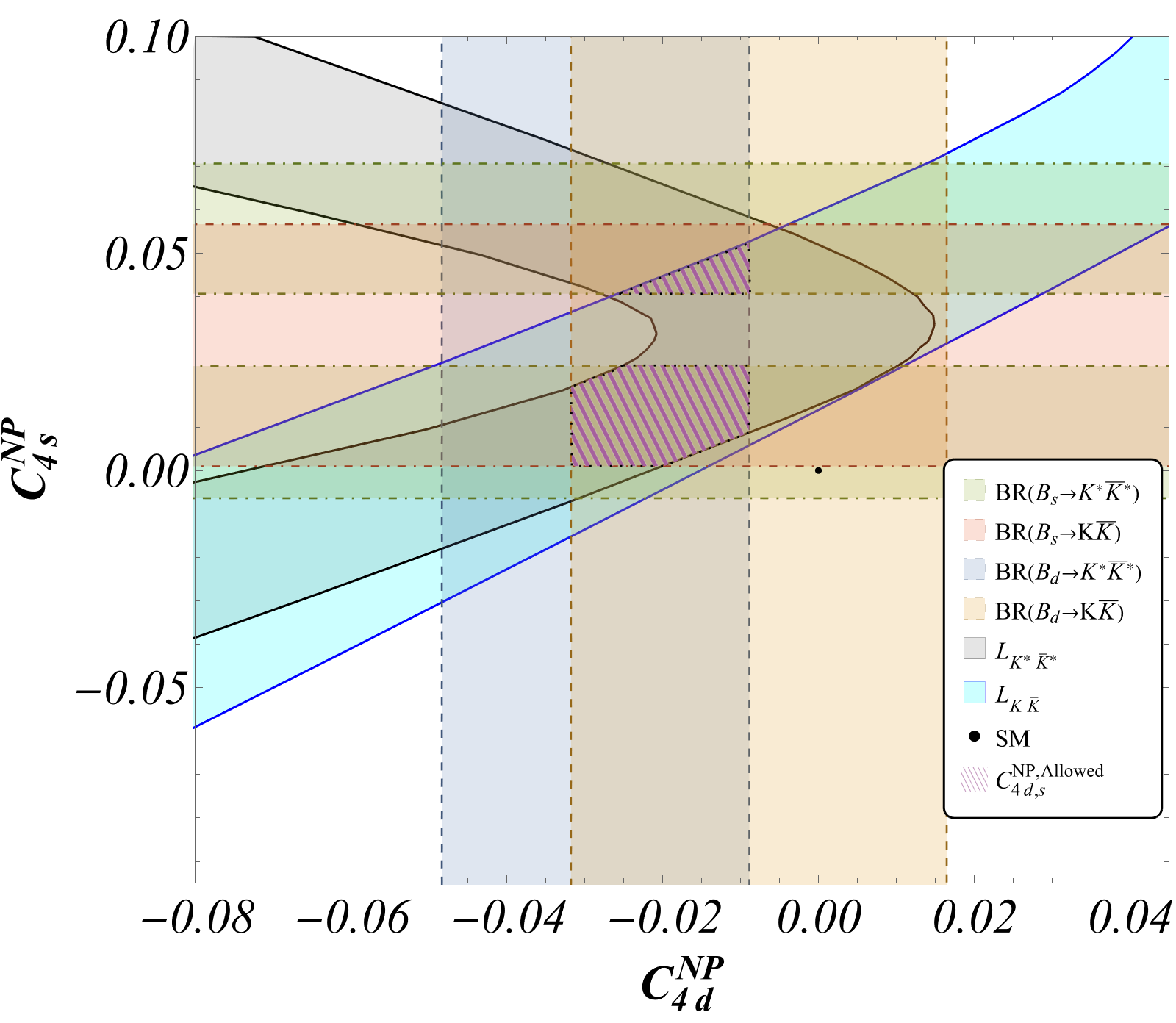}
\includegraphics[width=0.5\textwidth,height=0.44\textwidth]{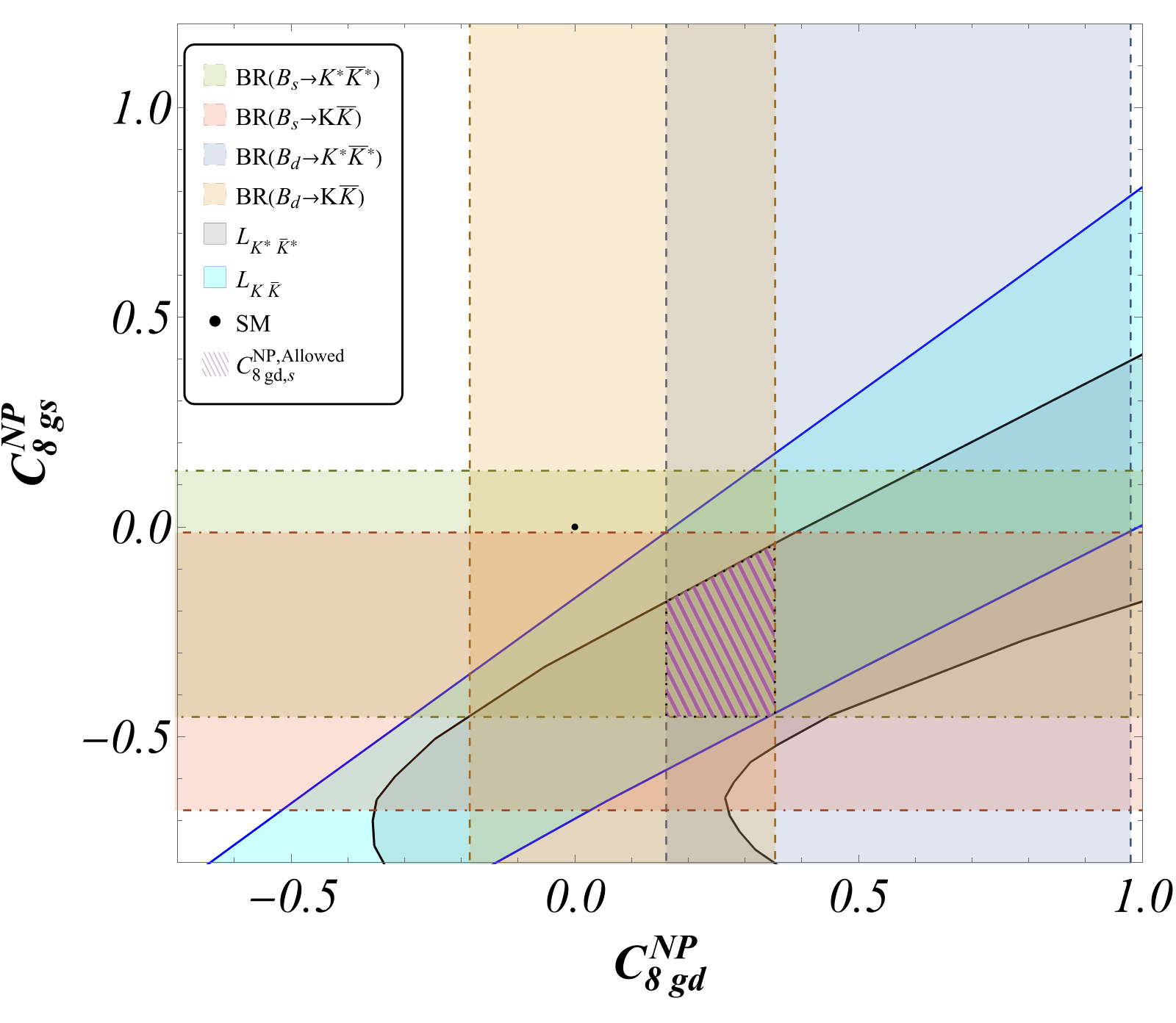}\\
\includegraphics[width=0.5\textwidth,height=0.44\textwidth]{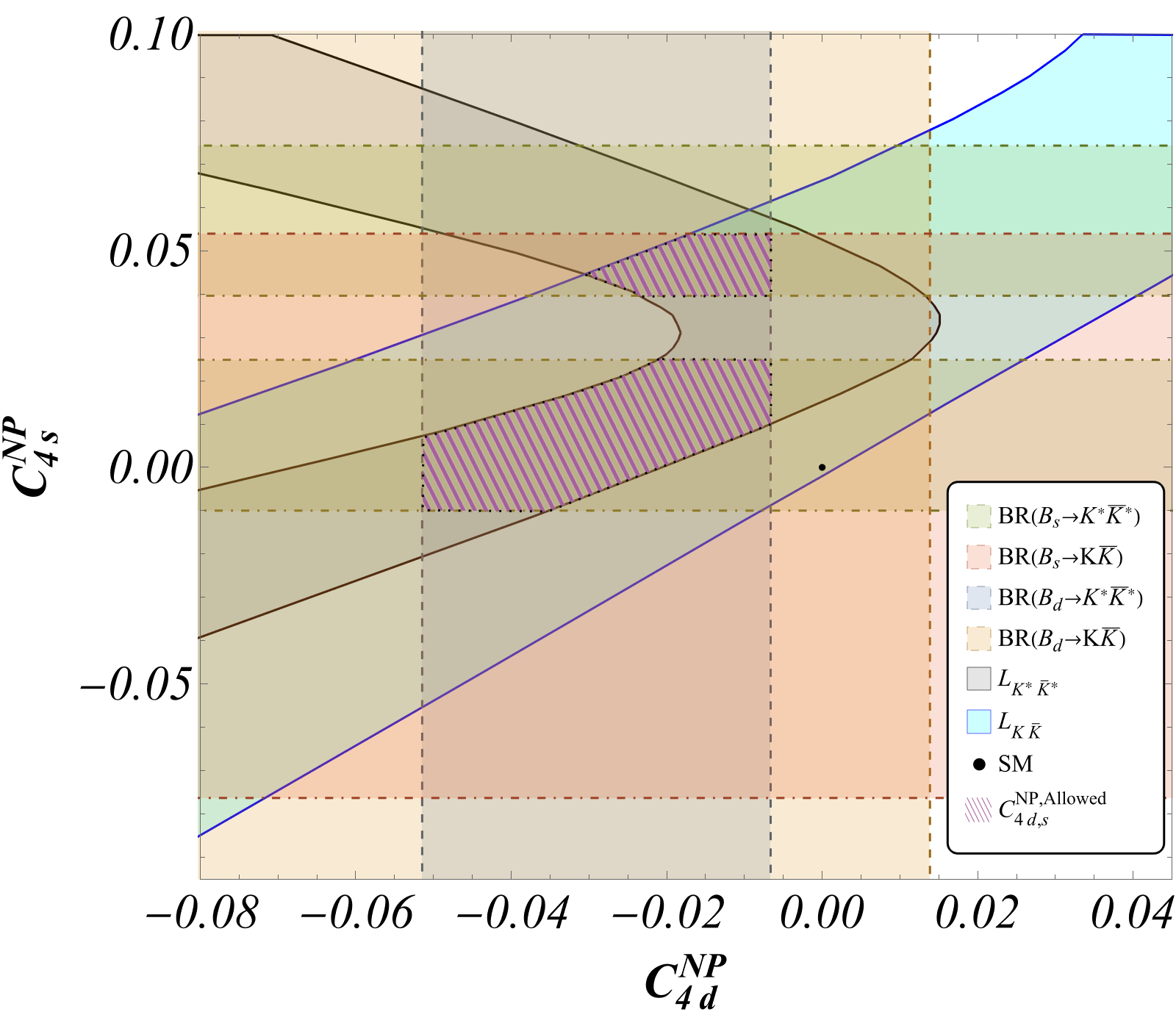}
\includegraphics[width=0.5\textwidth,height=0.44\textwidth]{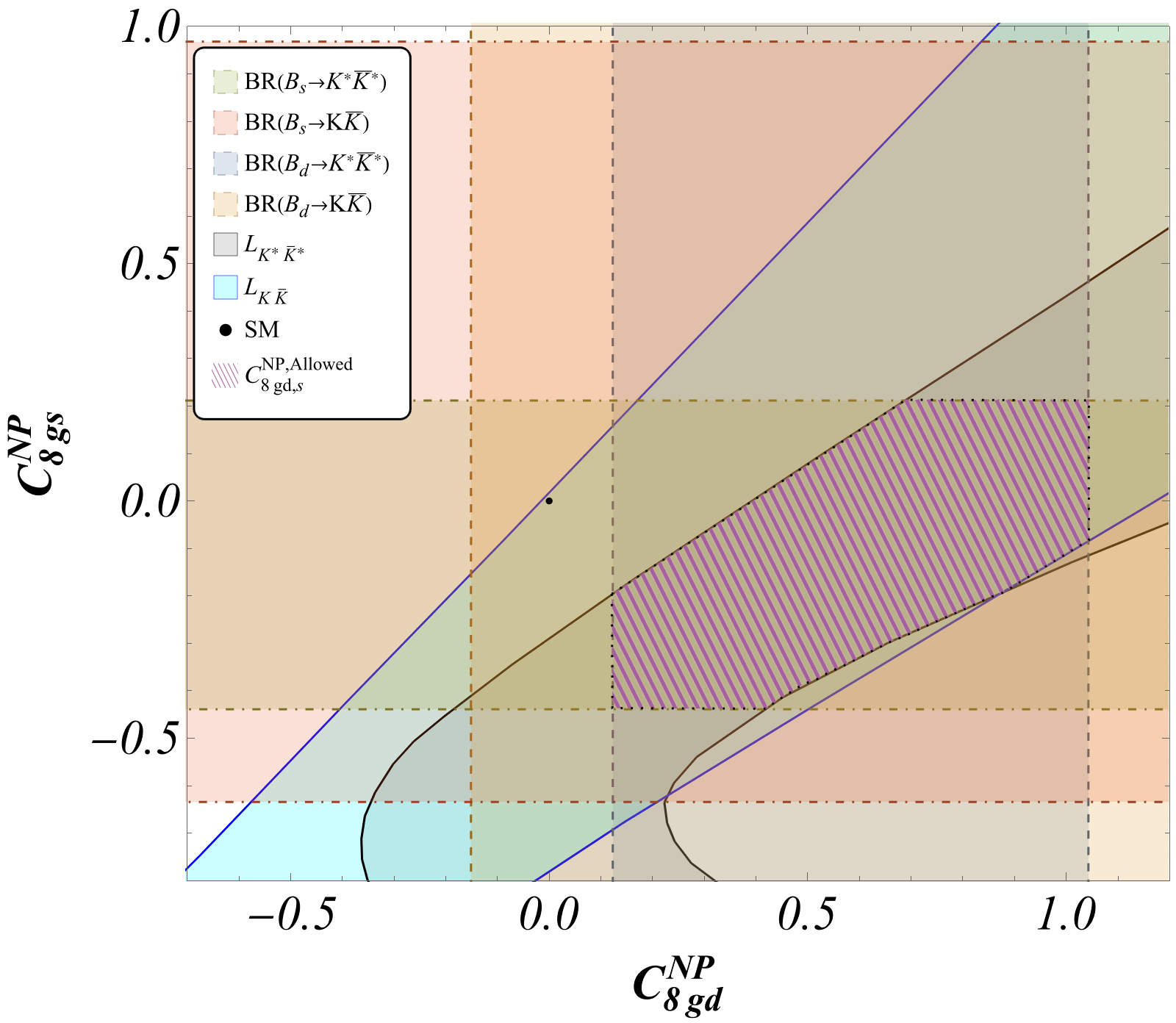}\\
\includegraphics[width=0.5\textwidth,height=0.44\textwidth]{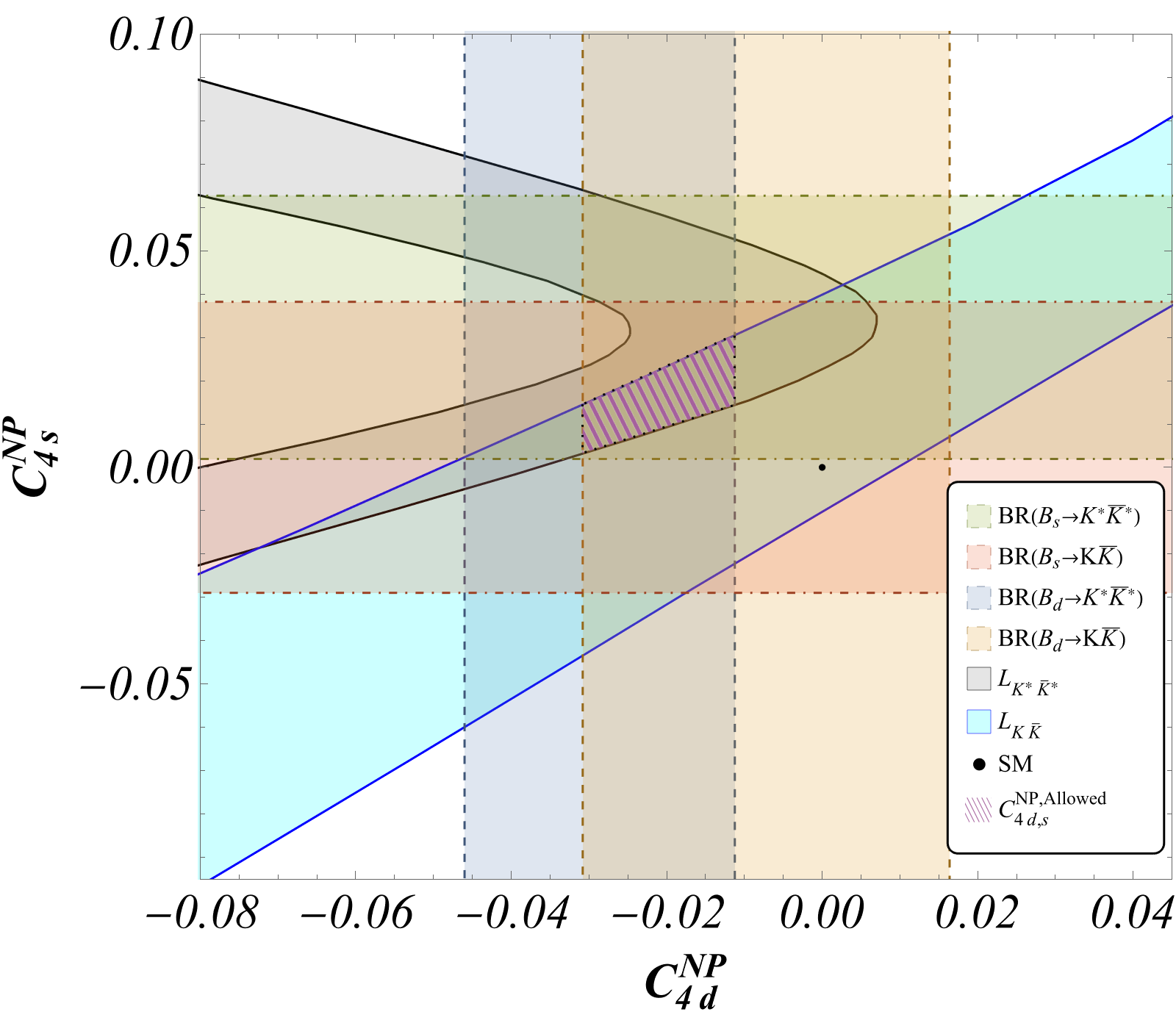}
\includegraphics[width=0.5\textwidth,height=0.44\textwidth]{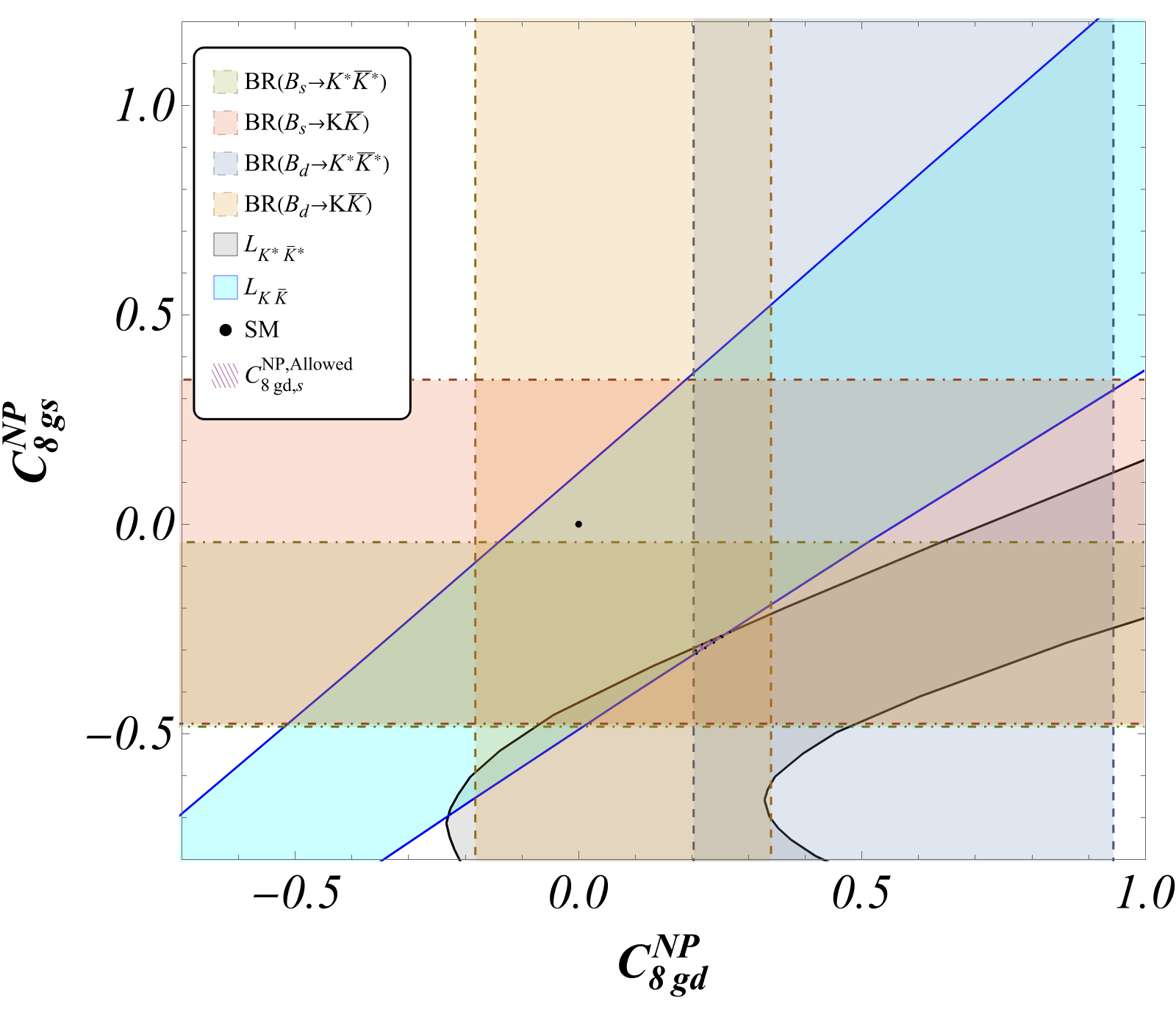}
\caption{Allowed regions for the relevant  Wilson coefficient pairs $\ccal_{4d}-\ccal_{4s}$ (left) and $\ccal_{8gd}-\ccal_{8gs}$ (right) considering $L_{\bar{K}^*K^*}$ and $L_{KK}$ observables together with the constraints from the individual branching ratios using 
our previous set of form factors but updated $\lambda_{B_d}$ (top), the 
new LCSR only (mid)  %(left) $C_{4d}-C_{4s}$ and (right) $C_{8gd}-C_{8gs}$ (bottom)
 and combined LCSR+Lattice (bottom).
 % : (left) $C_{4d}-C_{4s}$ and (right) $C_{8gd}-C_{8gs}$}
}
%\caption{Only LCSR.}
\label{fig:figallowed}
\end{figure}

%\begin{eqnarray}
%\tilde{L}_{K^*}&=&..... f(R_d,L_K^*,L_K) \nonumber \cr
%\tilde{L}_K&=&..... f(R_d,L_K^*,L_K) \end{eqnarray}

%\bigskip

%
%\item $L_{K^*\phi}$. This case contrary to the previous ones is not constructed from a ratio of exact U-spin partners. Here a discussion on the form factors is more cumbersome. {\bf We shall discuss if we want to include these observables in this paper}

\subsection{$\lambda_{B_d}$ and $\lambda_{B_s}$ dependencies}\label{sec:lambdas}

The inverse moments of the Light Cone Distribution Amplitude (LCDA) $\lambda_{B_s}$ and $\lambda_{B_d}$ suffer from large uncertainties. Different determination procedures are available in the literature. This includes for instance the usage of the decay mode $B\rightarrow \gamma \ell \nu_{\ell}$ \cite{Beneke:2018wjp, Wang:2016qii, Beneke:2011nf,Janowski:2021yvz}, QCD sum rules \cite{ Braun:2003wx, Khodjamirian:2020hob}, using $B\rightarrow \pi$ \cite{Wang:2015vgv} and $B\rightarrow \rho$ \cite{Gao:2019lta} transitions and indirect constraints from lattice \cite{Mandal:2023lhp}.  Experimentally the following $90\%~\rm{C.L.}$ lower limit is available:  $0.24~{\rm GeV} <\lambda_{B_d}$.  The theoretical determinations are affected by uncertainties that vary between  $10\%$ to $40\%$. And the extracted value is in general model dependent. Given the potential range of variation, we have considered important to assess the sensitivity of our observables with respect to this input parameter. Thus, we consider the range $0.24~{\rm GeV}<\lambda_{B_d}< 0.68~{\rm GeV}$ (2$\sigma$ range from Eq.\ref{lambdaBdeq})
and evaluate the effect on $\alpha_4$ which is dominant in our observables. We find that the variation on this amplitude is at most $2\%$  implying a small impact in most $L$-observables except for $R_d$ that depends on $\lambda_{B_d}$ on both numerator and denominator and  exhibits a large sensitivity to this parameter. %Thus we can conclude a low sensitivity towards $\lambda_{B_d}$ for our optimized observables as well.
Notice that in this paper we kept the form factor inputs fixed; however, it is important to emphasize that the LCSR predictions themselves also depend on this parameter.

\section{Sensitivity to New Physics}\label{sec:np}

In this Section, we analyze the sensitivity of the set of $L$-observables to NP contributions, considering the various form factor determinations.
Particular attention is paid to the consistency of NP interpretations between the observables $L_{K^*\bar{K}^*}$, $L_{K\bar{K}}$, and the branching ratios ${\cal B}(\bar{B}{s} \to {K}^{*0} \bar{K}^0)$ and ${\cal B}(\bar{B}_{d} \to \bar{K}^{*0} {K}^0)$, as a function of the chosen form factors inputs.
 
 We also explore in \refsubsec{enhancement} if  the enhancement mechanism that was found in Ref.~\cite{Biswas:2023pyw} for the pair ${\hat L}_{K^{*}}$-${\hat L}_K$ is also at work for the newly introduced pair of  $\tilde{L}_{K^*}$-$\tilde{L}_K$ observables.

\subsection{Comparison of allowed regions in $\ccal_{4d}-\ccal_{4s}$ and $\ccal_{8gd}-\ccal_{8gs}$ planes}

Ref.~\cite{Biswas:2023pyw} demonstrated that the tensions observed between data and the observables 
$L_{{K}^*\bar{K}^*}$-$L_{K\bar{K}}$ could be coherently explained by introducing NP in three scenarios:
\begin{itemize}
\item[i)] NP in the tree-level coefficient $\ccal_{1q}$, however this solution is disfavoured
 because the required NP amount to explain the $L$-observables is in conflict with bounds from \cite{Lenz:2019lvd}.
\item[ii)] NP in the Wilson coefficient of the QCD penguin operator $Q_{4f}$.
\item[iii)] NP in the Wilson coefficient of the chromomagnetic operator $Q_{8gf}$.
\end{itemize}
 See \refapp{WET} for the definition of the operators.
In this Subsection, we compare the allowed (or overlapping) regions in the $\ccal_{4d}$--$\ccal_{4s}$ and $\ccal_{8gd}$--$\ccal_{8gs}$ planes, as constrained by the observables $L_{{K}^*\bar{K}^*}$, $L_{K\bar{K}}$, and the individual branching ratios, under three different form factor scenarios. These are illustrated in Fig.~\ref{fig:figallowed}. The three cases shown are:

\begin{itemize} 

\item Previous form factor determination: 
As a reference we reproduce here our results from Ref.~\cite{Biswas:2023pyw} but taken the updated value for $\lambda_{B_d}$. 
This yielded two broad regions of compatibility in the $\ccal_{4d}$--$\ccal_{4s}$ plane and one in the $\ccal_{8gd}$--$\ccal_{8gs}$ plane.

\item New LCSR-only determination: Using only the LCSR form factors, which carry larger uncertainties, especially for $f_0^{B_q \to K}$, significantly enlarges the overlap regions among the six observables. Consequently, the allowed regions in both planes are broader than in the previous analysis. 
%However, some regions previously allowed in the $\ccal_{4f}$ plane (notably around $\ccal_{4s} \sim 0.03$) are now excluded. 
%In contrast, the allowed region in the $\ccal_{8gf}$ plane is fully retained and even expanded.

\item Combined LCSR+Lattice QCD determination: 
This scenario, which drastically changes the SM predictions for $L_{{K}^*\bar{K}^*}$ and $L_{K\bar{K}}$, leads to a very different picture. The overlap region in the $C_{4f}$ plane becomes much smaller.
%and is not entirely contained within the LCSR-only region, particularly due to the reappearance of allowed values near $\ccal_{4s} \sim 0.03$. 
The situation is even more constrained in the $\ccal_{8gf}$ plane, where only a very marginal narrow region remains viable (nearly invisible) around $\ccal_{8gd} \sim +0.22$ and $\ccal_{8gs} \sim -0.30$.

\end{itemize}

In summary, given the very different regions of overlap identified depending on whether the LCSR alone or a combined LCSR + lattice approach is used, we conclude that it is necessary to measure additional observables while simultaneously improving the accuracy  of the $L$-observable measurements. These observables can serve as consistency tests to help discriminate between the two form factor determinations, particularly in cases where one of the two shows no region of overlap for any NP Wilson coefficient assuming no other solution within the SM is found.  The $\tilde{L}$-observables introduced in Section~\ref{sec:subdef} once measured, are excellent candidates for this purpose.

\subsection{Enhancement Mechanism in the $\tilde{L}_K$ and $\tilde{L}_{K^*}$ pair}\label{sec:enhancement}

Using the pair of observables $\tilde{L}_{K^{*}}$--$\tilde{L}_K$, we have identified an enhancement mechanism that is partially inherited from the mechanism observed between the $\hat{L}_{K^*}$-$\hat{L}_K$ observables in~\cite{Biswas:2023pyw}. 
%depending on which Wilson coefficient receives the NP contribution, namely $C_{4s}$ and/or $C_{4d}$ ($C_{8gd}$ and/or $C_{8gs}$). 
This mechanism can provide a strong signal for NP in the $\tilde{L}$ observables  when combined with the tensions observed in $L_{{K}^*\bar{K}^*}$ and $L_{K\bar{K}}$. On the one hand, they can help confirm or dismiss the tensions in $\bar{K}^{(*)}K^{(*)}$, and on the other, offer a discriminating method between the two form factor determinations if no common solution is found in one form factor determination once all other SM explanations have been excluded. 

The core of the mechanism can be understood as follows: The dominant contribution to both $\tilde{L}$-observables arises from $\alpha_{4}^{c}$, with the signs of the prefactors multiplying the Wilson coefficient $\ccal_{4d}$ being opposite in the two $\tilde{L}$ observables\footnote{This can be seen in \refeq{RdI} or \refeq{RdII} considering that $\tilde{L}_{K^*}\propto numerator (R_d)$ and $\tilde{L}_{K}\propto denominator (R_d)$.}. This can be traced to the expression of $\alpha_4^c$ for the decay $B_q \to M_1 M_2$ (see \cite{Biswas:2023pyw} for definitions):
\begin{eqnarray}
\label{eq:alpha4generic}
\alpha^c_4=\begin{cases}
    a^c_4  + r^{M_2}_{\chi} a^c_6  & \text{if $M_1 M_2 =PP, PV$},\\
    a^c_4  - r^{M_2}_{\chi} a^c_6  & \text{if $M_1 M_2 = VV, VP$},
  \end{cases}
\end{eqnarray}
with $r_\chi^K=2 m_K/(m_b (m_s+m_d))\simeq 1.41$ and $r_\chi^{K^*}=2 m_{K^*} f_K^\perp/(m_b f_K^*) \simeq 0.29$ and where
the relative sign between the two terms  in \refeq{alpha4generic} depends on the meson ($M_1$) that collects the spectator quark being a pseudoscalar or a vector (therefore is opposite in the decay of the numerator of  $\tilde{L}_K$ compared to $\tilde{L}_{K^*}$). 
%
%%%%
%
\begin{figure}[!th]%\centering
\includegraphics[width=0.5\textwidth,height=0.435\textwidth]{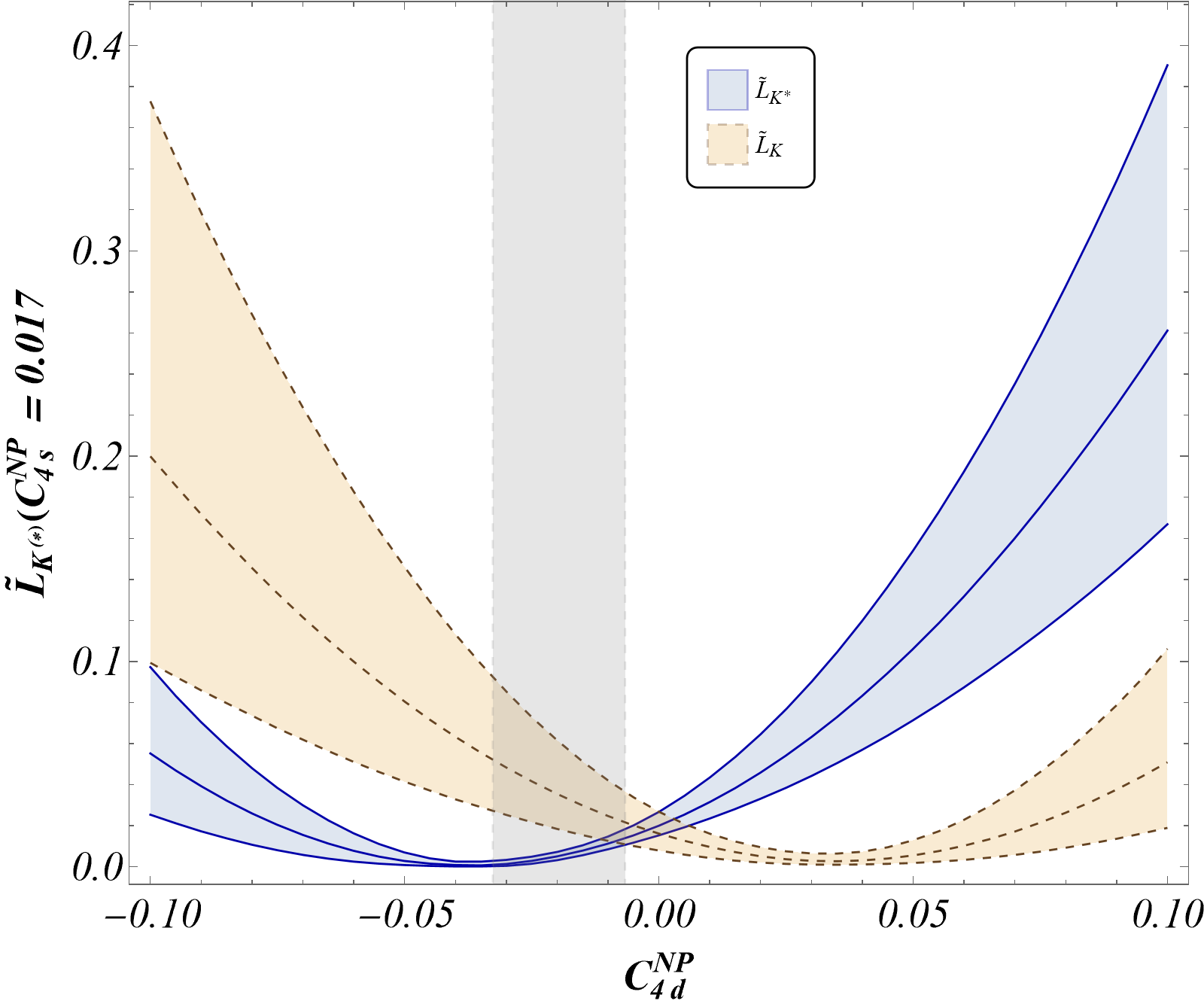}
\includegraphics[width=0.5\textwidth,height=0.435\textwidth]{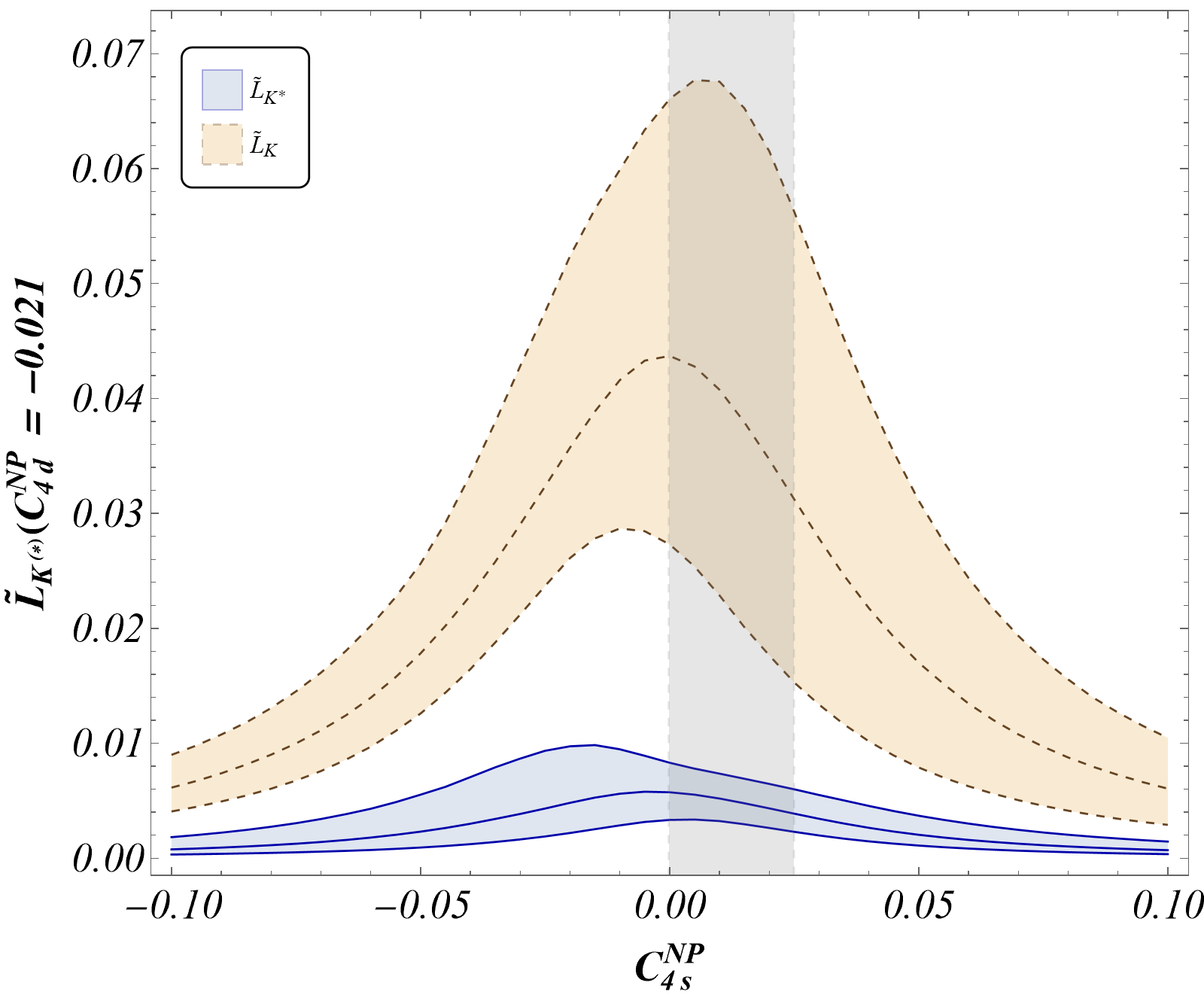}\\
\includegraphics[width=0.5\textwidth,height=0.435\textwidth]{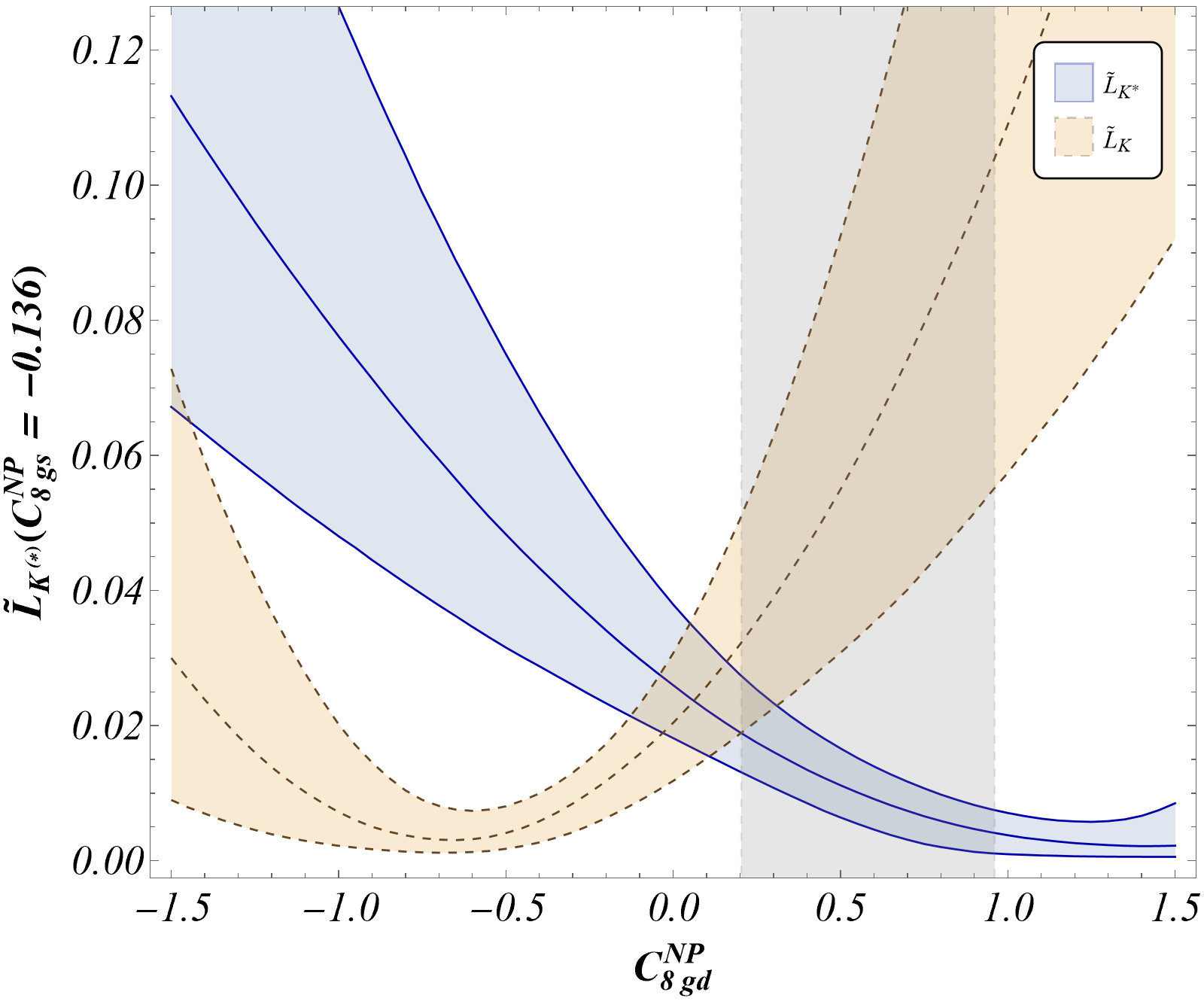}
\includegraphics[width=0.5\textwidth,height=0.435\textwidth]{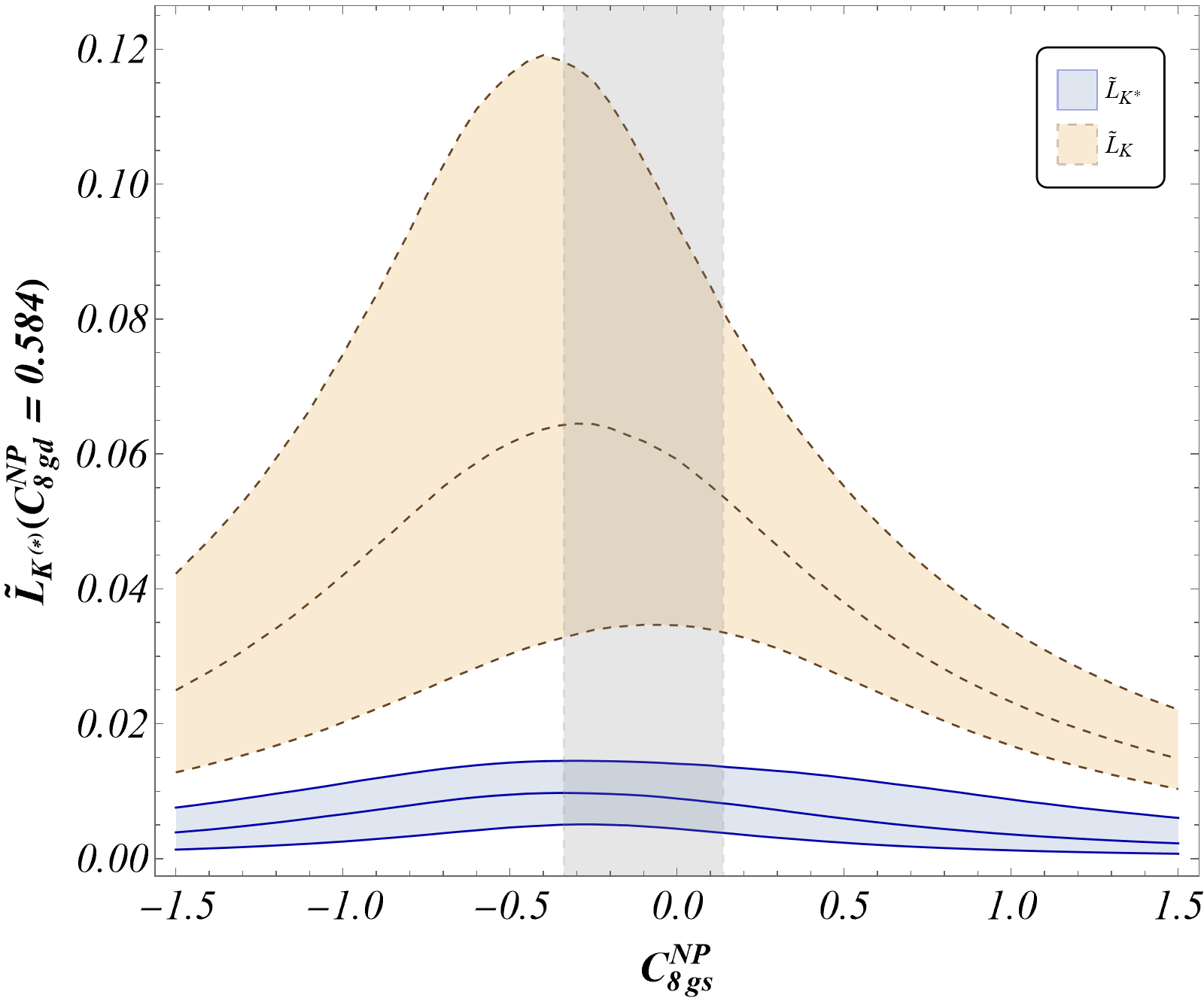}\\
\caption{$\tilde{L}_{K^{(*)}}$ observables as a function of  the NP coefficients $\ccal_{4d,4s,8gd,8gs}$. The top row displays the variation of $\tilde{L}_{K^{(*)}}$ w.r.t $\ccal_{4d}$ for a fixed value of $\ccal_{4s}$  (left), and vice versa (right) corresponding to the form factor values provided in rows 3  and 6 of \reftab{table6} (Only LCSR case). The second row follows the same order corresponding to the same form factors, but for $\ccal_{8gd,s}$. 
%The third row displays the dependence of these observables on $C_{4d,s}^{NP}$ in the same order, but corresponding to the form factor values provided in rows 4 and 7 of table~\ref{table2} (Lattice + LCSR) case. 
%The corresponding plots for $C_{8gd,s}^{NP}$ have not been displayed since there is practically no common region of overlap in the $C_{8gd,s}^{NP}$ plane for the Lattice + LCSR case (as can be seen from fig.~\ref{fig:fig2}). 
The value at which the non-varying NP Wilson coefficient is fixed in each case is displayed in the label for the vertical axis, and is chosen from the hatched magenta regions in Fig.~\ref{fig:figallowed}. Shaded regions correspond to the allowed values considering the constraints from the other observables for a fixed value of $\ccal_{4d,4s,8gd,8gs}$.}
%\caption{Only LCSR.}
\label{fig:fig3a}
\end{figure}
%%%%
%
But an important role is also played by a term inside the chirally enhanced term $a_6^c=\left({\cal C}_{6q}+{\cal C}_{5q}/3\right)\, \delta_{M_2P}+...$ that only contributes if the second meson $M_2$ (the one that does not collect the spectator quark) is a pseudoscalar. The SM contribution and the coefficient of $\ccal_{4d}$ has the same sign in $\tilde{L}_{K^*}$ this is because the chirally enhanced term $a_{6c}$ dominates the SM contribution of $a_{4c}$ and being subtracted from $a_{4c}$ interferes destructively with the SM contribution. As a consequence the SM and the prefactor of $\ccal_{4d}$ that had opposite signs inside $a_{4}^{c}$, due to the dominant $a_{6c}$ contribution becomes of the same sign. This does not happen for $\tilde{L}_K$. It is clear from this discussion that for the $\tilde{L}$ observables, the Wilson coefficients $\ccal_{4d,8gd}$ play a central role (compared to $\ccal_{4s,8gs}$)
and define the size of the enhancement, namely for $\ccal_{4d,8gd}\to 0$ the two $\tilde{L}$ observables tend to approach each other.
%
%
%%%%%%%% aqui 
%
%
\begin{figure}[!t]%\centering
\includegraphics[width=0.5\textwidth,height=0.435\textwidth]{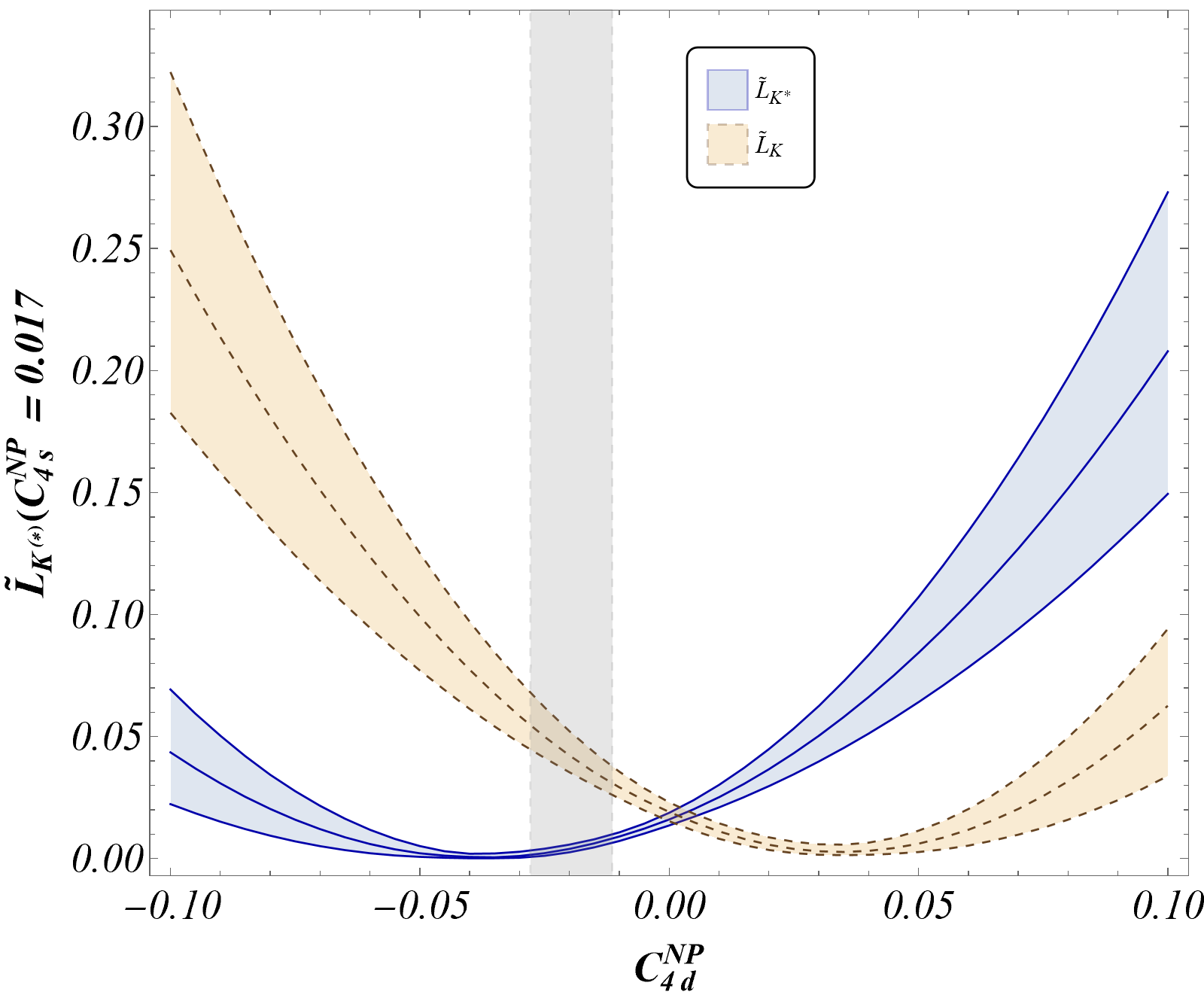}
\includegraphics[width=0.5\textwidth,height=0.435\textwidth]{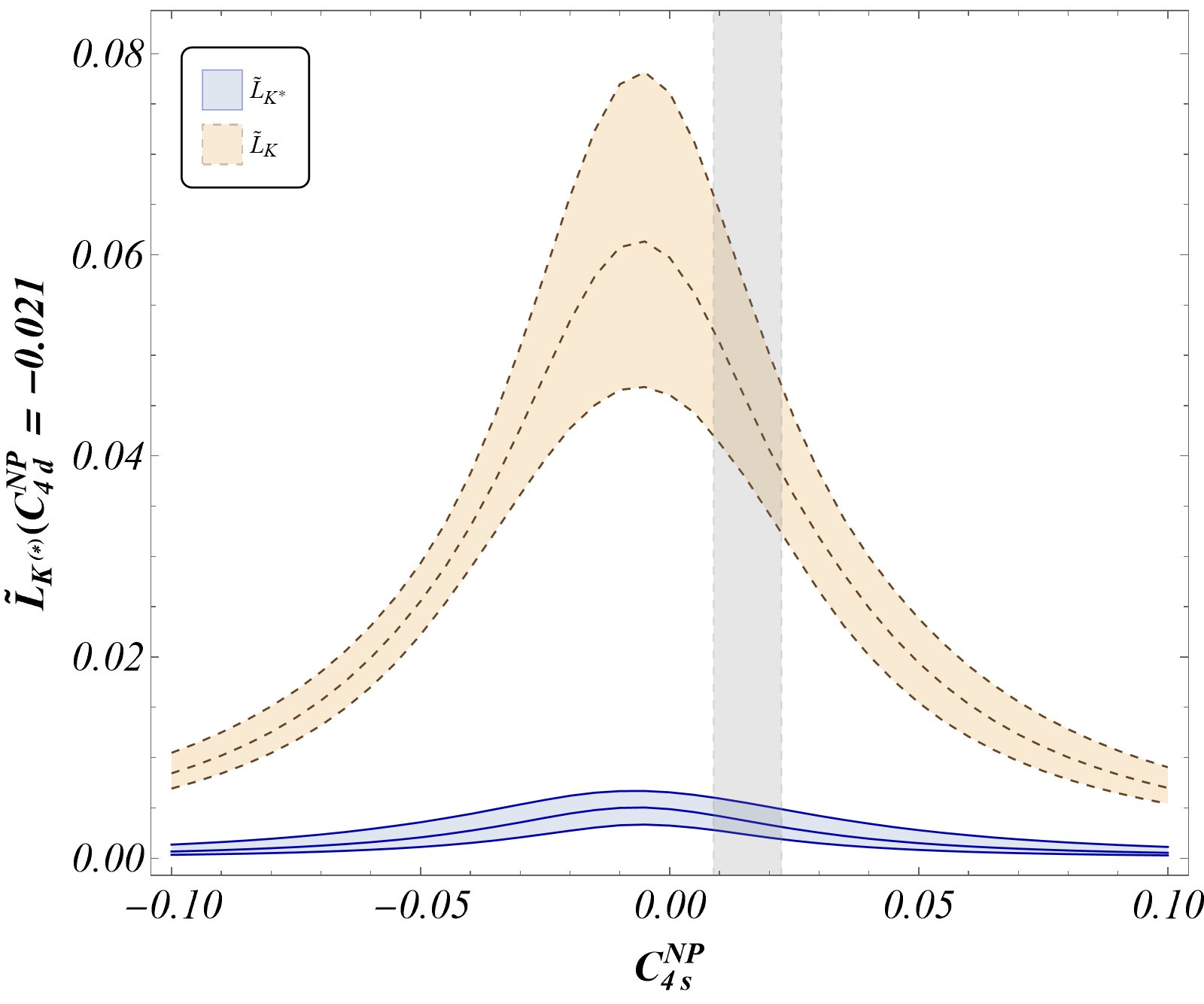}
\includegraphics[width=0.5\textwidth,height=0.435\textwidth]{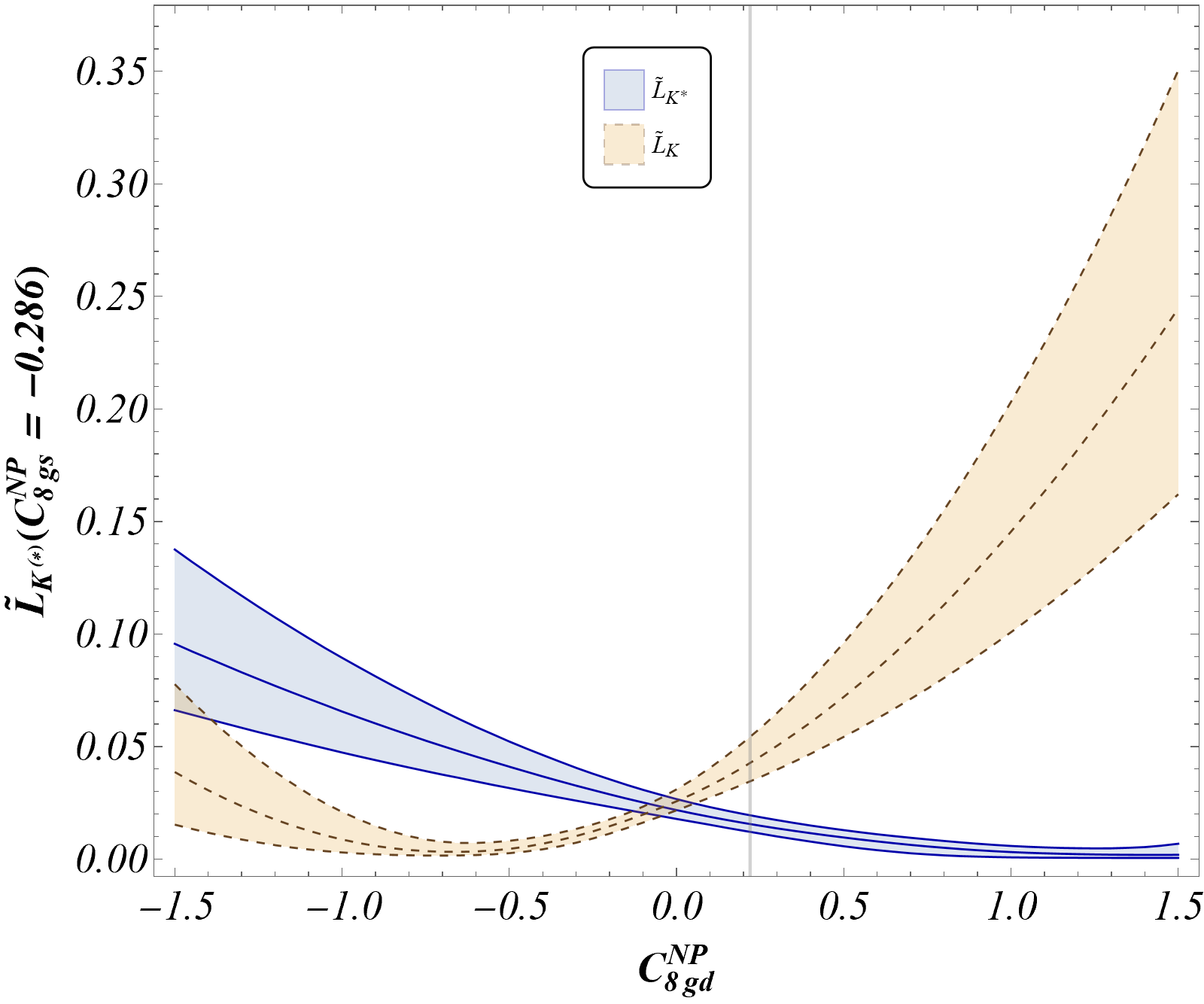}
\includegraphics[width=0.5\textwidth,height=0.435\textwidth]{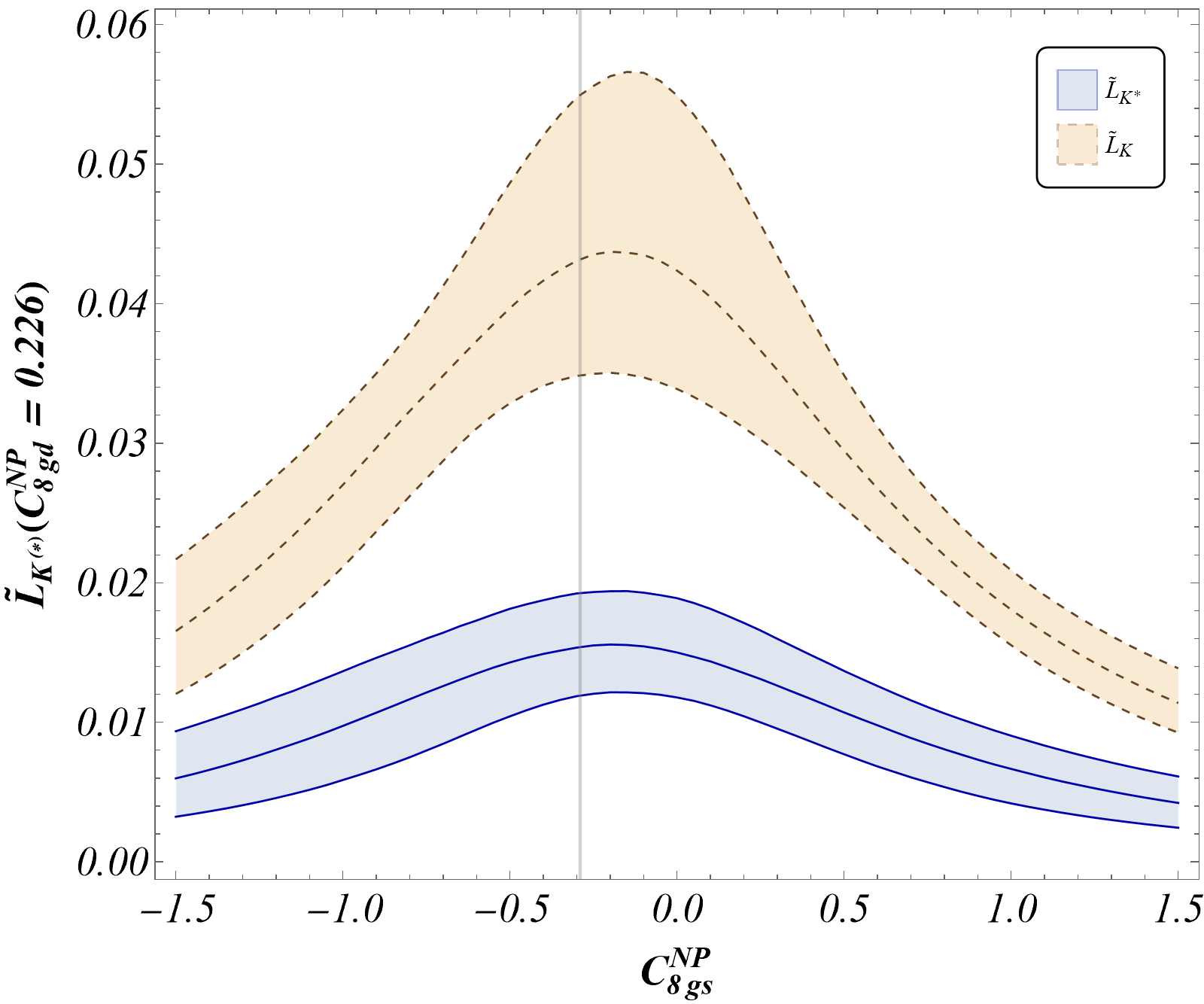}
\caption{$\tilde{L}_{K^{(*)}}$ observables as a function of the NP coefficients $\ccal_{4d,4s,8gd,8gs}$ using the LCSR+lattice determination. Same conventions as in Fig.\ref{fig:fig3a}.
%The value at which the non-varying NP Wilson coefficient is fixed in each case is displayed in the label for the y-axis, and is chosen from the hatched magenta regions in Fig.~\ref{fig:fig2}.
}
%\caption{Only LCSR.}
\label{fig:fig3b}
\end{figure}
In the following we explore the mechanism varying the value of $\ccal_{4d}$ ($\ccal_{4s}$) for fixed $\ccal_{4s}$ ($\ccal_{4d}$) and similarly for $\ccal_{8gd,8gs}$ given the very different role played by the $b\to d$ versus the $b\to s$ Wilson coefficients.
This is illustrated in Fig.~\ref{fig:fig3a} (using determination ``I") and Fig.~\ref{fig:fig3b} (using determination ``II"):

\begin{itemize}

\item \textbf{Mechanism seen as a function of $\ccal_{4d}$ ($\ccal_{8gd}$):} This is shown in the plots of $\ccal_{4d}$ versus the $\tilde{L}$-observables in Fig.~\ref{fig:fig3a} and Fig.~\ref{fig:fig3b}. 
The minima in the plots of $\ccal_{4d}$ in Fig.~\ref{fig:fig3a} and Fig.~\ref{fig:fig3b} correspond to the position of the minimum of $R_d/L_{\rm total}$ for $\tilde{L}_{K^*}$ and the minimum of the product of the denominators of $R_d$ and $L_{\rm total}$ for $\tilde{L}_K$. As a result, $\tilde{L}_K$ ($\tilde{L}_{K^*}$) is suppressed by one (three) order(s) of magnitude relative to the SM prediction when $\ccal_{4d} \sim -{\cal C}_{4d}^{\rm SM} \sim +0.04$ ($\ccal_{4d} \sim +{\cal C}_{4d}^{\rm SM} \sim -0.04$), due to cancellation with the SM contribution.
For other values of $\ccal_{4d}$, the $\tilde{L}$-observables follow approximately parabolic curves, and since the minima are nearly symmetric around zero, they intersect near $\ccal_{4d} \approx 0$ (in agreement with the explanation of the mechanism discussed above). In summary, this mechanism implies:
\begin{enumerate}
    \item If $\ccal_{4d} \simeq 0$, both $\tilde{L}$-observables approximately coincide.
    \item If $\tilde{L}_{K^*} \ll \tilde{L}_K$, then $\ccal_{4d} < 0$, while $\tilde{L}_{K^*} \gg \tilde{L}_K$ implies $\ccal_{4d} > 0$. This can be easily understood in the following way: given that the prefactor of $\ccal_{4d}$ and the SM contribution has the same (opposite) sign in $\tilde{L}_{K^*}$ ($\tilde{L}_K$), the sign of $\ccal_{4d}$ determines which observable is enhanced and which one is suppressed. 
\end{enumerate}
%Following the discussion on the mechanism it is clear that the sign of $C_{4d}^{\rm NP}$ 
%decides which observables is enhanced in front of the other.
Taking into account the data on $L_{{K}^*\bar{K}^*}$--$L_{K\bar{K}}$, the preferred scenario is $\ccal_{4d} < 0$, implying $\tilde{L}_{K^*} \ll \tilde{L}_K$. The enhancement can be substantial, especially when $\tilde{L}_{K^*}$ is strongly suppressed. We also observe that the closer $\ccal_{4s}$ is to zero for a non-zero $\ccal_{4d}$, the stronger the enhancement. Additionally, the second form factor determination yields more precise $\tilde{L}$-observables with significantly reduced uncertainties. A similar behaviour is observed for $\ccal_{8gd,s}$, with two relatively symmetric minima. However, due to the subdominant impact of $\ccal_{8gs}$, the minima are not aligned with the SM value of $\ccal_{8gd}$ but occur at a much larger $|\ccal_{8gd}|$ value because larger values of $|\ccal_{8gd}|$ are needed to counterbalance the SM contribution.
%%%%%%%%%%

%%%%%%%%%%%%
%

\item \textbf{Mechanism seen as a function of $\ccal_{4s}$ ($\ccal_{8gs}$):} Illustrated in the plots of $\ccal_{4s}$ versus the $\tilde{L}$-observables in Fig.~\ref{fig:fig3a} and Fig.~\ref{fig:fig3b}. Due to the asymmetric role played by 
$\ccal_{4d}$
  and 
$\ccal_{4s}$, the enhancement mechanism is at work only if 
$C_{4d}^{\rm NP}$
  is non-zero.
For each fixed $\ccal_{4d}<0$, both $\tilde{L}$-observables exhibit a maximum at similar values of $\ccal_{4s}$, typically around $\ccal_{4s} \sim 0$, though not exactly zero. At this point, the enhancement between $\tilde{L}_K$ and $\tilde{L}_{K^*}$ is maximal. For allowed values of $\ccal_{4d} < 0$, we find $\tilde{L}_K \gg \tilde{L}_{K^*}$, but the two maxima converge as $\ccal_{4d} \to 0$\footnote{Note that the SM predictions ($\ccal_{4d} = \ccal_{4s} = 0$) for the two $\tilde{L}$-observables
differ for a given choice of the form factor inputs. As a result, they cannot be expected to coincide at this point.}, as we have seen above.
For $\ccal_{8gs}$, a similar pattern is observed for fixed $\ccal_{8gd}$, though the enhancement is significantly smaller. Only for $\ccal_{8gd}$ values several times larger than the SM value does a notable enhancement appear; otherwise, the two observables are difficult to distinguish. For allowed values of $\ccal_{8gd}$--$\ccal_{8gs}$ (up to 300\% of the SM value), constrained by $L_{{K}^*\bar{K}^*}$--$L_{K\bar{K}}$, we find $\tilde{L}_K > \tilde{L}_{K^*}$. In the second form factor determination, which yields only a marginal set of allowed points for $\ccal_{8gd,s}$, a very specific prediction emerges (see Fig.~\ref{fig:fig3b}).

\end{itemize}

\section{Summary and conclusions}
\label{sec:conl}

This paper, organized in two main parts, explores the impact  of two different form factor computations, either including or excluding lattice QCD data, on a set of optimized non-leptonic observables, so called $L$-observables.

In the first part of this paper, 
we perform a novel combined LCSR analysis of the $B_d \to K^{(*)}$ and $B_s \to K^{(*)}$ form factors.
A key feature of this analysis is the inclusion of cross-correlations between these form factors, which are typically sizeable.
We also note that these results represent the first determinations for the $B_s \to K^{(*)}$ form factors obtained using LCSRs with $B$-meson distribution amplitudes.

We then combine our LCSR results with available lattice QCD computations using a simplified series expansion, namely the BSZ parametrization.
While the impact of the LCSR input on the $B_{d,s} \to K$ form factors is marginal (though not negligible), it is crucial for the $B_{d,s} \to K^*$ form factors in order to achieve precise predictions.
All results of our analyses are provided as ancillary machine-readable files attached to the arXiv version of this paper.

In the second part of the paper we explore the impact of the two different form factor determinations, using only LCSR (determination ``I") or combining LCSR with lattice QCD (determination ``II"), on a selected group of optimized non-leptonic observables. 
{For completeness, we also present results based solely on lattice QCD form factors (determination ``III"). While the outcomes that rely on the $B_{d,s} \to K$ form factors are aligned with the ones of determination ``II", those dependent on the $B_{d,s} \to K^*$ lattice form factors are subject to large uncertainties and cannot be fully trusted. Therefore, our analysis primarily focuses on determinations ``I" and ``II".}

In particular, we compute the $L_{{K}^*\bar{K}^*}$ and $L_{K\bar{K}}$ observables associated with the $\bar{B}_{d,s} \to \bar{K}^{(*)} K^{(*)}$ decays, along with two new observables, $\tilde{L}_{K^*}$ and $\tilde{L}_K$, which may become accessible in the near future. Within the SM framework, we explicitly demonstrate how the  $L_{{K}^*\bar{K}^*}$ and $L_{K\bar{K}}$ observables serve as probes of U-spin breaking, whose magnitude is strongly dependent on the chosen form factor determination.

Our main conclusions are as follows. The updated form factor determination  using only LCSR yields SM predictions for the $L_{{K}^*\bar{K}^*}$ and $L_{K\bar{K}}$ observables that are
in good agreement with our previous results. However, 
a significant increase in uncertainty is observed for $L_{K\bar{K}}$, where the uncertainty doubles compared to our earlier prediction that used lattice input for $f_0^{B_d \to K}$. In contrast, combining LCSR with lattice data leads to a dramatic shift in the SM predictions. 
Specifically, the tension between data and the SM prediction for $L_{K^*\bar{K}^*}$ increases to 4.4$\sigma$, while the tension for $L_{K\bar{K}}$ drops below 1$\sigma$. A summary of the SM predictions for the $L_{{K}^*\bar{K}^*}$ and $L_{K\bar{K}}$ observables using the different form factor determinations is provided in Table~\ref{tab:table6}.

Moreover, the particular choice of form factor determination has a significant impact on the sensitivity of these observables to NP. The allowed regions in the $\ccal_{4d} - \ccal_{4s}$ and $\ccal_{8gd} - \ccal_{8gs}$ parameter spaces shrink substantially when using the LCSR+lattice determination, to the extent that the $\ccal_{8gd} - \ccal_{8gs}$ solution becomes marginal.

We also identify a mechanism that leads to a relative enhancement between the newly defined observables $\tilde{L}_{K^*}$ and $\tilde{L}_K$ in the presence of NP. This enhancement depends on the specific values of $\ccal_{4d}$, $\ccal_{4s}$, $\ccal_{8gd}$, and $\ccal_{8gs}$ realized in Nature. The measurement of these new observables can not only help confirm or refute the picture suggested by the $L_{{K}^*\bar{K}^*}$ and $L_{K\bar{K}}$ observables, but also serve as a guide to discern which form factor determination is more appropriate. For example:
\begin{itemize}
    \item[i)] If $\tilde{L}_{K^*}$ is found to be smaller (larger) than $\tilde{L}_K$, this would point to $\ccal_{4d} < 0$ ($\ccal_{4d} > 0$) and/or $\ccal_{8gd} > 0$ ($\ccal_{8gd} < 0$), in agreement (tension) with the findings from $L_{{K}^*\bar{K}^*}$ and  $L_{K\bar{K}}$.
    \item[ii)] If both observables are measured to be nearly equal and deviate from the SM prediction  this would suggest $\ccal_{4d} \sim 0$ ($\ccal_{8gd} \sim 0$) and a non-zero $\ccal_{4s}$ ($\ccal_{8gs}$), but in tension with the $L_{K^*\bar{K}^*}$ and $L_{K\bar{K}}$-observables when combined with their corresponding branching ratios.
   
\end{itemize}

In summary, it is essential that experimental collaborations provide precise measurements of the new $\tilde{L}$-observables. These measurements are crucial not only for confirming or refuting the tensions observed in the $L_{{K}^*\bar{K}^*}$ and $L_{K\bar{K}}$ observables, but also for guiding the choice of the most reliable form factor determination, whether LCSR only or LCSR combined with lattice QCD.

Precise measurements of the new $\tilde{L}$-observables by experimental collaborations are crucial for two main reasons.
First, they are essential for confirming or refuting the tensions observed in the $L_{{K}^*\bar{K}^*}$ and $L_{K\bar{K}}$ observables, thereby serving as a valuable probe of potential NP.
Second, a comparison between predictions and measurements of the $\tilde{L}$-observables provides a meaningful test of our theoretical understanding of QCDF and form factor predictions at low $q^2$.
In particular, current and forthcoming lattice QCD results offer highly stringent predictions for the $\tilde{L}$-observables, which can be validated experimentally.

\section*{Acknowledgements}

The authors are grateful to M\'eril Reboud for helpful discussion about the \texttt{EOS} software and Davide Lancierini for enlightening discussions on the experimental prospects.
J.M. gratefully acknowledges the financial support from ICREA under the ICREA Academia programme 2018 and to AGAUR under the Icrea Academia programme 2024 and from Departament de Recerca i Universitats de la Generalitat de Catalunya. J.M. also received financial support from the Spanish Ministry of Science, Innovation and Universities (project PID2020-112965GB-I00) and from the IPPP Diva Award 2024.
\sloppy This work has been partially supported by STFC consolidated grants ST/T000694/1 and ST/X000664/1. G.T.X. received support for this project from the European Union’s
Horizon 2020 research and innovation programme under the Marie Sklodowska-Curie grant
agreement No 945422. This research was supported by the Deutsche Forschungsgemein-
schaft (DFG, German Research Foundation) under grant 396021762 - TRR 257.

%\item Can we get a handle on $A_0$ of $B_d$ and $B_s$ from a difference source? For instance, from semileptonic $B_{d,s} \to K^*\ell\ell$ in the SCET limit? 

\appendix
\section{Weak Effective Theory}\label{app:WET}

The effective Hamiltonian at the $m_b$ scale that describes the $b\to s,d$ transitions in non-leptonic B decays discussed in this work is
\begin{equation}\label{eq:wet}
H_{\rm eff}=\frac{G_F}{\sqrt{2}}\sum_{p=c,u} \lambda_p^{(s,d)}
 \Big({\cal C}_{1s,d}^{p} Q_{1s,d}^p + {\cal C}_{2s,d}^{p} Q_{2s,d}^p+\sum_{i=3 \ldots 10} {\cal C}_{is,d} Q_{is,d} + {\cal C}_{7\gamma s,d} Q_{7\gamma s,d} + {\cal C}_{8gs,d} Q_{8gs,d}\Big) \,,
\end{equation}
where $\lambda_p^{(s,d)}=V_{pb}V^*_{ps,d}$.  The relevant operators (with the corresponding Wilson coefficients) for the discussion in this paper are:
\begin{align}
 Q_{1f}^p &= (\bar p b)_{V-A} (\bar f p)_{V-A} \,, % & Q_{7s} &= (\bar f b)_{V-A} \sum_q\,\frac{3}{2} e_q (\bar q q)_{V+A} \,, \nonumber \\[-2.2mm]
% Q_{2f}^p &= (\bar p_i b_j)_{V-A} (\bar f_j p_i)_{V-A} \,, & Q_{8f} &= (\bar f_i b_j)_{V-A} \sum_q\,\frac{3}{2} e_q (\bar q_j q_i)_{V+A} \,, \nonumber \\[-2.2mm]
% Q_{3f} &= (\bar f b)_{V-A} \sum_q\,(\bar q q)_{V-A} \,, &Q_{9f} &= (\bar f b)_{V-A} \sum_q\,\frac{3}{2} e_q (\bar q q)_{V-A} \,, \nonumber \\[-2.2mm]
 &Q_{4f} &= (\bar f_i b_j)_{V-A} \sum_q\,(\bar q_j q_i)_{V-A} \,, %& & %Q_{10f} &= (\bar f_i b_j)_{V-A} \sum_q\,\frac{3}{2} e_q (\bar q_j q_i)_{V-A} 
 \,
 \nonumber\\[-2.2mm]
% Q_{5f} &= (\bar f b)_{V-A} \sum_q\,(\bar q q)_{V+A} \,, &Q_{7\gamma f} &= \frac{-e}{8\pi^2}\,m_b\bar f\sigma_{\mu\nu}(1+\gamma_5) F^{\mu\nu} b \,,\nonumber \\[-2.2mm]
 Q_{6f} &= (\bar f_i b_j)_{V-A} \sum_q\,(\bar q_j q_i)_{V+A} \, , &Q_{8gf} &= \frac{-g_s}{8\pi^2}\,m_b\, \bar f\sigma_{\mu\nu}(1+\gamma_5) G^{\mu\nu} b \,, \nonumber
%\label{operators}
\end{align}
with $f=s,d$ and where $Q_{1f}^p$ is a current-current operator, $Q_{4f,6f}$ are QCD penguin operators and $Q_{8gf}$ is the chromomagnetic dipole operator. 
In the above $(\bar q_1 q_2)_{V\pm A}=\bar q_1\gamma_\mu(1\pm\gamma_5)q_2$, 
$i,j$ are colour indices, $e_q$ are the electric charges of the quarks in units of $|e|$ and 
a summation over $q=u,d,s,c,b$ is implied. 
We follow the conventions and definitions of Ref.~\cite{Beneke:2001ev} (see in this reference the complete set of operators).

\bibliographystyle{JHEP}
\bibliography{references}

\end{document}